\documentclass[pdflatex,sn-mathphys]{sn-jnl}

\jyear{2022}%

\usepackage{epstopdf}
\usepackage{longtable}
\usepackage{blindtext,alltt}
\usepackage{graphicx,psfrag,epsf}
\usepackage{amsmath,amssymb,graphicx,multirow}
\usepackage{bbm}
\usepackage{mathrsfs}
\usepackage{booktabs}
\usepackage{color}
\usepackage{xcolor}
\usepackage{comment}
\usepackage{adjustbox}
\usepackage{mathtools}
\usepackage{subcaption}
\allowdisplaybreaks

\theoremstyle{plain}

\theoremstyle{definition}

\theoremstyle{remark}

\theoremstyle{plain}
\newtheorem{Assumption}{Assumption}[section]

\definecolor{mypink}{rgb}{0.858, 0.188, 0.478}

\DeclarePairedDelimiter{\ceil}{\lceil}{\rceil}

\begin{document}

\title[mCube: Multinomial Micro-level reserving Model.]{mCube: Multinomial Micro-level reserving Model.}

\author*[1]{\fnm{Emmanuel Jordy} \sur{Menvouta}}\email{emmanueljordy.menvoutankpwele@kuleuven.be}

\author[2]{\fnm{Jolien} \sur{Ponnet}}\email{jolien.ponnet@nbb.be}
\equalcont{These authors contributed equally to this work.}

\author[3]{\fnm{Robin} \sur{Van Oirbeek}}\email{robin.vanoirbeek@gmail.com}
\equalcont{These authors contributed equally to this work.}

\author[1,4]{\fnm{Tim} \sur{Verdonck}}\email{tim.verdonck@uantwerpen.be}
\equalcont{These authors contributed equally to this work.}

\affil*[1]{KU Leuven, Department of Mathematics, Section of Statistics and Data Science, Leuven, Belgium}

\affil[2]{National Bank of Belgium}

\affil[3]{DKV Belgium}

\affil[4]{University of Antwerp, Department of Mathematics, Section of Applied Mathematics, Antwerp, Belgium}

\abstract{This paper presents a multinomial multi-state micro-level reserving model, denoted mCube. We propose a unified framework for modelling the time and the payment process for IBNR and RBNS claims and for modeling IBNR claim counts. We use multinomial distributions for the time process and spliced mixture models for the payment process. We illustrate the excellent performance of the proposed model on a real data set of a major insurance company consisting of bodily injury claims. It is shown that the proposed model produces a best estimate distribution that is centered around the true reserve.}


\maketitle
\section{Introduction}
\label{sec:intro}
A central part of an insurance company is the management of its future cash flows and  solvency capital. To this end, insurers have to set aside reserves to cover outstanding claims liabilities. After an insured event has occurred, it always takes some time to settle the final payment of a claim. Taking into account two different sources of delay in general insurance, insurers typically set aside separate reserves for Incurred But Not Reported claims (IBNR) and Reported But Not Settled claims (RBNS). The number of RBNS claims is known, together with specific information relative to each claim. To determine the reserve for RBNS claims, the aim is to predict the amount that still needs to be paid. For IBNR claims, the insurer does not have information about each specific claim or the total number of these claims. The goal when modelling IBNR reserves is firstly to estimate the number of these claims, and secondly, to estimate the cost for each claim. 

 New regulations such as the Solvency II and IFRS frameworks guide insurance companies towards best practices for the calculation of their reserves. Following these regulations, it is important that models for claims reserving not only predict accurately the ultimate reserve amount but also the distribution of future cash-flows conditional on currently available information. More information on the calculation of insurance reserves can be found in \cite{solv2}.

A first class of models developed for the task of claims reserving are collective or macro-level models that focus on aggregated data organized in a so-called run-off triangle (often on an annual or quarterly basis). Popular macro-level models include the chain-ladder method \cite{mack1993} and the Bornhuetter–Ferguson method \cite{BornFerg1972}. These methods have been successfully applied for decades due to their ease of use and sound theoretical foundations (see for example \cite{englandverrall2002,wuthrich2008stochastic}). 
Furthermore, many extensions have been developed to produce more realistic results (e.g. \cite{liu2009predictive,merz2010paid,QuargMack2008,verdonck2011influence}). 

However, aggregating data may lead to several problems and yields a loss of information. Therefore, micro-level or individual claims reserving models focus on granular or claim-specific data of the individual claims.
They aim to model two processes: a time process representing the individual states occupied by a claim, and a payment process which represents the amount paid for a claim in a particular state. The earliest articles on micro-level reserving include \cite{arjas_1989} and \cite{Norberg1993}. The modelling ideas from the early papers have been extended in a (semi-)parametric form \cite{pigeon2013,antonioplat2014}, as well as a non-parametric form \cite{duval2019,delong2020}. 

In this paper, we adopt the multi-state approach to loss reserving proposed by \cite{Hach1980} and further considered by \cite{hesselager1994}, \cite{Antonio2016}, and \cite{Bettonville2020}. In particular, we
\begin{itemize}
    \item introduce a new model for IBNR claim counts, based on the multinomial distribution;
    \item make a connection between the time process modelling and the multi-state competing risk framework;
    \item use a semi-parametric modelling of the payment distribution, through a mixture distribution with a Generalised Pareto Distribution (GPD) for the tails;
    \item include practical recommendations on how to apply the proposed method to any micro-level data set. 
    \item { compare the predictive capabilities of our proposed method with other micro-reserving models.}
\end{itemize}
\vspace{\baselineskip}

The remainder of this paper is organised as follows. Section \ref{sec:claimsreservingproblem} presents the claim reserving problem and in  Section \ref{sec:IBNR} we construct a model for IBNR claims based on the multinomial distribution. The models for the time and payment processes based on the multinomial distribution, are the subject of Section \ref{sec:RBNS}. In Section \ref{sec:hyperparameteropt}, the hyper-parameter tuning for the models is presented and a numerical example on real data is investigated in Section \ref{sec:casestudy}. Finally, the main findings and suggestions for further research are summarized in Section \ref{sec:conclusion}.

\section{The claims reserving problem}
\label{sec:claimsreservingproblem}
\begin{figure}[h]
\centering
\includegraphics[width=0.7\textwidth]{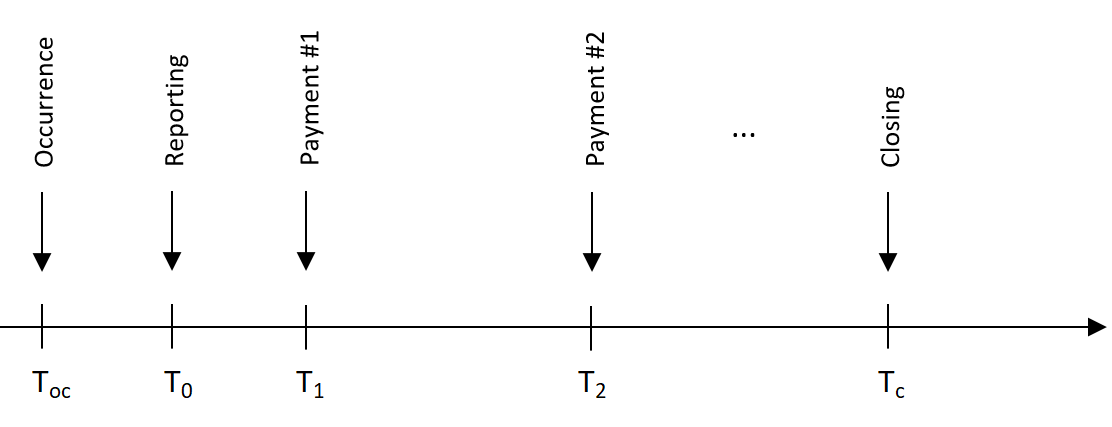}
\caption{Claim development process.
}
\label{fig:payCycles}
\end{figure}

The development of a non-life insurance claim is presented in Figure \ref{fig:payCycles}. 
The occurrence date, $T_{oc}$,  is the date at which the claim event occurs and the reporting date, $T_{0}$, refers to the date at which the claim is reported to the insurer.
Once the insurer is aware of the claim and accepts the claim for reimbursement, some payments at different moments (here represented by $T_{1}$ and $T_{2}$) follow to compensate the insured for their loss. Once the insurance company reimburse the complete loss covered by the policy, the claim closes, which is represented by $T_{c}$. Note that we assume that once a claim is closed, it cannot be reopened. 

At the moment of evaluation (commonly: end of a quarter, mid year or end of book year), denoted $\tau$, insurance companies have to set reserves aside to cover their future liabilities. These liabilities can come from three sources, which are enumerated below. 
\begin{itemize}
    \item Claims that have occurred before the evaluation period but which are not yet reported to the insurer, i.e. $T_{oc} \leq \tau < T_{0}$. These are called Incurred But Not Reported (IBNR) claims.  
    \item  Claims that have occurred and have been reported before the evaluation period, but  which are still open at evaluation date, i.e. $T_{0} \leq \tau < T_{c}$. These are called Reported But Not Settled (RBNS) claims.
    \item Claims that have been closed before the evaluation date, but might get reopened. 
\end{itemize}
\vspace{\baselineskip}

Let us assume that we work on a sufficiently rich probability space $(\Omega, \mathcal{F}, \mathbb{P})$ and denote $C_{k,t}$ as the random variable representing the cumulative amount paid for claim $k$ at time $t$. Next, we want to predict the reserve at evaluation time $\tau$, or in other words, the remaining amount to be paid for the claim until it is closed. 
This can be estimated by  $\hat{\mathbb{E}}[{R}_{k,\tau} \mid \mathcal{F}_{\tau}]$, denoted by $\hat{R}_{k,\tau}$, which is obtained by the difference of $\hat{C}_{k,T_{c}}$ and {$C_{k,\tau}$}. Here, ${F}_{\tau}$ represents the information available at time $\tau$  and $\hat{C}_{k,T_{c}}$ equals the estimated total cost of the claim at closing time, which can be obtained by $\hat{\mathbb{E}}[{C}_{k,T_{c}} \mid \mathcal{F}_{\tau}]$.

Note that for IBNR claims, we have that ${C}_{k,\tau}$ equals zero and hence, $\hat{R}_{k,\tau} = \hat{C}_{k,T_{c}}$. Finally, the estimated reserve for the whole portfolio is calculated as follows: 
\begin{equation}
    \hat{R}_{\tau} = \sum_{k_{1} =1}^{ \mathbf{n}^{RBNS}} \hat{R}_{k_1,\tau} + \sum_{k_{2} =1}^{ \hat{\mathbf{N}}^{IBNR}} \hat{R}_{k_2,\tau},
    \label{eq: best estimate cost}
\end{equation}
with $\mathbf{n}^{RBNS}$ representing the number of RBNS claims and $\hat{\mathbf{N}}^{IBNR}$
the estimated number of IBNR claims. The goal of this paper is to determine $\hat{R}_{\tau}$, the estimated reserve of the whole portfolio, which is also called the best estimate cost of the portfolio.

\section{Multinomial IBNR model}
\label{sec:IBNR}
Suppose an insurer has aggregated data on reported claims that occurred in accident year $i = 1,\ldots,I$, with $I$ the year of evaluation, and reported in development year  $j= 0, \ldots,J-1$. This data is then typically represented in a table with the accident years as rows and  the development years as columns. From the reported claims at time $\tau = I$, we obtain the upper triangle generating information $\mathcal{N}_{\tau} = \sigma\{ N_{i,j}; 0 \leq i\leq \tau, j \geq 0, i+j \leq \tau\}$. The goal of this section is to develop a model to determine $\hat{\mathbf{N}}^{IBNR}$, which represents the estimated number of IBNR claims based on the information $\mathcal{N}_{\tau}]$. In order to define this model, some notation needs to be introduced first. %

Denote $\pi_{i,j}$ as the probability that a claim occurred in year $i$, will be reported $j$ years later. By assuming that no accident is reported beyond the last development year (no tail factor), we have the following multinomial probability vector for each accident year $i$: $\pi_i = (\pi_{i,0}, \ldots, \pi_{i,J-1})$. Next, $N_{i}$ represents the number of claims that occurred in accident year $i$ and $N_{i,j}$ the number of these claims that have been reported after $j$ years. Note that since we assume that no claims will be reported after the last development year, we have $N_{i} = \sum_{j=0}^{J-1}N_{i,j}$. The number of observed claims that occurred in year $i$ is written by $N_i^{obs}$, and since a claim can only be observed once reported, we have that $N_{i}^{obs} = \sum_{j=0}^{I- i} N_{i,j}$. Finally, the number of IBNR claims that occurred in year $i$ is given by $N_i^{IBNR} = \sum_{j=I-i+1}^{J-1} N_{i,j}$, such that the number of IBNR claims over all accident years is given by
\begin{equation}
    N^{IBNR} = \sum_{i=1}^{I} N_{i}^{IBNR} = \sum_{i=1}^{I} \sum_{j=I-i+1}^{J-1} N_{i,j}
    \label{eq: N_IBNR}
\end{equation} 
Hence, $N_i^{IBNR}$ equals $N_i - N_i^{obs}$. 

\begin{Assumption}
\label{assumptionIBNR}

 We assume that stationarity holds, e.g. $\pi_{1} = \pi_{2} = \ldots = \pi_{I}$. Moreover, we assume that conditional on $N_{i}$, we have that $(N_{i,0}, \ldots, N_{i,J-1})$ follows a multinomial distribution with event probabilities $\pi_{i}$ and $N_{i}$ number of trials. 
A final assumption is that, conditional on the number of observed claims, the number of IBNR claims follows a negative binomial distribution with parameters $p_{i} = \sum_{j=0}^{I-i} \pi_{i,j}$ and $r_{i} = N_{i}^{obs} = \sum_{j=0}^{I-i} N_{i,j}$. In other words, $N_{i}^{IBNR} \mid N_{i}^{obs} \sim \text{NegBinom} (r_{i},p_{i})$, such that $E[N_{i}^{IBNR}\mid N_{i}^{obs}] = r_{i}(1-p_{i})/p_{i}$ and $V[N_{i}^{IBNR}\mid N_{i}^{obs}] = r_{i}(1-p_{i})/p_{i}^2$. Note that the negative binomial distribution expresses the distribution of the number of failures in a sequence of Bernoulli trials before $r_i$ successes are reached, with $p_i$ being the probability of success. In this case, a success coincides with reporting.
\end{Assumption}

As shown by  \cite{schmidt1998}, these specifications are consistent with the chain-ladder  \cite{mack1993} since the predicted number of IBNR claims have the same  expected value. We can build a predictive distribution for the number of yearly IBNR claims by repeating the following steps a sufficiently large number (e.g., 100) of times:
\begin{enumerate}
    \item Estimate $\pi_{1} = (\pi_{1,0}, \pi_{1,1},…, \pi_{1,J-1})$ by its maximum likelihood estimator, denoted by $\hat{\pi}_1$. The other probabilities ${\pi}_k$, $k\in \{2,\ldots,I\}$, are by the stationary assumption also estimated as $\hat{\pi}_k=\hat{\pi}_1$. 
    \item Standardise the empirical multinomial probabilities such that the probabilities for the  unobserved development years sum to 1. This implies that if there are $k$ unobserved development years for accident year $i$, their corresponding multinomial probabilities are given by
    
    \begin{equation}
    (\tilde{\pi}_{i,J-k},\ldots, \tilde{\pi}_{i,J-1}) = \left(\frac{\hat{\pi}_{i,J-k}}{\sum_{l=1}^{k} \hat{\pi}_{i,J-l}},\ldots, \frac{\hat{\pi}_{i,J-1}}{\sum_{l=1}^{k} \hat{\pi}_{i,J-l}} \right).
\end{equation}

    \item Sample the estimated yearly number of IBNR claims, ${\hat{N}_{i}^{IBNR}}$, using each accident year specific negative binomial distribution. 
    \item To obtain the estimated number of IBNR claims  for the unobserved development years and accident year $i$, $ \{\hat{N}_{i,I-i+1}, \ldots, \hat{N}_{i,J-1}\} $, sample from  a multinomial distribution with parameters $n=N_{i}^{IBNR}$ and  $p=(\tilde{\pi}_{i,I-i+1},\ldots, \tilde{\pi}_{i,J-1})$.
\end{enumerate}

Following Assumption \ref{assumptionIBNR}, it holds that $\hat{N}^{IBNR} = \sum_{i=1}^{I} r_{i}(1-\hat{p}_{i})/\hat{p}_{i}$. The simulations are only used to build a predictive distribution or to construct confidence intervals. We note that the aim of this model for IBNR claim count is not to propose a model that captures aspects that are not captured in the framework of Over-Dispersed Poisson for the chain-ladder. Rather, the aim is to propose a parametrization which uses the multinomial distribution, as this multinomial distribution is the central theme of the current article.

\section{Models for individual claims}

\label{sec:RBNS}
 In this section, the details of the proposed multinomial micro-level reserving model are discussed. This model will be used to determine the estimated reserves $\hat{R}_{k,\tau}$ for each claim $k$, such that we are able to obtain the best estimate cost of the portfolio based on Equation \eqref{eq: best estimate cost}. We start by a detailed explanation of the multi-state approach, followed by a discussion of the models that are selected for the time and payment process. 

\subsection{Multi-state approach}
\begin{figure}[htbp!]
\centering
\includegraphics[width=\textwidth]{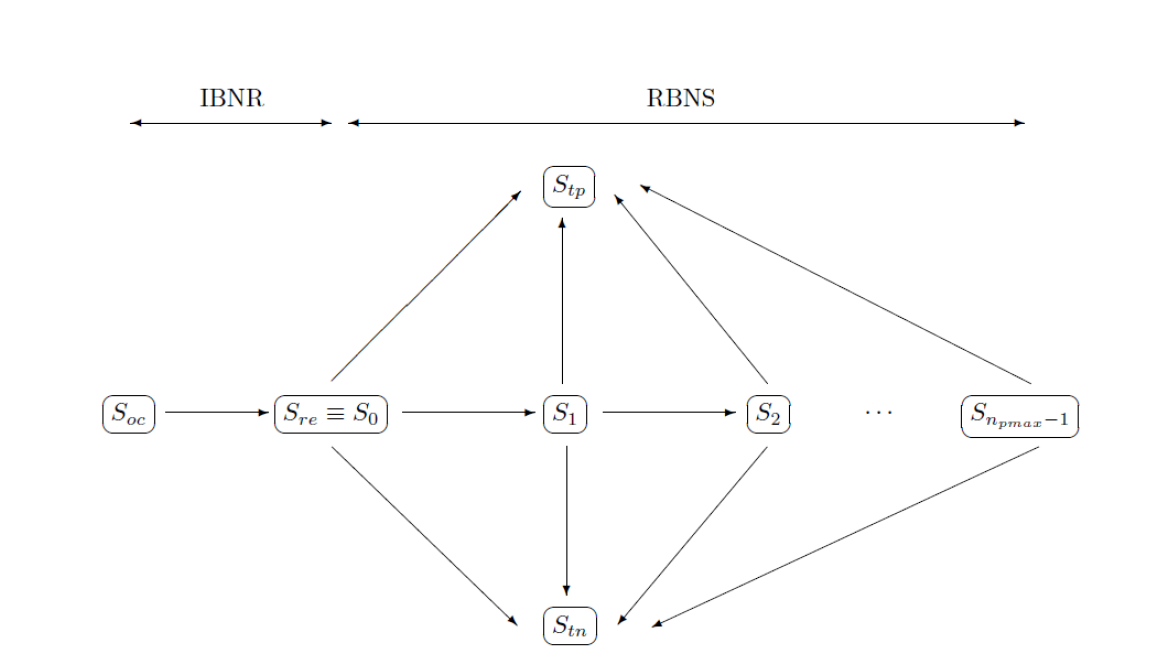}
\caption{Discrete time multi-state model, source:  \cite{Antonio2016}.}
\label{fig:RBNSTrans}
\end{figure}

This section uses the multi-state approach of  \cite{Bettonville2020} represented in Figure \ref{fig:RBNSTrans}. In this approach, an RBNS claim occured in state $S_{oc}$ and is reported in state $S_{0}$. Once reported, either a first payment can occur, implying a transition from state $S_{0}$ to state $S_{1}$, or the claim can go to one of two absorbing states, $S_{tn}$ or $S_{tp}$.  Here, $S_{tn}$ represents the fact that the claim went to a terminal state without payment, and $S_{tp}$ represents that the claim went to a terminal state with payment. In this framework, the states $S_{j}$, $ \{j \in \{0, n_{pmax}-1\}\}$ are all strictly transient and moreover, for $j>0$, state $S_j$ implies $j$ payments were made prior to the current time point. The integer $ n_{pmax}$ represents the maximum number of transitions a claim is allowed before being forced to an absorbing state. More specifically, we represent the multi-state model $(\mathscr{S},\mathscr{T})$ with state space $\mathscr{S}=\{S_{0},S_{1},\ldots,S_{n_{pmax}-1},S_{tn},S_{tp}\}$ and set of direct transitions $\mathscr{T}$. An event corresponds to the transition from one state in $\mathscr{S}$ to another. The set of direct transitions $\mathscr{T}$ defines all possible transitions in the multi-state model, indicated by arrows in Figure \ref{fig:RBNSTrans}. One advantage of individual claims reserving models is that they allow us to take into account covariate information such as the history of incremental payments, line of business, reporting delay, as well as any other type of information which is available to describe individual claims and their development. We denote by $\mathcal{F}_{k,T}$ the filtration containing all the information concerning claim $k$ at time $T$. In a multi-state framework, $C_{k,T_{c}}$ can be computed by determining the next states the claim will occupy and  summing the amount paid in these future states together with the amount already paid. 

\subsection{ Multinomial model for the time process}
\label{sec:timeprocess}
We use the methodology from discrete time survival analysis, and model the time  until an event or transition, from one state to the other. We define an event as being the occurrence of a payment or as the transition to a terminal state without payment as in  \cite{hesselager1994} and  \cite{Bettonville2020}. Furthermore, we say that a claim is censored or open when it is not in one of the two absorbing states at the moment of evaluation. The choice of discretization for  the multi-state model of the previous section is arbitrarily chosen  to be monthly, with one month corresponding to 30 days. The proposed model can be adapted to work with any discretization but the number of parameters increase with the granularity considered for the data.

 Our goal is to model  the time $T_{k,j}$ of the transition of claim $k$ from a state $S_{j}$ to a state $S_{j+1}$, $ j \in \{  0,1,\ldots,n_{pmax}-2\}$ or to a terminal state, using covariate information included in $\mathcal{F}_{k,t}$, through the covariate vector $\mathbf{x_{k,t}}$. If we denote by $\Delta_{k,j}$ the random and discrete censoring time of claim $k$ in state $S_{j}$, we have the following assumption: 
 \begin{Assumption}
 \label{ass:censoring}
 We assume that $T_{k,j}$ and $\Delta_{k,j}$ are independent and that the censoring mechanism is non-informative. 
 \end{Assumption}

\begin{figure}[htbp!]
\centering
\includegraphics[width=\textwidth]{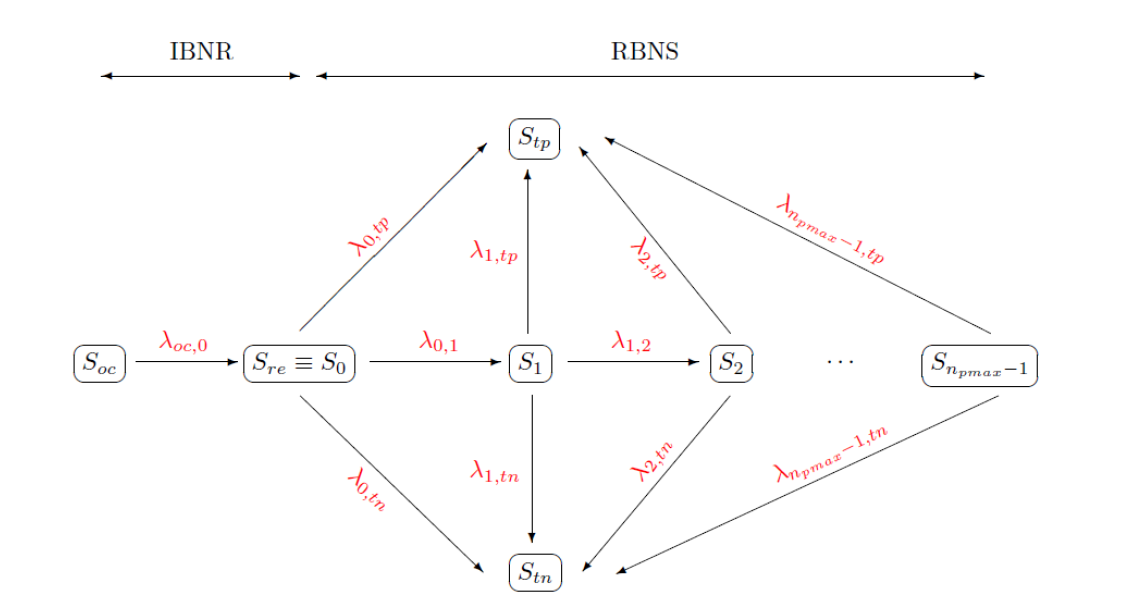}
\caption{Representation of transition probabilities in the multi-state model, source:  \cite{Antonio2016}.}
\label{fig:transprobs}
\end{figure}

Based on discrete-time competing risks literature  \cite{Allison1982DiscreteTimeMF}, we represent the event type as a random variable $\epsilon_{k,j}$, taking values $P$, $TP$, $TN$ corresponding respectively to a transition due to a payment, a terminal payment, or a termination without payment. The discrete-time cause-specific hazard functions are then given by: 
 \begin{align}
 \label{timeprocmultinom}
        \lambda_{j,j+1}(t \mid \mathbf{x_{k,t}}) &= \mathbb{P}(T_{k,j} = t, \epsilon_{k,j} = P \mid T_{k,j} \geq t,\mathbf{x_{k,t}} ) \\
        &= \frac{\exp(\alpha_{j,j+1} + \mathbf{\beta}_{j,j+1}^{T} \mathbf{x_{k,t}})} {1 + \sum_{e} \exp(\alpha_{j,e} +\mathbf{\beta}_{j,e}^{T} \mathbf{x_{k,t}})}, \nonumber\\
         \lambda_{j,{tp}}(t \mid \mathbf{x_{k,t}}) &= \mathbb{P}(T_{k,j} = t, \epsilon_{k,j} = TP \mid T_{k,j} \geq t, \mathbf{x_{k,t}}) \nonumber\\
        &= \frac{\exp(\alpha_{j,TP} + \mathbf{\beta}_{j,TP}^{T} \mathbf{x_{k,t}})} {1 + \sum_{e} \exp(\alpha_{j,e} +\mathbf{\beta}_{j,e}^{T} \mathbf{x_{k,t}})} ,\nonumber\\
          \lambda_{j,{tn}}(t \mid \mathbf{x_{k,t}}) &= \mathbb{P}(T_{k,j} = t, \epsilon_{k,j} = TN \mid T_{k,j} \geq t, \mathbf{x_{k,t}}) \nonumber\\
        &= \frac{\exp(\alpha_{j,TN} + \mathbf{\beta}_{j,TN}^{T} \mathbf{x_{k,t}})} {1 + \sum_{e} \exp(\alpha_{j,e} +\mathbf{\beta}_{j,e}^{T} \mathbf{x_{k,t}})} \nonumber,\\
         \lambda_{j,{j}}(t \mid \mathbf{x_{k,t}}) &= \mathbb{P}(T_{k,j} > t \mid T_{k,j} \geq t,\mathbf{x_{k,t}} )\nonumber\\
         &= 1 -  \lambda_{j,j+1}(t \mid \mathbf{x_{k,t}}) -  \lambda_{j,{tp}}(t \mid \mathbf{x_{k,t}}) - \lambda_{j,{tn}}(t \mid \mathbf{x_{k,t}}), \nonumber 
    \end{align}
where $e$ iterates over all event types. $\alpha_{j,e}$ and $\mathbf{\beta}_{j,e}^{T}$ are  the parameters in the multinomial regression model relating to event type $e$. Following Assumption \ref{ass:censoring}, the parameters are estimated by their maximum likelihood estimator\footnote{using the function \texttt{multinom} in the package \texttt{nnet}  \cite{nnet2002} in \texttt{R}}. For a claim that has occured but has not yet been reported, only one event in the multi-state process is possible, namely reporting. Similarly to  \cite{Bettonville2020}, we estimate the monthly probability of reporting using a binomial Generalized Linear Model (GLM):
\begin{align}
\label{eq:repDel}
\lambda_{oc,0}(t \mid \mathbf{x_{k,t}}) &= \mathbb{P}(T_{k,oc} = t \mid  T_{k,oc} \geq t, \mathbf{x_{k,t}})\\
&= \frac{1}{ 1 + \exp(\alpha_{oc} +\mathbf{\beta}_{oc}^{T} \mathbf{x_{k,t}})}, \nonumber
\end{align}
with $\alpha_{oc}$  and $\mathbf{\beta}_{oc}^{T}$ the logistic regression parameters estimated by their maximum likelihood estimator.
Note that, $T_{k,j}$ denotes the time period at which the claim moves out of state $S_{j}$ and is reset to 0 each time the claim enters a new non-absorbing state. We treat each discrete time unit as a separate observation in the data set. Hence, for a claim $k$ in state $S_{j}$, there are as many lines as the number of  time units the claim is in this state. A representation of the different cause-specific hazards is shown in Figure \ref{fig:transprobs}. 

\subsection{Modelling of the payment distributions}
\label{sec:paymentmodels}
The difficulties in modelling the payment distribution arise from some stylised properties. First of all, negative payments can be present. For example, when the insurance company has to pay a third party and the insured has an insurance policy with a per-loss deductible of $d$, she has to pay $d$ to the insurance company.   Moreover, the payment distribution consists of a high number  of small incremental payments, and a small number of very large payments in absolute value. Multiple models have been proposed to overcome these issues:  \cite{antonioplat2014} use a lognormal distribution to model payments,  \cite{Bettonville2020} use a mixture of  lognormal distributions for the first payment and a mixture of lognormal or lognormal and pareto distributions for the link ratios,  \cite{FreesValdez2008} use a generalised beta distribution of the second kind, and  \cite{pigeon2013} use a multivariate extension of the univariate skew normal distribution.  \cite{Reynkens2017} propose to model censored losses using a mixed Erlang distribution for the body of the distribution and a Generalised Pareto Distribution (GPD) for the tail. The authors also propose to use the mean excess plot  \cite{beirlant2005} to assess when to split the body and the tail of the distribution. This estimation procedure has the advantage of taking into account both random censoring and truncation.
We propose to model the payment distribution  using a data-driven modification of  \cite{montuelle2014}, which allows for the inclusion of covariate information and which can model the skewness of the distribution. Let $Y^{j}$ denote the random variable representing the payment size for a claim in state $S_{j}$. Moreover, we make the following assumption on the conditional distribution of $Y^{j}$ given $\mathbf{x_{t}}$.

\begin{figure}[h]
\centering
\includegraphics[width=0.8\textwidth]{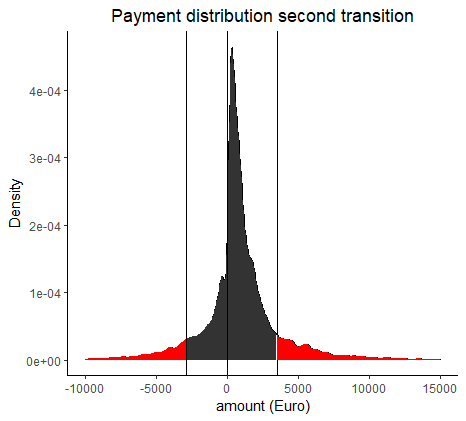}
\caption{Spliced payment distribution with three splitting points. } 
\label{fig:splicing}
\end{figure}

\begin{Assumption}
\label{ass:paydist}
We assume that the density of $Y^{j}$ conditional on $\mathbf{x_{t}}$ is a $L$-component mixture i.e. $f(y \mid x)= \sum_{l=1}^{L} \pi_{l}^{j}(x) f_{l}^{j}(y)$, where $\pi_{l}^{j}(x)$ is the $lth$ element of the covariate-dependent vector of multinomial logistic mixture weights and $f_{l}^{j}$ are the densities of the mixture components. We further assume that $L$ is known, $f_{1}^{j}$ and $f_{L}^{j}$ are  densities of a Generalized Pareto Distribution (GPD) and $f_{l}$ for $ l \in \{2,L-1\}$ are truncated normal distributions on the interval $[b_{l-1}^{j}, b_{l}^{j}[$. In this case, $b_{1},\ldots, b_{L-1}$ represent the splitting points separating the  density into bins $\mathcal{B}_1^{(j)}, \ldots, \mathcal{B}_L^{(j)}$.
\end{Assumption}
 A representation of Assumption \ref{ass:paydist} is shown in Figure \ref{fig:splicing} with $L = 4$ bins.

 The splitting points $b_{2}^{j},\ldots, b_{L-2}^{j}$ can be chosen freely so that each bin has some interpretation. In practice, four splitting points can be chosen as shown in figure \ref{fig:splicing} to represent small or large negative payments, as well as small or large positive payments. The leftmost and rightmost splitting points, respectively $b_{1}^{j}$ and  $b_{L-1}^{j}$,  need to be well chosen in order to assure that the observations of these bins can be considered to be a sample from a GPD with $b_{1}^{j}$, $\iota_{1}^{j}$ and $\varphi_{1}^{j}$ ($b_{L-1}^{j}$, $\iota_{L}^{j}$ and $\varphi_{L}^{j}$) as the location, scale and shape parameter for $ \mathbf{B}_{1}^{j}$ ($ \mathbf{B}_{L}^{j}$). To this end, we can use tools from extreme value theory such as the mean excess plot or the Gerstengarbe plot  \cite{Beirlant2004}. Following Assumption \ref{ass:paydist}, the expected payment for a claim $k$ in state $S_{j}$ conditional on its covariate vector $\mathbf{x_{k,t}}$ is given by 
 
 \begin{align}
 \label{eq:splicing}
     \mathbb{E}[Y^{j} \mid \mathbf{x_{k,t}}] &= \sum_{l = 1}^{L} \pi_{l}^{j}(\mathbf{x_{k,t}}) \mu_{l}\\
     \pi_{l}(\mathbf{x_{k,t}}) &=  \frac{\exp({\gamma}_{0,l}^{(j)} +  \mathbf{x_{k,t}}^{T}{\gamma}_{l}^{(j)})}{1 + \sum_{m=1}^{L} \exp({\gamma}_{0,m}^{(j)} +  \mathbf{x_{k,t}}^{T}{\gamma}_{m}^{(j)}) }, \nonumber
 \end{align}
 
 with $\mu_{l}$ the mean of the $lth$ component in the mixture. The parameter vectors ${\gamma}_{0}^{(j)}$ and ${\gamma}^{(j)}$ are estimated using maximum likelihood\footnote{using the function \texttt{multinom} in the package \texttt{nnet}  \cite{nnet2002} in \texttt{R}}. Also the parameters $\mu_{l}, \ldots, \mu_{L}$ are obtained using maximum likelihood estimation. Hence, we can express the expected cumulative amount paid for claim $k$ at closure as
\begin{equation}
\label{totalPay}
  {C}_{k,T_{c}} = \sum_{j: S_{j} \notin \{ S_{tn}, S_{tp}\} }  \mathbb{E}[Y^{j} \mid \mathbf{x_{k,T_{k,j}}}].
\end{equation}

\subsection{Total cost simulation for RBNS claims}
\label{sec:openclaims}
Similarly to  \cite{Delong2021}, we choose a one-period-ahead forecast to determine the estimated reserves $\hat{R}_{k, \tau}$ for each open claim $k$ as it allows us to intervene during the settlement process and helps us keep interpretability in the claims development process. In the modelling of the time process \eqref{timeprocmultinom}, the next state to visit is decided by a multinomial probability vector.  Instead of assigning  an open   claim to the state with the highest multinomial probability, we take a sample of size one from the estimated multinomial distribution. Next, the expected payment for a claim in that state and with a given covariate vector is simulated using \eqref{eq:splicing}. All this has the advantage of adding variability to the predictive distribution. We note that using the expected value of the payments results in smoothed inputs in the simulation of RBNS total cost as in \cite{Delong2021} and \cite{Gabrielli2021}. However, this is acceptable in our case since we are mostly interested in expected future payments and the expected total cost of the claim at closure as explained in the introduction. Another strategy could be to take into account the full distribution of future payments, by simulating an observation from a mixture component chosen based on the probabilities in \eqref{eq:splicing}. 

Repeating these multinomial samplings  ${N_{sim}}$ times, we obtain a predictive distribution for $C_{k,T_C}^{r}$, the total payment of claim $k$, and its reserve ${R}_{k,\tau}^{r}= C_{k,T_C}^{r} - C_{k,\tau}  $ with $r \in \{1, \ldots, N_{sim}\}$. 

\begin{figure}[h]
\centering
\includegraphics[width=0.8\textwidth]{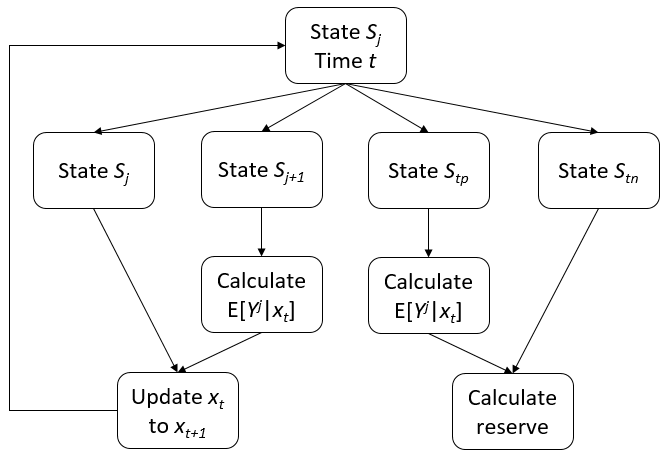}
\caption{RBNS total cost simulation strategy. }
\label{fig:rbnsstrat}
\end{figure}

\subsection{Total cost simulation for IBNR claims}
\label{sec:IBNRclaims}
For IBNR claims, an extra step is added compared to RBNS claims, since we need to take into account the reporting delay. For each IBNR claim, this extra step consists in taking a sample of size one from a Bernoulli distribution using the estimated reporting probabilities \eqref{eq:repDel}. If this Bernoulli sample is 1, the RBNS total cost simulation strategy is applied. If the sample equals 0, then the covariate vector is updated and the claim remains in state $S_{oc}$. The Bernoulli sampling procedure is repeated until the claim leaves the state $S_{oc}$.

\section{Hyper-parameter optimization}
\label{sec:hyperparameteropt}
The flexibility of the proposed models lead to multiple hyper-parameters that need to be set prior to fitting the time and payment model. In this section, we explain the role of each hyper-parameter, as well as our tuning strategy. Furthermore, we want to alleviate the linearity assumption of the continuous variables that are used in the various regression models of mCube. To this end, each continuous variable will be binned into categories using the strategy that will be explained in the next section. A binning strategy is applied such that the proposed non-complex model can still capture complicated patterns.

\subsection{Feature engineering}
\label{sec:feature engineering}
Individual claims data contain both static information such as the line of business, claim type and injured body part as well as dynamic information such as the cumulative payments and dates, that evolve throughout the claim's life. Table \ref{table:dynamicInfo} shows a sample of the dynamic information available in an individual claim. 

\begin{table}[!htbp]
\centering
\begin{tabular}{ccccccc}
\hline\hline
 PolNumb& cumPay & bookDate & accDate & repDate & Status & closedDate  \\
 \hline
  2640440& 4,087.61& 09-01-2012& 01-01-2012& 02-01-2012& O &28-08-2012\\
  2640440& 4,127.11 & 10-01-2012 & 01-01-2012& 02-01-2012& O &28-08-2012\\
  2640440 & 7.12 & 02-02-2012 & 01-01-2012& 02-01-2012& O& 28-08-2012\\
  2640440& 297.12& 07-02-2012& 01-01-2012& 02-01-2012& O& 28-08-2012\\
  2640440 & 297.12& 28-08-2012& 01-01-2012& 02-01-2012& C& 28-08-2012\\
  \hline\hline
\end{tabular}
\caption{Example of the dynamic claim information available from the database. \label{table:dynamicInfo} }
\end{table}

Since mCube requires that all time varying variables are transformed into their respective values at the end of the fixed discrete time steps, the raw data set shown in Table \ref{table:dynamicInfo} needs to be processed. More specifically, in our case, we have set 30 days as our fixed time step ('perLen') and for example for the cumulative payment we record its value at the end of the subsequent time step. If this value differs in absolute value more than \textit{"minPayVal"} from previously recorded payment, we consider that a payment was performed and the claim will experience a transition from the current state. In case that the claim resides in $S_0$, the previously recorded payment equals 0. As a result, transType will be set to P, TN or TP, depending on the transition type. In case that no payment has occurred, transType will be equal to N. See Table 2 for how the information in Table 1 will be transformed given the above described rules.

Next we will work out how all commonly present time varying variables are transformed to comply to mCube. These variables can be considered as the minimal set of variables that are present in data set used for any micro-level reserving model and this set can be augmented by other time-varying variables that happen to be recorded by the insurance company at hand. As we work. As we work in a  discrete time setting with a chosen period length (\textit{"perLen"}), we denote the event times for claim $k$, as $t_{0}^{k} \leq t_{1}^{k}<t_{2}^{k} \ldots \leq t_{Q}^{k}$ with $t_{0}$ representing the accident date, $t_{1}$ the reporting date, $t_{2}, \ldots, t_{Q-1}$ the payment dates and  $t_{Q}$ representing the closing date. If  $i \geq 1$ and  $t$ is such that $t_{i}^{k} \leq t < t_{i+1}^{k}$, we create the following variables for claim $k$: 
\begin{itemize}
    \item $ x_{k,t}^{1} = \max\left( 1, \ceil*{\frac{t_{1}^{k} - t_{0}^{k}}{perLen}} \right)$,  to represent the reporting delay (deltRep). 
    \item $x_{k,t}^{2} = \mathbbm{1}_{t_{1}^{k} = t_{0}^{k}}$,  to a represent a fast reporting indicator (fastRep).
    \item $x_{k,t}^{3} = \max\left( 1, \ceil*{\frac{t - t_{1}^{k}}{perLen}} \right)$, to represent the time since reporting (inProcTime). 
    \item $x_{k,t}^{4}  = y_{i-1}^{k}$, the payment at time $t_{i-1}^{k}$, with $i \geq 3$ (delt1Pay). 
    \item $x_{k,t}^{5}  = \ceil*{\frac{t- t_{i-1}^{k}}{perLen}}$, with $ i \geq 3$ for the time since the previous payment (delt1PayTime). 
    \item $x_{k,t}^{6} = \sum_{\{s : t_{s}^{k}< t\}} y_{s}^{k}$, for the cumulative payments up to time $t$ (cumDel1tPay). 
    \item $x_{k,t}^{7}  =  x_{k,t}^{3} \mathbbm{1}_{\{i = 1\}} + x_{k,t}^{5} \mathbbm{1}_{\{i>1\}}$ for the time spent in the current state (inStateTime). 
\end{itemize}
Hence, at each time $t$, we have the claim's feature information vector $\mathbf{x_{k,t}} = (x_{k,t}^{1}, x_{k,t}^{2}, x_{k,t}^{3}, x_{k,t}^{4}, x_{k,t}^{5}, x_{k,t}^{6}, x_{k,t}^{7}, \mathbf{x}_{base})$ with $\mathbf{x}_{base}$ representing the remaining static information of the claim. When we want to model the payment distribution, we also add as covariate, an indicator of if it is a terminal payment or not as this information is always available before estimating a payment. Using the accident, reporting, booking or settlement date, additional features such that the day, month, quarter or financial year during which any of these events have occurred can be engineered. These would then help us capture calendar effects more easily. For the sake of simplicity, none of these seasonal variables were included in our analysis, even if there is no technical limitation to not do so. Table \ref{table:dynamicInfoUpdate} contains the transformed database which contains any of the proposed transformations of this section.
\begin{table}[!htbp]
\centering
\begin{tabular}{ccccccc}
\hline\hline
 \texttt{polNumb}& cumPay & bookDate & accDate & repDate & transType & closedDate  \\
 \hline
  2640440& 4,127.11 & 10-01-2012 & 01-01-2012& 02-01-2012& P &28-08-2012\\
  2640440& 297.12& 07-02-2012& 01-01-2012& 02-01-2012& P& 28-08-2012\\
   2640440& 297.12& 08-03-2012& 01-01-2012& 02-01-2012& N& 28-08-2012\\
    2640440& 297.12& 07-04-2012& 01-01-2012& 02-01-2012& N& 28-08-2012\\
     2640440& 297.12& 07-05-2012& 01-01-2012& 02-01-2012& N& 28-08-2012\\
      2640440& 297.12& 06-06-2012& 01-01-2012& 02-01-2012& N& 28-08-2012\\
    2640440& 297.12& 06-07-2012& 01-01-2012& 02-01-2012& N& 28-08-2012\\
  2640440 & 297.12& 28-08-2012& 01-01-2012& 02-01-2012& TN& 28-08-2012\\
   & & & & & &\\
    deltRep& fastRep & procTime & deltPay & cumDeltPay & stateTime & state\\
    \hline
      1& 0 & 1 & NA& NA& 1 &$S_{0}$\\
       1& 0 & 2 & 4,127.11& 4,127.11& 1 &$S_{1}$\\
        1& 0 & 3 & -3829.99& 297.12& 1 &$S_{2}$\\
         1& 0 & 4 & -3829.99& 297.12& 2 &$S_{2}$\\
          1& 0 & 5 & -3829.99& 297.12& 3 &$S_{2}$\\
           1& 0 & 6 & -3829.99& 297.12& 4 &$S_{2}$\\
            1& 0 & 7 & -3829.99& 297.12& 5 &$S_{2}$\\
             1& 0 & 8 & -3829.99& 297.12& 6 &$S_{2}$\\
  \hline\hline
\end{tabular}
\caption{Example of the dynamic claim information available from the database. \label{table:dynamicInfoUpdate} }
\end{table}

\subsection{Binning continuous predictors}
We use a modified version of the data-driven strategy to bin continuous variables of  \cite{Henckaerts2018}, where the authors propose to fit a Generalized Additive Model (GAM) where the covariate effects of the continuous variables are fitted using cubic splines. Next, the spline estimates of the continuous predictors are binned using an evolutionary regression tree  \cite{JSSv061i01}. Therefore, we make the following adaptation to the algorithm of  \cite{Henckaerts2018} : 

\begin{enumerate}
    \item First a sufficiently large bootstrap sample is taken from the data set. We recommend to sample between 50,000 to 100,000 observations for each bootstrap sample and to only take a limited number of bootstrap samples. In our case study, we have used a sample size per bootstrap of 100,000 and 10 bootstraps samples were taken.
    \item In each bootstrap repetition, and for each continuous predictor, we split each continuous variable into 40 groups where the split points are the 0.025, 0.05,\ldots, 0.95, 0.975 quantiles. 
    For each group, the median value of the continuous variable of interest is chosen as the group representative or mediod.
    \item We then fit a multinomial regression similar to \eqref{timeprocmultinom} in which the variable $x_{k,t}^{7}$ (inStateTime), and the medioids of interest obtained in the previous step are used as the predictors and the transition type is chosen as a response.
    \item For each hazard function, the corresponding multinomial parameter estimates for each group representative are used as responses in a local regression (loess). The predictors of the local regression are the medioids. As such, a covariate estimate is obtained for each value of the considered continuous variable, instead of just its mediods only. Note that the approach described from step (2) to (4) can also be replaced by using a spline estimate for the considered continuous variable in the multinomial model of step (3), instead of what is currently proposed, however this requires a much longer fitting time than the current proposal and the end result of both approaches was found to be relatively similar.
    \item For each hazard function, a regression tree is then fitted on the obtained parameters to obtain a set of \textit{"nGroups"} -1 splitting points for the continuous variable.
    \item In the final step, the splitting points of the different hazard functions are merged, as to obtain a single set of splitting points that is used to bin the considered continuous variable in the same way for each considered transition function. As such, we don't need to define a separate binned version of the considered continuous variable for each hazard function. We proceed by merging and ordering the splitting points of all 3 hazard functions.
\end{enumerate}

Note that we impose that each bin obtained by this method has at least \textit{"nMinLev"} observations. 
The choice of the hyper-parameters \textit{"nGroups"}, \textit{"nGroupsFin"} and  \textit{"nMinLev"} is discussed in appendix D.

\subsection{Hyper-parameters} 

Due to the high flexibility of the time and payment models of mCube, multiple hyper-parameters need to be set prior to fitting any of the models. Due to the computational complexity of mCube and the large number of hyper-parameters and possible values for these hyper-parameters, a search in the hyper-parameter space is not feasible. We propose to choose values for the hyper-parameters based on patterns present in the data and business logic. We refer to Appendix C for more details on this matter. 

\section{Case study}
\label{sec:casestudy}

In this section, the mCube  is applied on a real data set of a major insurance company in order to explore its performance.

\subsection{Data}
\label{sec:data}
The data used in this section is a random sample of the set of claims obtained obtained from a European insurer and resulting in a total of 25,821 body injury claims occurring between 2006 and 2012, of which the latest evaluation moment was December 31, 2018, some of the claims still being opened at that time. To anonymize the payment data, we have multiplied all payment by a non-disclosed constant value. Furthermore, we adjust all payments for inflation. All the information that is present prior to and including December 31, 2012, is equal to the training set. Our aim is to predict the remaining cost of every open claim at the end of December 31, 2012 until December 31, 2018. The set of all such open claims corresponds to our entire test set. Since we have all information up to December 31, 2018 and since we have 6 years of information for each claim, the true cost evaluated at December 31, 2012, is known for all claims in our test set. 

\begin{figure}[!htbp]
\begin{subfigure}{.45\textwidth}
    \centering
    \includegraphics[width=.85\linewidth]{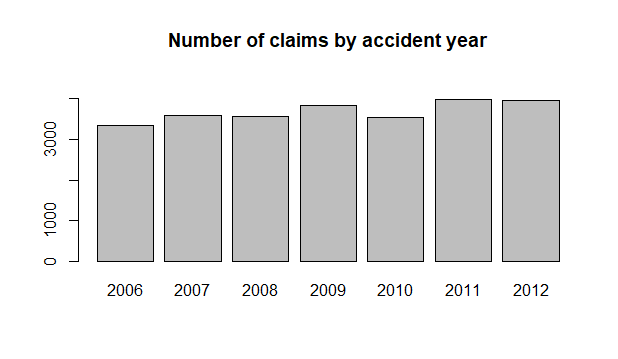}
    \caption{}
    \label{fig:AY}
\end{subfigure}
\begin{subfigure}{.45\textwidth}
    \centering
    \includegraphics[width=.85\linewidth]{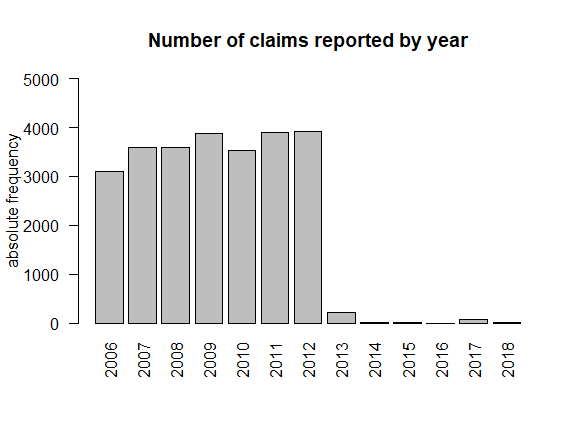}
    \caption{}
    \label{fig:repYear}
\end{subfigure}
\newline
\begin{subfigure}{.45\textwidth}
    \centering
    \includegraphics[width=.85\linewidth]{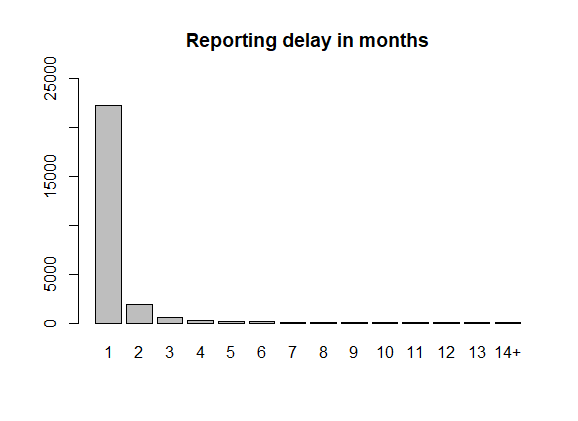}
    \caption{}
    \label{fig:repDel}
\end{subfigure}
\begin{subfigure}{.45\textwidth}
    \centering
    \includegraphics[width=.85\linewidth]{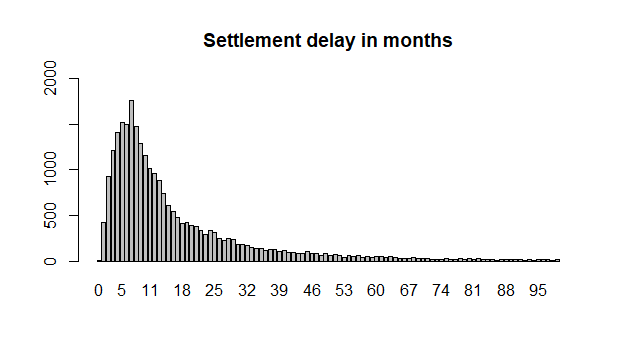}
    \caption{}
    \label{fig:setDel}
\end{subfigure}
\caption{ Distribution of the number of claims per accident and reporting year, together with the distribution of the reporting and settlement delay.} 
\end{figure}

From Figures \ref{fig:AY}, \ref{fig:repYear}, \ref{fig:repDel} and \ref{fig:setDel}, obtained from the entire data set, we observe that the number of accidents is stable over the different accident years and most claims are reported in the month in which they happened. The claim settlement distribution is right skewed with most claims being settled within 2 years. 
Table \ref{table:absrelfreq} shows the distribution of the total number of transitions for claims in the data set. We observe that only around 13\% of claims have more than 6 transitions. 
Furthermore, we observe that most claims have 2 or 3 transitions in the multi-state process and about 64\% of all selected claims have 3 transitions or less, hence a substantial part of the data set consists of claims with a long and relatively complicated development pattern, which is rather typical for bodily injury claims. Table \ref{table:timeInState} shows the distribution of the time spent in each state. We note that observations for states greater than $5$ are lumped together in a state $S_{5+}$ due to the small number of observations in each of the individual data sets and given that only a small percentage of the claims have more than 6 transitions (see also Table \ref{table:absrelfreq}). We observe that claims spend on average between 5 and 7 months (30-day periods) in each state. Furthermore, states $S_{0}$ and $S_{5+}$ are the states  with the lowest average time spent and also the most skewed. The reason being that these data sets have the most diverse types of claims.

\begin{table}[!htb]
\centering
\begin{tabular}{c c c c c c c c }
\hline\hline
& \multicolumn{7}{c}{\hspace{3ex} \textbf{Total number of transitions}}\\
\cline{2-8}
& $1$ & $2$  & $3$  & $4$  & $5$ & 6& $>6$\\
\hline
abs.freq & 2,275& 7,429&  7,250&  3,125&  1,481&  848& 3,413\\
rel.freq & 8.81 \% & 28.77 \%&  28.08 \%&  12.10 \% &  5.74 \%&  3.28\% & 13.22\%\\
  \hline\hline
\end{tabular}
\caption{Absolute and relative frequencies of the total number of transitions for each claim.\label{table:absrelfreq}}
\end{table}

\begin{table}[!htbp]
\centering
\begin{tabular}{c c c c c c c c }
\hline\hline
State & Min & Median & Mean & Max & IQR & Skewness & Kurtosis\\
\hline
$S_{0}$ &1.00  & 2.00 &  5.36 & 72.00 &  5.00 &  3.36 & 14.46  \\ 
  $S_{1}$ & 1.00 &  3.00 &  5.89 & 72.00 &  5.00 &  3.07 & 12.55  \\ 
  $S_{2}$& 1.00 &  4.00 &  7.02 & 72.00  & 6.00  & 2.82 & 10.14 \\ 
  $S_{3}$& 1.00  & 4.00 &  6.44 & 72.00 &  6.00 & 2.59 &  8.50 \\ 
  $S_{4}$ & 1.00 &  3.00 &  6.16 & 61.00 &  6.00 &  2.68 & 9.07 \\ 
  $S_{5+}$ & 1.00 &  3.00 &  4.64 & 61.00&   4.00 &  3.10 & 12.91 \\ 
  \hline\hline
\end{tabular}
\caption{Summary statistics for the time spent in each state.} \label{table:timeInState} 
\end{table}

Table \ref{table:payInState}  shows the payment distribution in each sate. From this table, we can see how complicated payment distributions are, due to the presence of, namely, the presence of negative payment amounts, small median payment amounts, right skewed distributions, and very large excess kurtosis. Furthermore, states $S_0$ and $S_{5+}$ have by far the highest values for skewness and kurtosis, which underlines yet again the heterogeneity present in the claim that are in both states.
\begin{table}[!htbp]
\centering
\begin{tabular}{c c c c c c c c }
\hline\hline
State & Min & Median & Mean & Max & IQR & Skewness & Kurtosis\\
\hline
$S_{0}$ & -18,365.69 &  1,149.30 &  2,808.84 & 343,463.26 &  2,360.00  &   22.35 &   679.28  \\ 
  $S_{1}$ & -45,968.06  &  662.06 &  1,390.16& 198,923.65 &  1,747.04  &   10.19 &   256.32\\
  $S_{2}$& -58,154.22 &   808.90 &  2041.66  &165,782.32 &  1,936.61 &     8.89  &  131.32\\
  $S_{3}$& -53,078.64  &  924.56 & 2,549.96 & 273,638.48  &  2,291.40  &    8.67  &  154.79 \\
  $S_{4}$ & -33,892.65  & 1,013.95 &  2,550.49 & 97,253.20  & 2,622.99  &    5.19  &   57.49 \\
  $S_{5+}$ & -123,556.32  &   938.34  &  4,088.08 & 586,846.69  &  2,932.84  &    16.84     &472.04\\ 
  \hline\hline
\end{tabular}
\caption{Summary statistics for the payment distribution in each state.} \label{table:payInState} 
\end{table}

\subsection{Time models evaluation}
We will now evaluate the performance of the time models on the training data. Table \ref{table:numObsTimeMod} shows the number of observations (rows) for each data set, used to estimate the time models. Since we have added one row for each time period that a claim has spend in the respective state, we observe a high number of observations in Table. To evaluate the accuracy of the time models, we perform for each considered state a 5-fold cross-validation where for each claim in a hold-out set, we predict the time that it takes to exit as well as the state it will exit to. For claim in a holdout set, we simulate 100 trajectories and for each trajectory, we record the time it took to exit the state and the transition type. For each claim, we define the final transition type as the transition type that was simulated the most in the 100 trajectories. Furthermore, we note that once a claim has stayed more than 24 months in a state, we force it to exit the state and once it has stayed more than 180 months in the process, we force the claim to an exit state. Table \ref{table:timeModEval} shows the  percentage of correctly predicted transitions and the mean bias time averaged for the 5 folds. We observe that for most states, we can predict around 80\% of the correct transitions. State $S_{5+}$ is more complicated as claims from different states are used to build that model hence we observe a drop in the performance.  We also observe that for states $S_{0}, \ldots, S_{4}$, the mean bias (predicted - true) exit time  is close to 0 and for state $S_{5+}$ it is negative, hence the next states as well as the time a claim stayed in a state are correctly estimated.
\begin{table}[!htbp]
\centering
\begin{tabular}{c c c c c c c }
\hline\hline
& \multicolumn{6}{c}{\hspace{3ex} \textbf{State}}\\
\cline{2-7}
& $S_{0}$ & $S_{1}$  & $S_{2}$  & $S_{3}$  & $S_{4}$ & $S_{5+}$\\
\hline
abs.freq & 64,253& 76,977&  58,342&  27,476&  15,462&  53,520\\
  \hline\hline
\end{tabular}
\caption{Number of observations in each state to model the time process.\label{table:numObsTimeMod}}
\end{table}

\begin{table}[!htbp] 
\centering
\begin{tabular}{c c c c c c c }
\hline\hline
&\multicolumn{6}{c}{\textbf{State}}\\
\cline{2-7}
TransType & $S_{0}$ & $S_{1}$  & $S_{2}$  & $S_{3}$  & $S_{4}$ & $S_{5+}$\\
\hline
N & 62 \%   & 73 \%   &77\%  & 75\%& 72\% & 65\% \\
P & 35 \%  & 19 \%  & 13\%  &22\% & 21\% & 31\% \\
TN & 0 \%  & 4 \%   & 6\%  &5\%  & 3\% & 1\% \\
TP & 3 \% & 4 \%   & 4\%  &4\%  & 3\% & 3\% \\
  \hline\hline
\end{tabular}
\caption{Percentage of transitions in the training data.\label{table:timeModFreq}}
\end{table}

\begin{table}[!htbp] 
\centering
\begin{tabular}{c c c c c c c }
\hline\hline
&\multicolumn{6}{c}{\textbf{State}}\\
\cline{2-7}
& $S_{0}$ & $S_{1}$  & $S_{2}$  & $S_{3}$  & $S_{4}$ & $S_{5+}$\\
\hline
\% correct transitions & 92\% & 83\%  & 82\%  &84\%  & 86\% & 68\% \\
mean bias time &0.14 & 0.09  & -0.23  & -0.08  & 0.04 & -14.09 \\
  \hline\hline
\end{tabular}
\caption{Percentage of correct transitions and mean bias time (months) for the time models.\label{table:timeModEval}}
\end{table}

\subsection{Interpreting marginal effects of covariates on the time models}
In this section, we investigate the marginal effect of the covariates on the transition probabilities using partial dependence plots (PDP) introduced by  \cite{friedman2001greedy} and implemented using the \texttt{iml}  \cite{iml} package in R . These PDP illustrate the marginal effect of a covariate (predictor) on the predictions made by the models. This marginal effect is marginalized over the other covariates meaning that to get the PDP for a categorical variable, we assign to each observation that same category for the variable of interest and average the predictions. Hence, as described by  \cite{molnar2019}, the PDP of a category represents the average prediction made by the model if we force each observation to have that category for the given categorical variable.

\begin{figure}[!htb]
\centering
\begin{subfigure}{.45\textwidth}
  \centering
  \includegraphics[width=.9\linewidth]{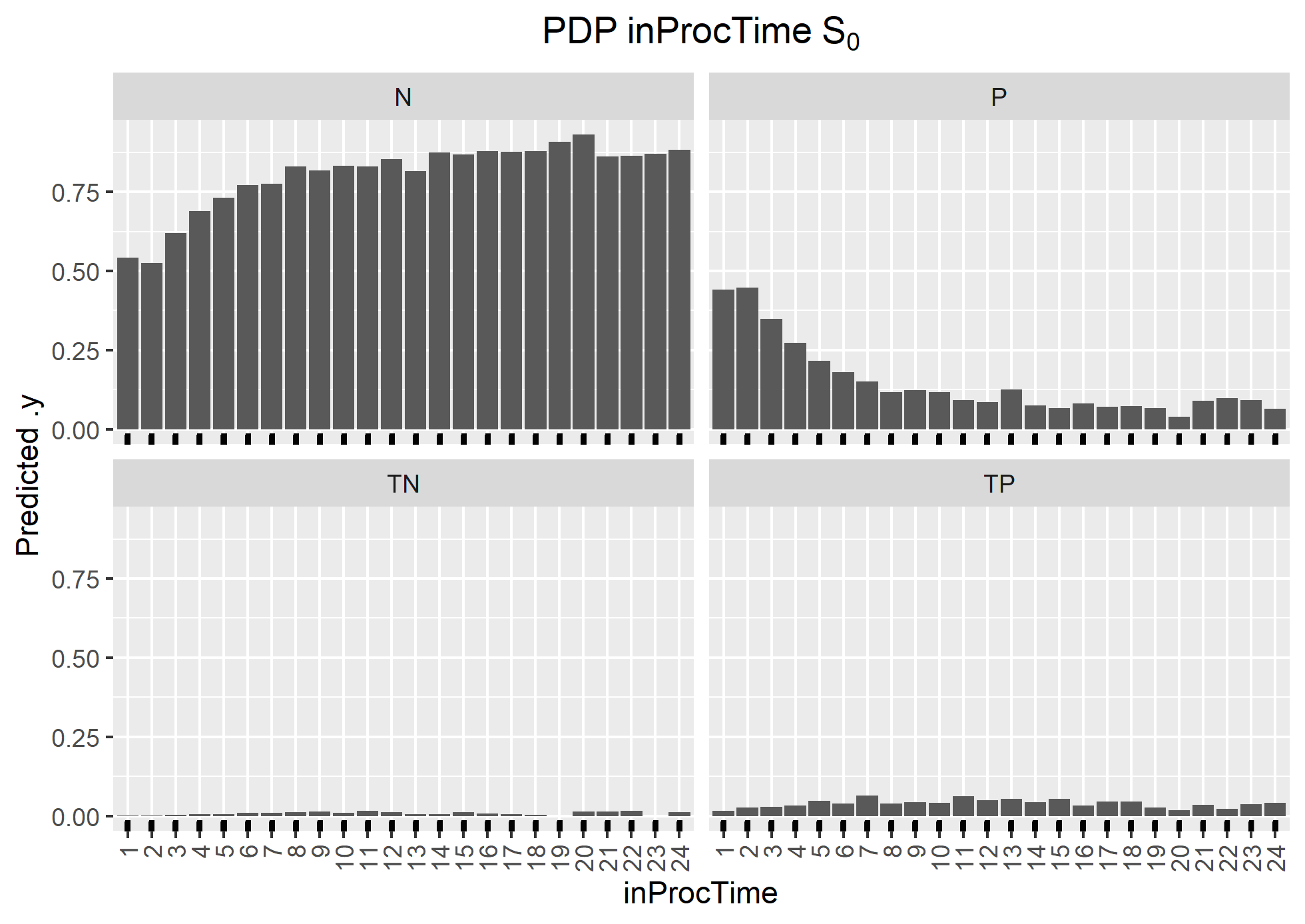}
  \caption{}
  \label{fig:pdp_proctime_s0}
\end{subfigure}%
\begin{subfigure}{.45\textwidth}
  \centering
  \includegraphics[width=.8\linewidth]{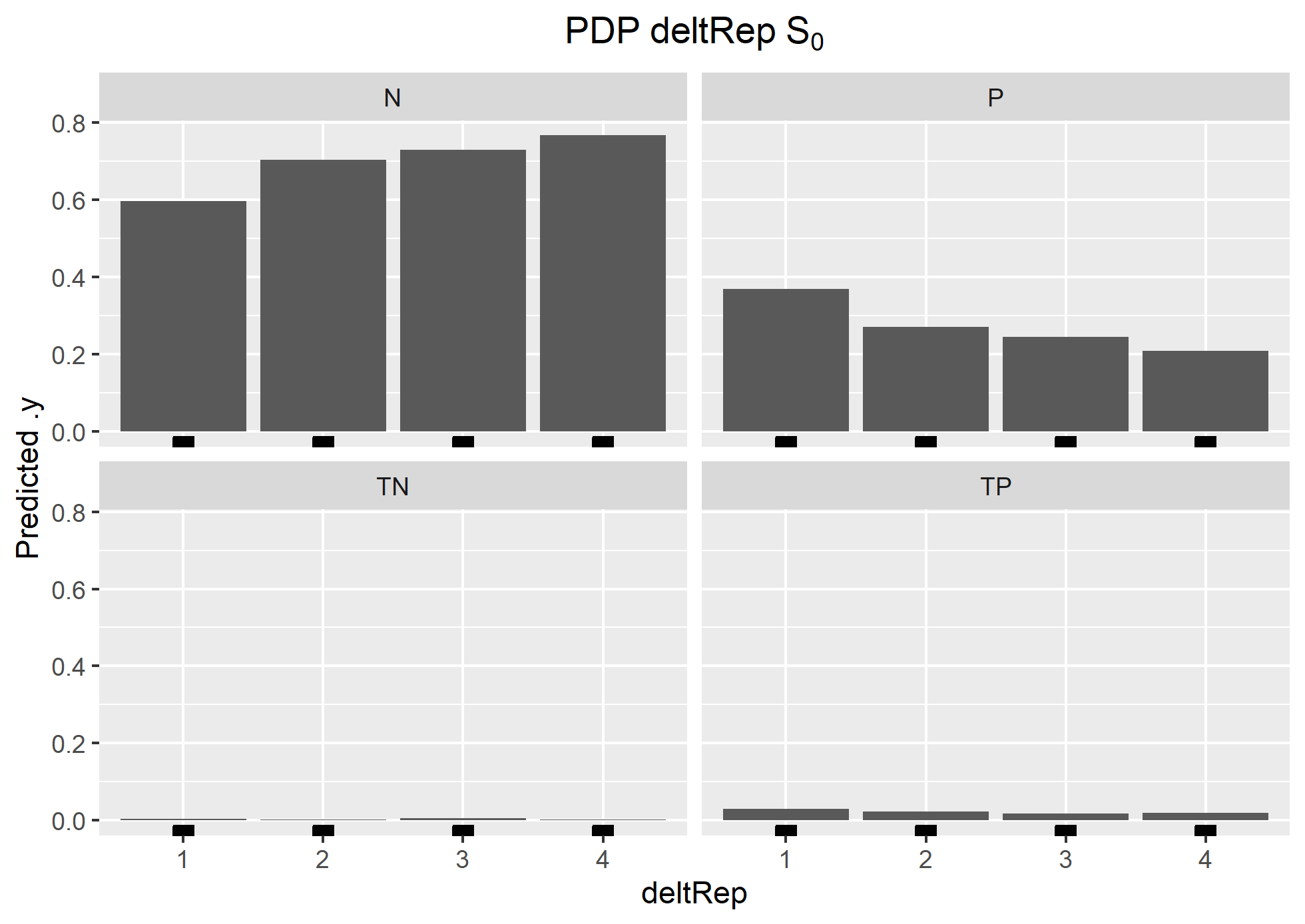}
  \caption{}
  \label{fig:pdp_deltrep_s0}
\end{subfigure}
\caption{Partial dependence plots representing the marginal effect on transition probabilities of the time spent in the process and the reporting delay for claims in $S_{0}$.}
\label{fig:pdp_s0}
\end{figure}

From Figure \ref{fig:pdp_s0}, we observe that for claims in $S_{0}$, an increase in the time spent in the process, and hence in the time spent in the state, decreases the probability to have a payment. Similarly, high values of the reporting delay decrease the probability of exiting the state.

For claims in $S_{1}$, we have two extra covariates, namely, the binned version of the size of the first (hence previous) payment, where the bins are given in Appendix E and the time spent in the current state. We observe that as for the model for claims in $S_{0}$, high values for the time spent in the process, the time spent in the state and the reporting delay decrease the probability of exiting the state. We also observe that a large cumulative previous payment increases the probability of having a payment.

\begin{figure}[!htb]
\begin{subfigure}{.45\textwidth}
    \centering
    \includegraphics[width=.8\linewidth]{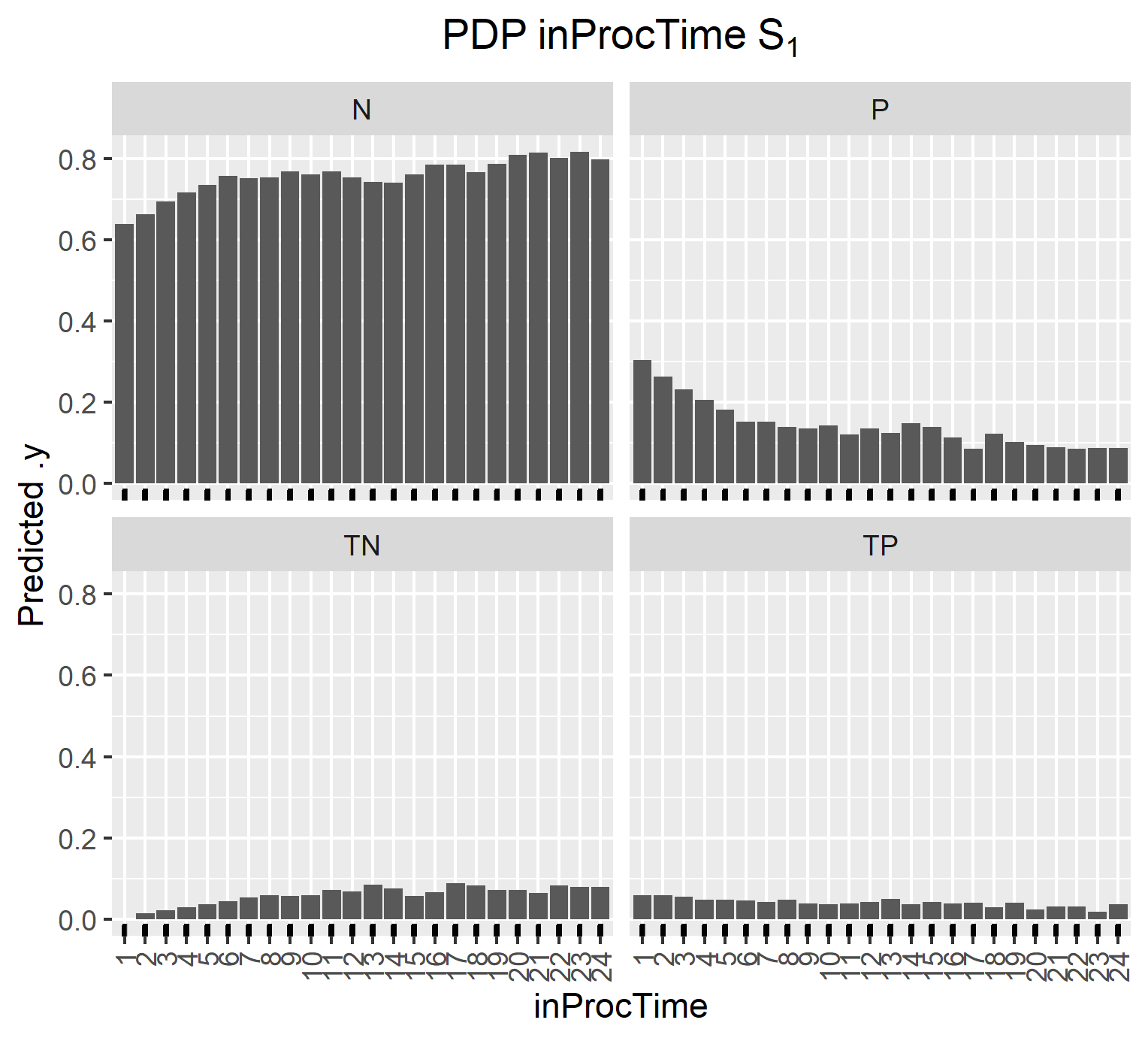}
    \caption{}
    \label{fig:pdp_proctime_s1}
\end{subfigure}
\begin{subfigure}{.45\textwidth}
    \centering
    \includegraphics[width=.8\linewidth]{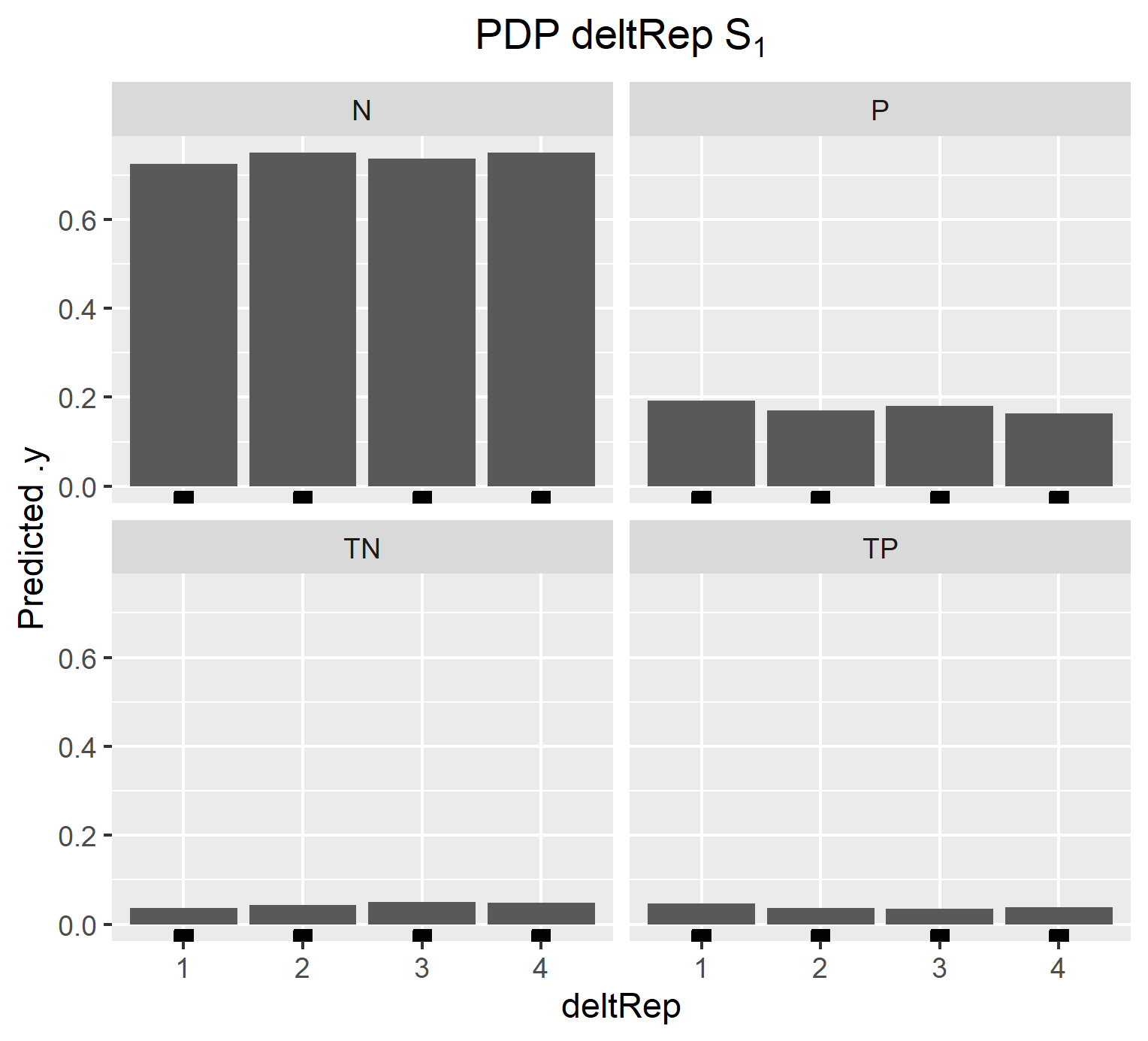}
    \caption{}
    \label{fig:pdp_deltrep_s1}
\end{subfigure}
\newline
\begin{subfigure}{.45\textwidth}
    \centering
    \includegraphics[width=.8\linewidth]{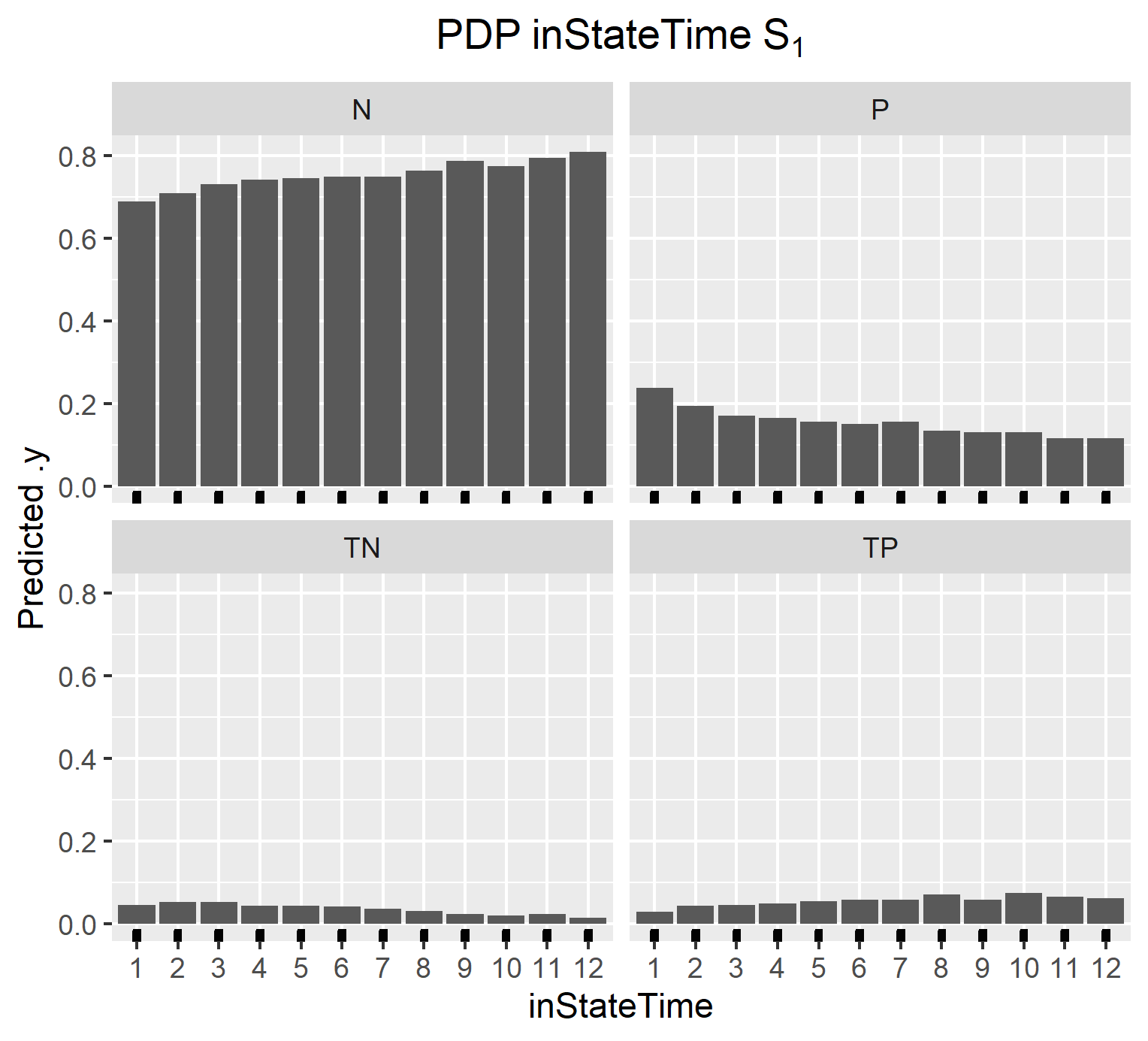}
    \caption{}
    \label{fig:pdp_statetime_s1}
\end{subfigure}
\begin{subfigure}{.45\textwidth}
    \centering
    \includegraphics[width=.8\linewidth]{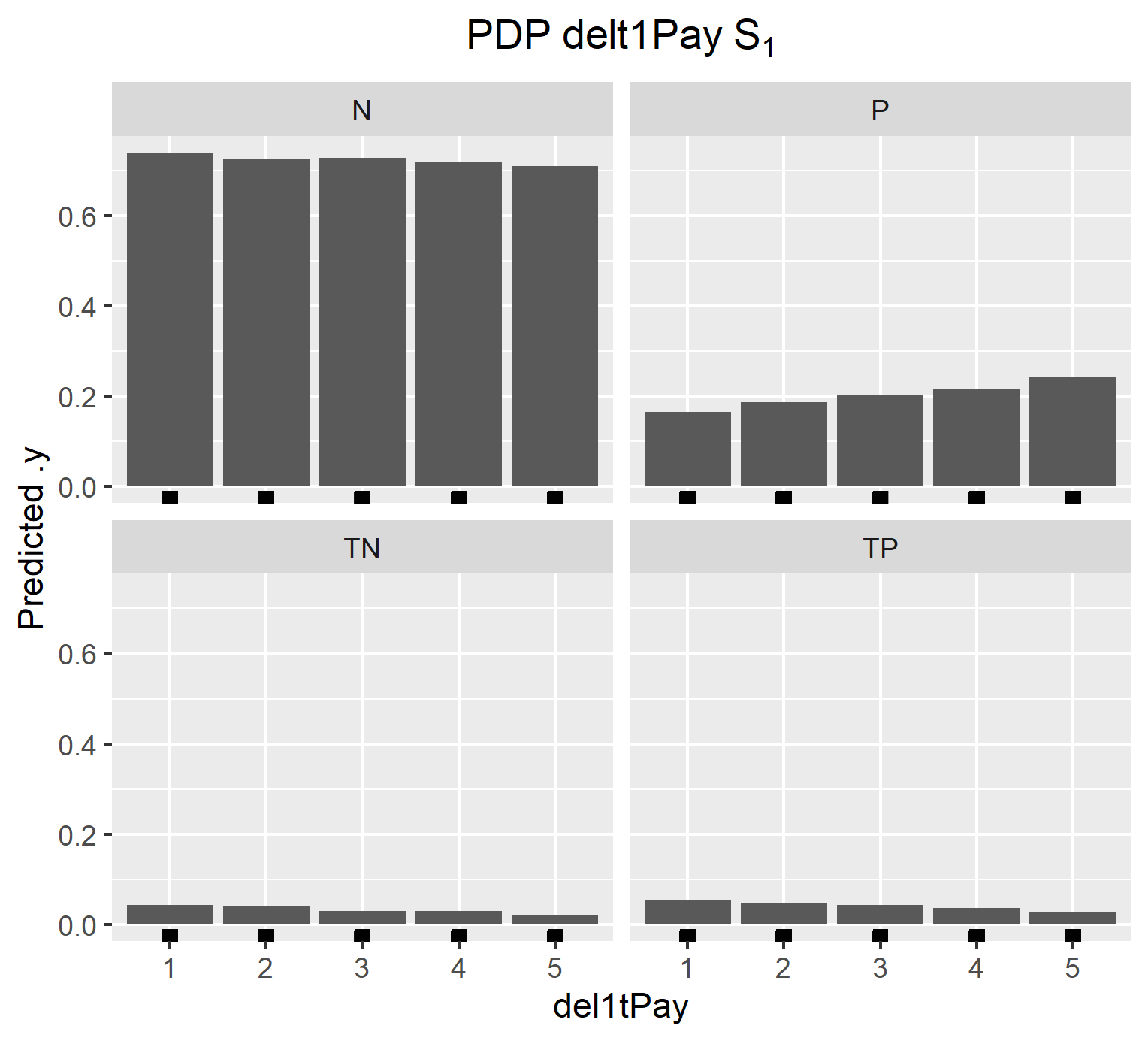}
    \caption{}
    \label{fig:pdp_cumdeltpay_s1}
\end{subfigure}
\caption{Partial dependence plots representing the marginal effect on transition probabilities of the time spent in the process (a), the reporting delay (b), the time spent in the state (c) and the previous payment size (d) for claims in $S_{1}$.}
\label{fig:pdp_s1}
\end{figure}

\begin{figure}[!htb]
\begin{subfigure}{.45\textwidth}
    \centering
    \includegraphics[width=.85\linewidth]{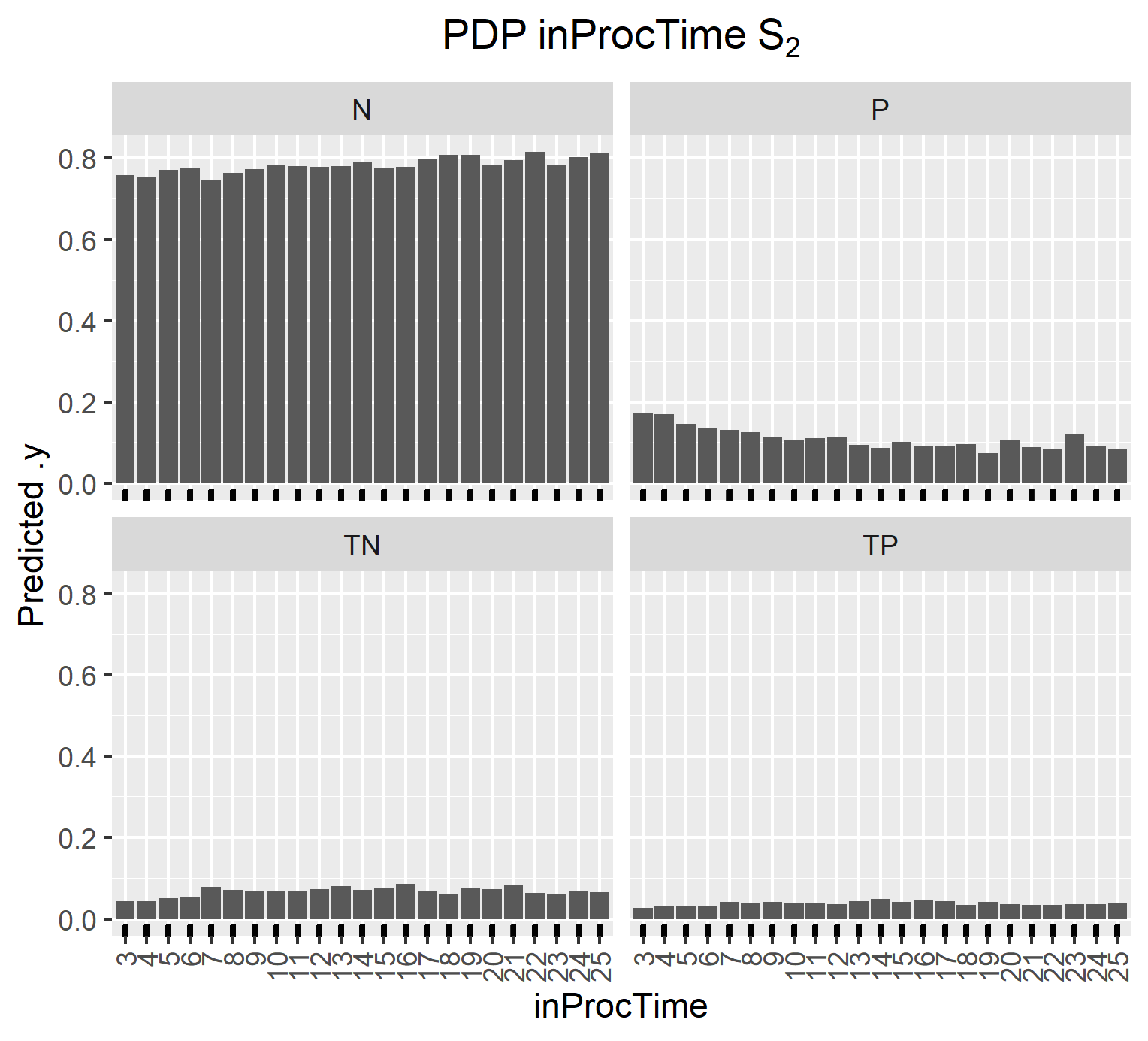}
    \caption{}
    \label{fig:pdp_proctime_s2}
\end{subfigure}
\begin{subfigure}{.45\textwidth}
    \centering
    \includegraphics[width=.85\linewidth]{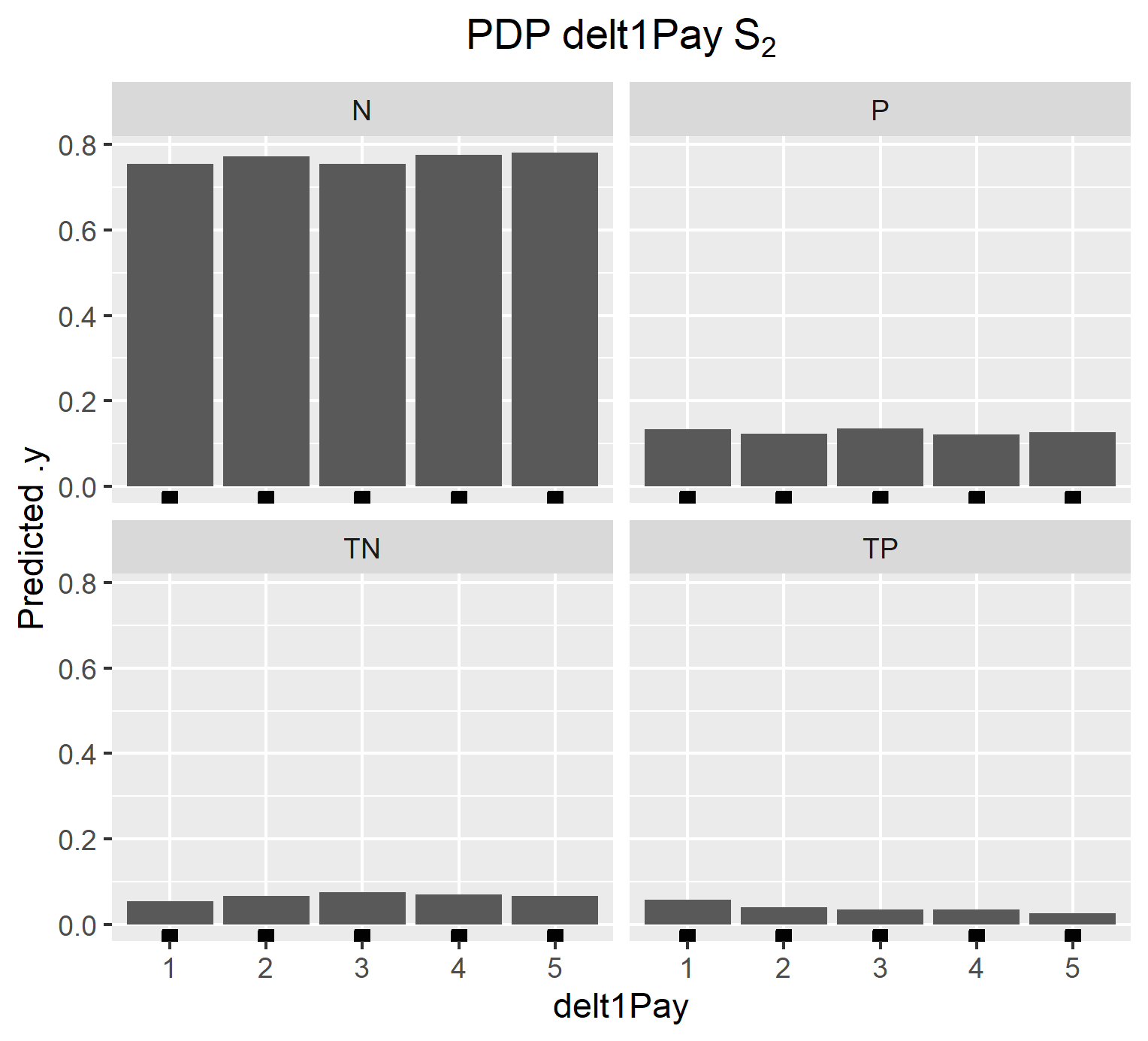}
    \caption{}
    \label{fig:pdpdeltpay_s2}
\end{subfigure}
\newline
\begin{subfigure}{.45\textwidth}
    \centering
    \includegraphics[width=.85\linewidth]{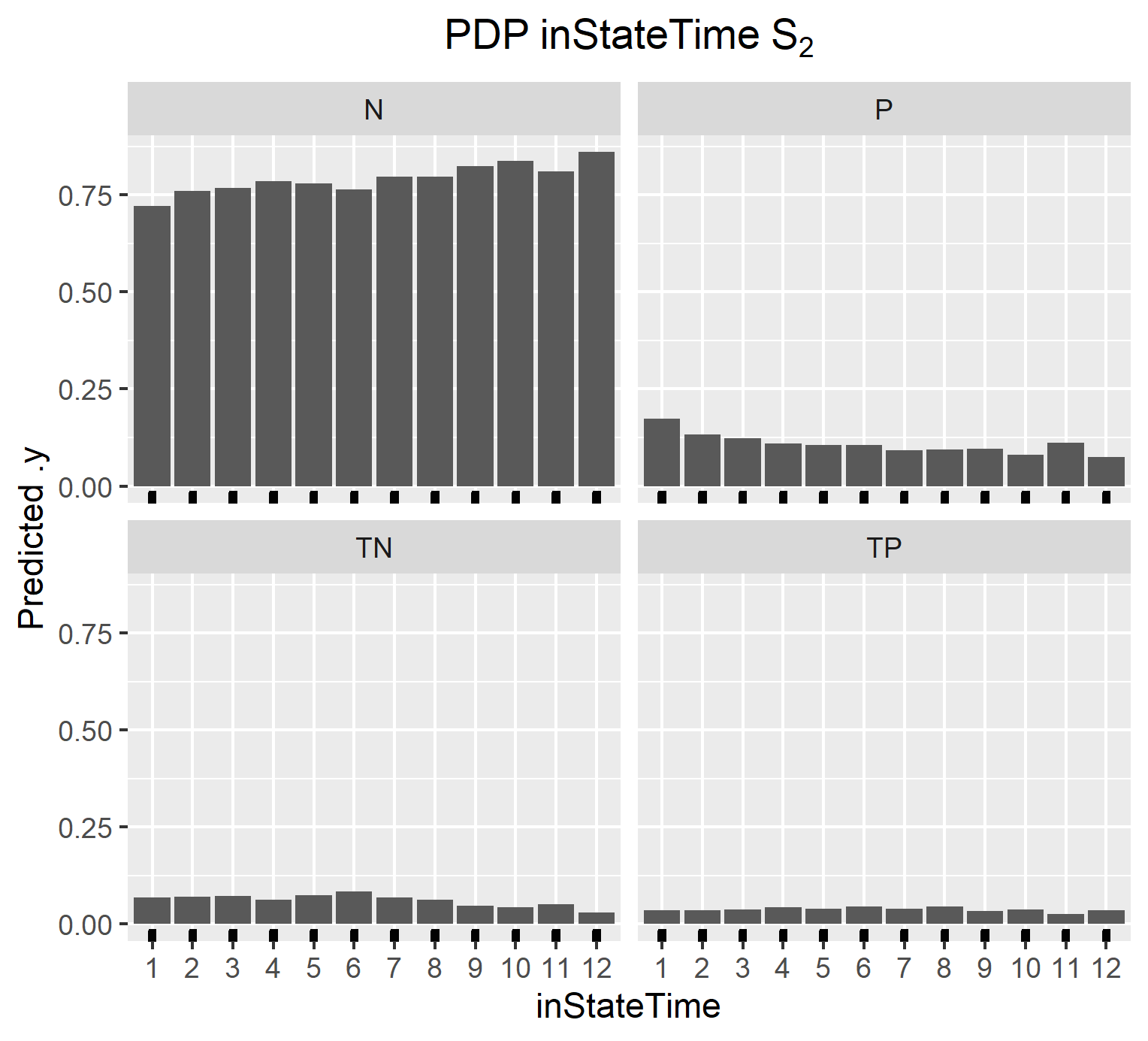}
    \caption{}
    \label{fig:pdp_statetime_s2}
\end{subfigure}
\begin{subfigure}{.45\textwidth}
    \centering
    \includegraphics[width=.85\linewidth]{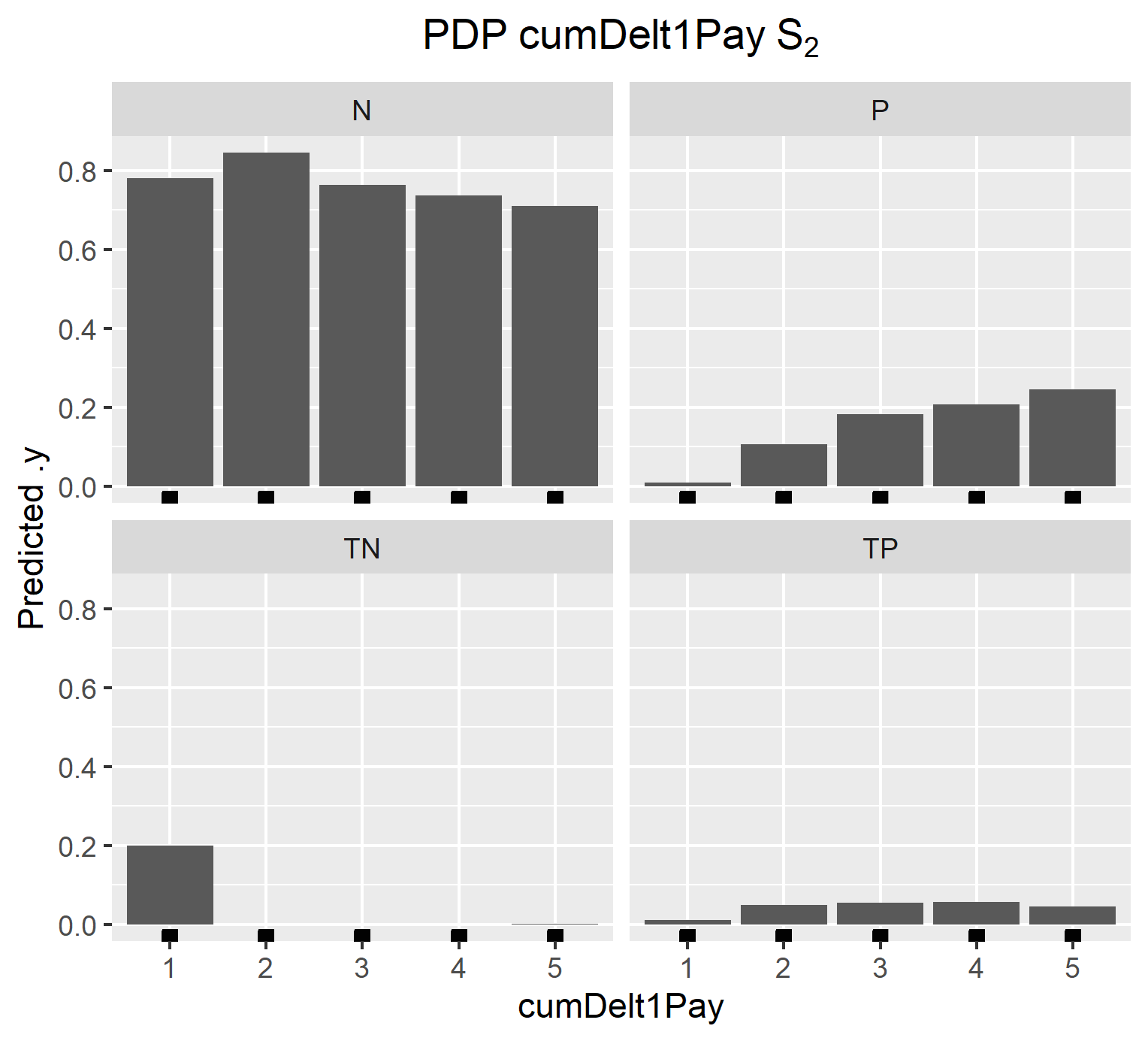}
    \caption{}
    \label{fig:pdp_cumdeltpay_s2}
\end{subfigure}
\caption{Partial dependence plots representing the marginal effect on transition probabilities of the time spent in the process (a), the previous payment size (b), the time spent in the state (c) and the cumulative previous payment size (d) for claims in $S_{2}$.}
\label{fig:pdp_s2}
\end{figure}

Figure \ref{fig:pdp_s2} shows the transition probabilities  $S_{2}$, we have one extra covariate as compared to $S_{1}$, namely, the binned size of the previous payment, where the bins are given in Appendix E and the time spent in the current state. Just like for the $S_1$ time model, we observe that, high values for the time spent in the process (inProcTime), and the time spent in the State (inStateTime), decrease the probability of a transition. We also observe that a large cumulative previous payment increases the probability of having a payment. However, the size of the previous payment only has a small impact on the transition probabilities. . Transition probabilities for claims in $S_{3}$, $S_{4}$ and $S_{5+}$ are shown in Figures \ref{fig:pdp_s3}, \ref{fig:pdp_s4}, \ref{fig:pdp_s5} from Appendix F, where we observe similar effects for the covariates as in $S_{2}$.

\subsection{Payment models evaluation}
In this section we evaluate the accuracy of the payment models used through 5-fold cross-validation. Table \ref{table:nObsPayMod} shows the number of observations in each state, which will be used to estimate the payment distributions. We observe that the number of observationsused to fit the payment model of a given transition is lower than the number of observations of the time fit of the same transition (see Table  \ref{table:nObsPayMod} versus Table \ref{table:numObsTimeMod}) , the reason being that we model a payment distribution conditionally on the occurrence of a payment. Hence, to model the payment distribution, we only use the lines in the data sets where the transition type contain a payment, i.e lines where transType is equal to 'P' or 'TP' in Table \ref{table:dynamicInfoUpdate}. During the 5-fold cross-validation, the payment models are  trained on the training portion and evaluated on each holdout set and we compute the Root Mean Square Error (RMSE) and Median Absolute Error (MAE) between the true and predicted payments. For each fold, we set $b_{2} = 0$ and other splitting points for the payment distribution are estimated using the Gerstengarbe plot as explained in Section \ref{sec:paymentmodels}. In this way, the four bins can be interpreted respectively as small and large negative or positive payments. From Table \ref{table:payModSummary}, we observe that the center of the payment distribution is well captured as the MAE is between 911 and 2,323 euro. We observe furthermore that the payment distributions in  data sets for states $S_{0}$ and $S_{5+}$ are the most difficult to model, given the non-homogeneity of claims in those data sets.

\begin{table}[!htbp]
\centering
\begin{tabular}{c c c c c c c }
\hline\hline
& \multicolumn{6}{c}{\hspace{3ex} \textbf{State}}\\
\cline{2-7}
& $S_{0}$ & $S_{1}$  & $S_{2}$  & $S_{3}$  & $S_{4}$ & $S_{5+}$\\
\hline
abs.freq & 24,480& 18,465&  10,062&  5,849&  3,941&  19,084\\
  \hline\hline
\end{tabular}
\caption{Number of observations in each data set to model the payment process.\label{table:nObsPayMod}}
\end{table}

\begin{table}[!htbp]
\centering
\begin{tabular}{c c c c c c c }
\hline\hline
&\multicolumn{6}{c}{\textbf{State}}\\
\cline{2-7}
& $S_{0}$ & $S_{1}$  & $S_{2}$  & $S_{3}$  & $S_{4}$ & $S_{5+}$\\
\hline
RMSE & 9,561& 4,946&4,494  & 5,238 & 4,391 & 8,008 \\
MAE &1,668 &1,060 & 911 & 1,193 & 1,478  & 2,323 \\
  \hline\hline
\end{tabular}
\caption{ 5-fold CV RMSE and MAE of the payment distributions in each data set.\label{table:payModSummary}}
\end{table}

\subsection{Interpreting marginal effects of covariates on the payment models}

We now look at the marginal effects of the covariates on the prediction of  either a large or small negative payment, represented by the bins $B_{1}$ and $B_{2}$ respectively. Similarly, we also look at the probabilities of obtaining  a small or large positive payment, represented by the bins $B_{3}$ and $B_{4}$. Table \ref{table:splitPointsPay} shows the splitting points for the payment distribution in each data set and in Table \ref{table:averagePayBin} the predicted mean  payment of each bin is shown, estimated as explained in Section \ref{sec:paymentmodels}.
\begin{table}[!htbp]
\centering
\begin{tabular}{c c c c c c c }
\hline\hline
& \multicolumn{6}{c}{\hspace{3ex} \textbf{State}}\\
\cline{2-7}
& $S_{0}$ & $S_{1}$  & $S_{2}$  & $S_{3}$  & $S_{4}$ & $S_{5+}$\\
\hline
$b_{1}$ & -1,230 & -3,000&  -2,310&  -1,780&  -2,140&  -2,690\\
$b_{2}$ & 0 & 0&  0&  0 &  0 &  0\\
$b_{3}$ &3,500 & 3,200&  2,970&  3,107&  2,500&  2,530\\
  \hline\hline
\end{tabular}
\caption{Splitting points ($b_{l}^{j}$) for the payment distribution in each data set.\label{table:splitPointsPay}}
\end{table}

\begin{table}[!htbp]
\centering
\begin{tabular}{c c c c c c c }
\hline\hline
& \multicolumn{6}{c}{\hspace{3ex} \textbf{State}}\\
\cline{2-7}
& $S_{0}$ & $S_{1}$  & $S_{2}$  & $S_{3}$  & $S_{4}$ & $S_{5+}$\\
\hline
$B_{1}$ &-6,043 & -11,725&  -10,644&  -8,024&  -10,047&  -13,725\\
$B_{2}$ & -475 & -1,248&  -800&  -670 & -788  &  -993\\
$B_{3}$ & 1,404&  1,152&  1,070&  1,094&948&  909\\
$B_{4}$ &  7,230&  7,510&  6,915&  7,400 &7,004&  9,914\\
  \hline\hline
\end{tabular}
\caption{Mean payment ($\mu_{l}$) in bins for each data set.\label{table:averagePayBin}}
\end{table}

\begin{table}[!htbp] 
\centering
\begin{tabular}{c c c c c c c }
\hline\hline
&\multicolumn{6}{c}{\textbf{State}}\\
\cline{2-7}
TransType & $S_{0}$ & $S_{1}$  & $S_{2}$  & $S_{3}$  & $S_{4}$ & $S_{5+}$\\
\hline
$B_{1}$ & 0.3 \%   & 6.9 \%   &3.5\%  & 3.5\%& 1.8\% & 1\% \\
$B_{2}$ & 0.4 \%  & 17.5 \%  & 9.3\%  &6.1\% & 4.8\% & 2.5\% \\
$B_{3}$ & 76.2 \%  & 60.9 \%   & 71.9\%  &72.7\%  & 70.2\% & 66.3\% \\
$B_{4}$ & 23.1 \% & 14.7 \%   & 15.3\%  &17.7\%  & 23.2\% & 30.2\% \\
  \hline\hline
\end{tabular}
\caption{Percentage of claims in each bin per state in the training data.\label{table:payModFreq}}
\end{table}

\begin{figure}[!htb]
\centering
\begin{subfigure}{.45\textwidth}
  \centering
  \includegraphics[width=.85\linewidth]{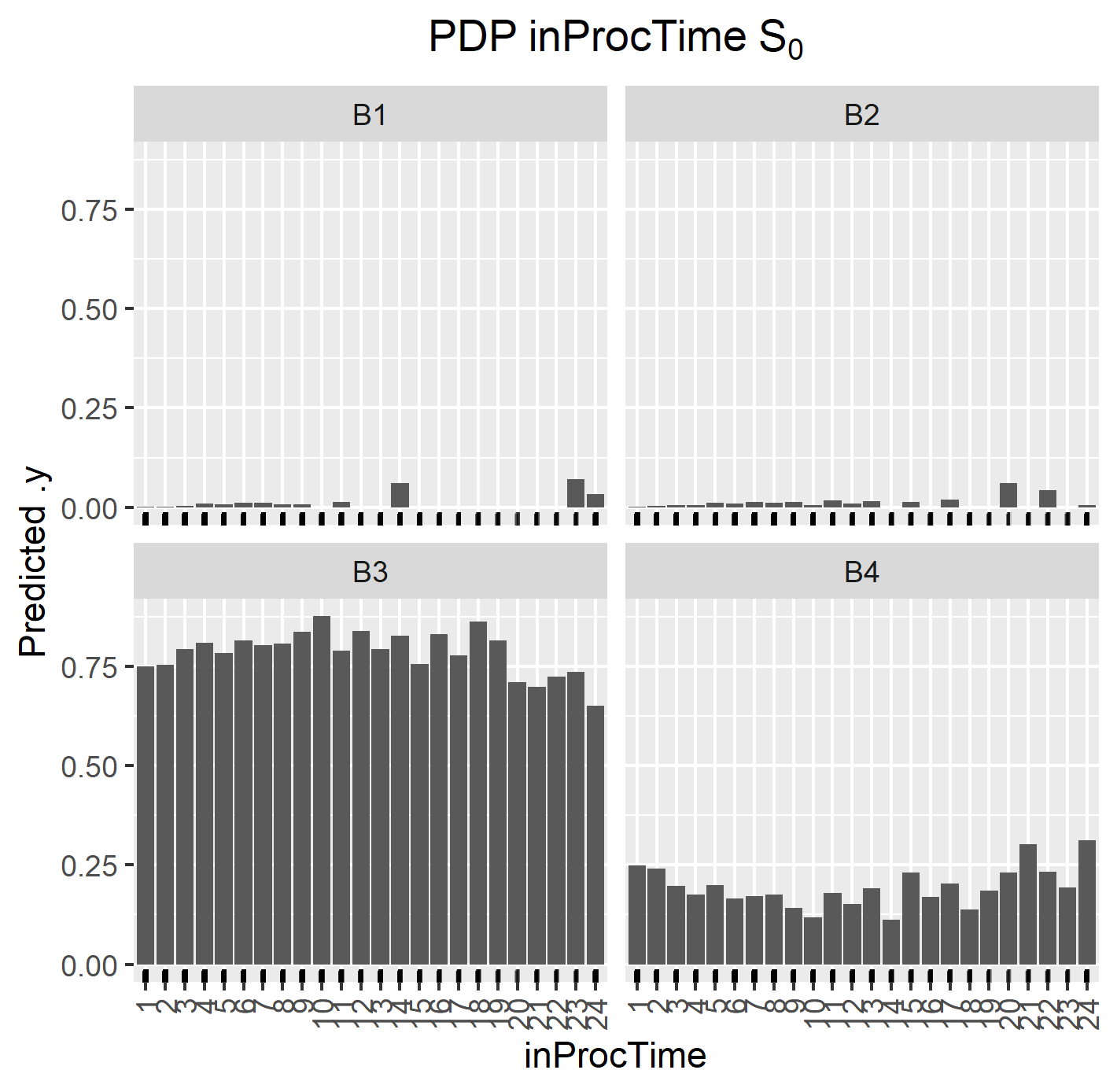}
  \caption{}
  \label{fig:pdp_proctime_s0_pay}
\end{subfigure}%
\begin{subfigure}{.45\textwidth}
  \centering
  \includegraphics[width=.85\linewidth]{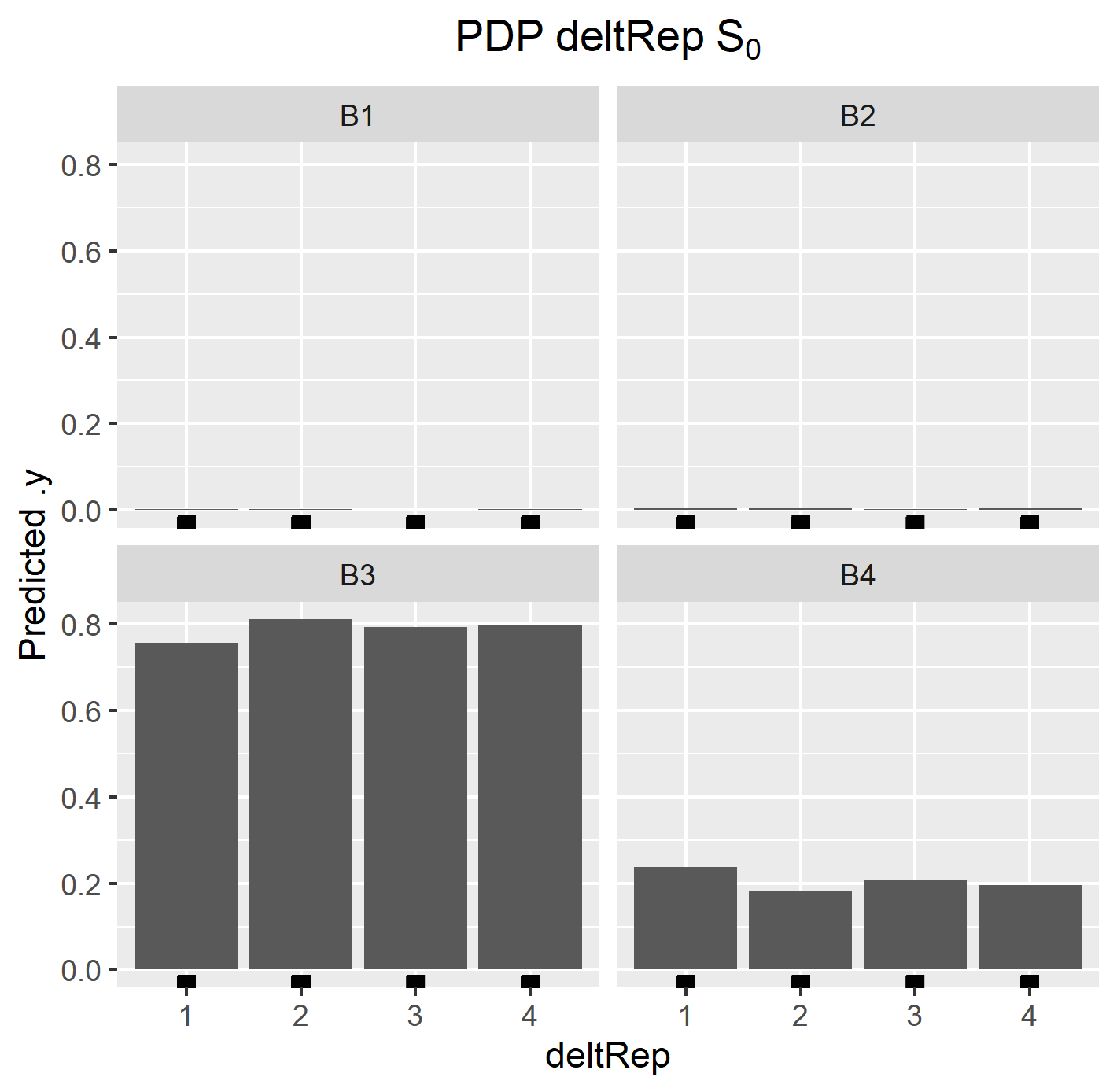}
  \caption{}
  \label{fig:pdp_deltrep_s0_pay}
\end{subfigure}
\caption{Partial dependence plots representing the marginal effect on probabilities to belong to a payment bin of the time spent in the process and the reporting delay for claims in $S_{0}$.}
\label{fig:pdp_s0_pay}
\end{figure}

From Figure \ref{fig:pdp_s0_pay}, we observe that the effect of the time spent in the process together with the reporting delay has little impact on the probability of having a small or large payment. Note in Table \ref{table:payModFreq} that almost all payments in $S_0$ are positive, hence pertaining to $B_3$ or $B_4$.

\begin{figure}[!htbp]
\begin{subfigure}{.45\textwidth}
    \centering
    \includegraphics[width=.85\linewidth]{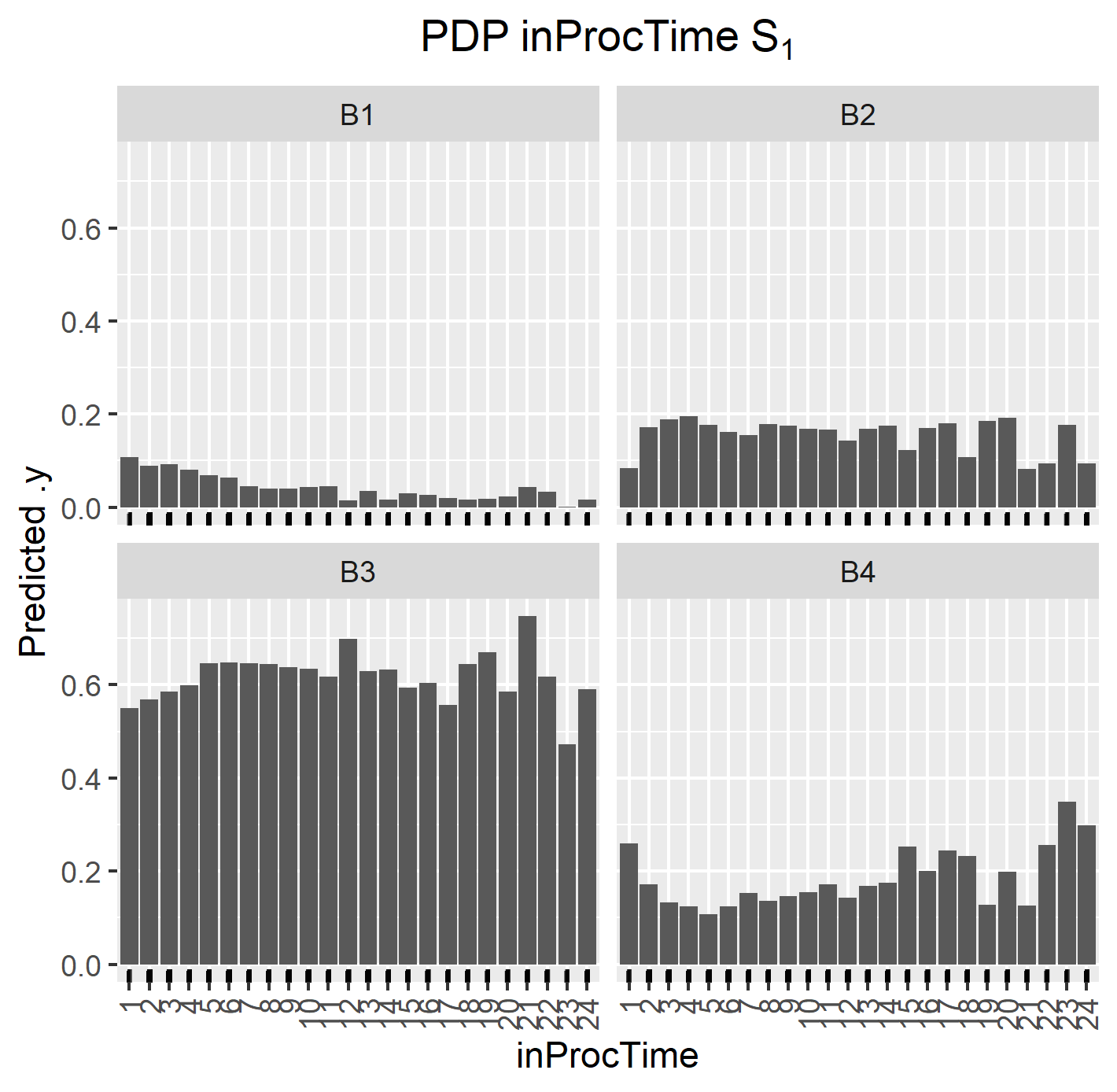}
    \caption{}
    \label{fig:pdp_proctime_s1_pay}
\end{subfigure}
\begin{subfigure}{.45\textwidth}
    \centering
    \includegraphics[width=.85\linewidth]{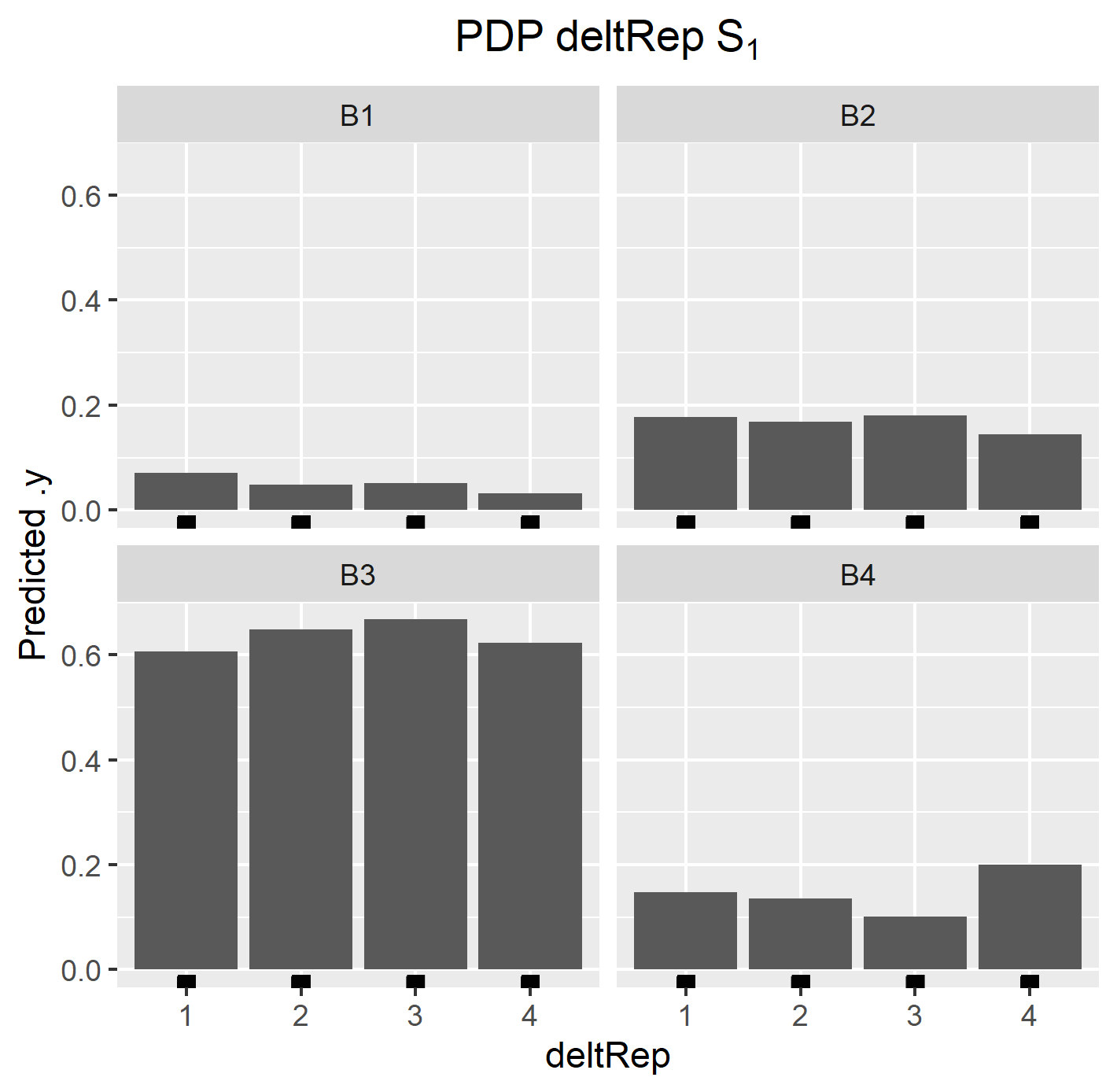}
    \caption{}
    \label{fig:pdp_deltrep_pay}
\end{subfigure}
\newline
\begin{subfigure}{.45\textwidth}
    \centering
    \includegraphics[width=.85\linewidth]{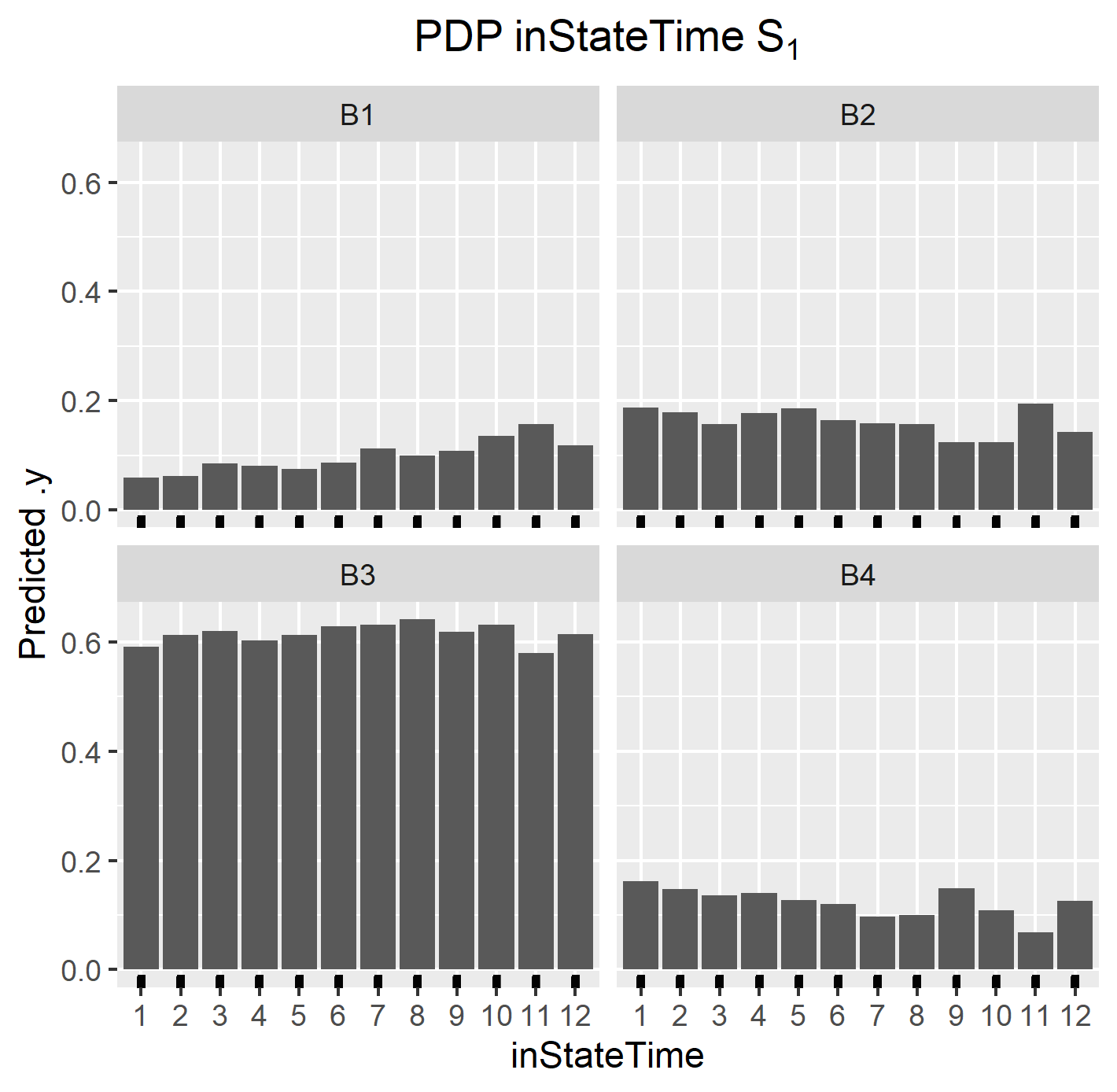}
    \caption{}
    \label{fig:pdp_statetime_s1_pay}
\end{subfigure}
\begin{subfigure}{.45\textwidth}
    \centering
    \includegraphics[width=.85\linewidth]{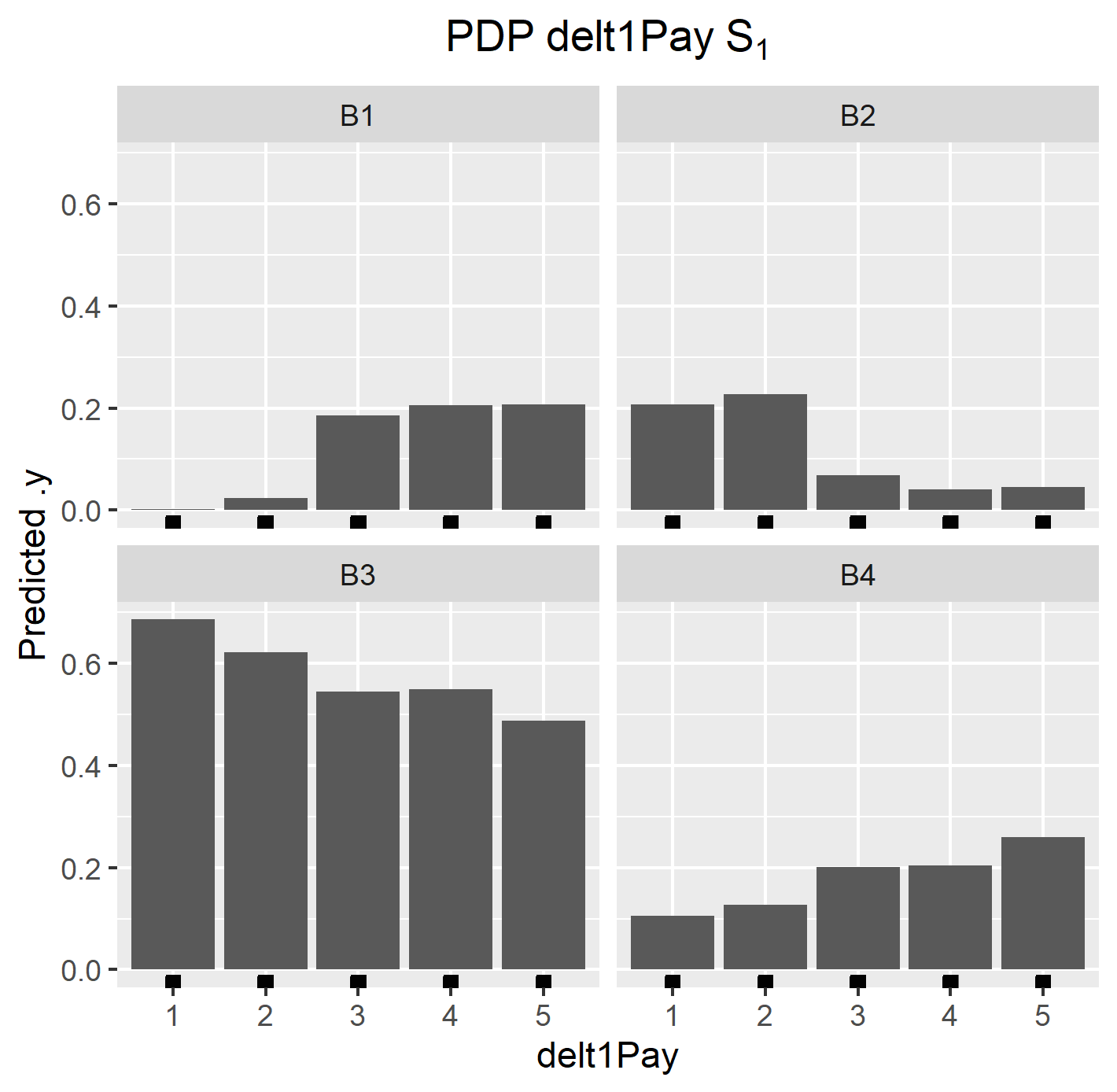}
    \caption{}
    \label{fig:pdp_cumdeltpay_s1_pay}
\end{subfigure}
\caption{Partial dependence plots representing the marginal effect on probabilities to belong to a payment bin of the time spent in the process (a), the reporting delay (b), the time spent in the state (c) and the previous payment size (d) for claims in $S_{1}$.}
\label{fig:pdp_s1_pay}
\end{figure}

In $S_1$ we see that about 25\% of all payments are negative. From Figure \ref{fig:pdp_s1_pay} we can see that the different covariates do have a small, yet much bigger effect than for $S_0$. The most likely payment is a small positive payment. Furthermore, we observe a probability of around 40\% for a large positive previous payment to lead to either a large payment of either sign.

\begin{figure}[!htb]
\begin{subfigure}{.45\textwidth}
    \centering
    \includegraphics[width=.85\linewidth]{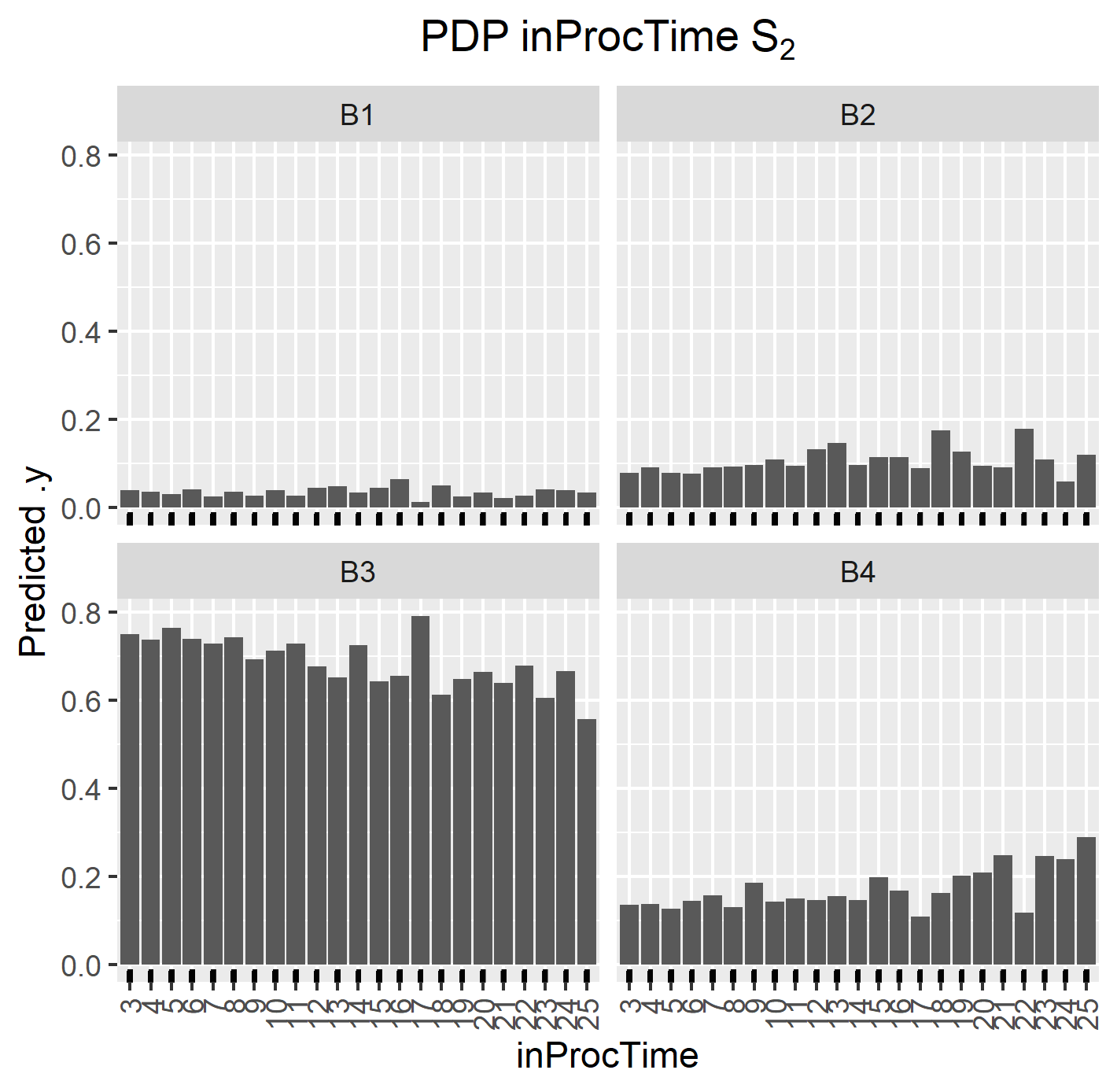}
    \caption{}
    \label{fig:pdp_proctime_s2_pay}
\end{subfigure}
\begin{subfigure}{.45\textwidth}
    \centering
    \includegraphics[width=.85\linewidth]{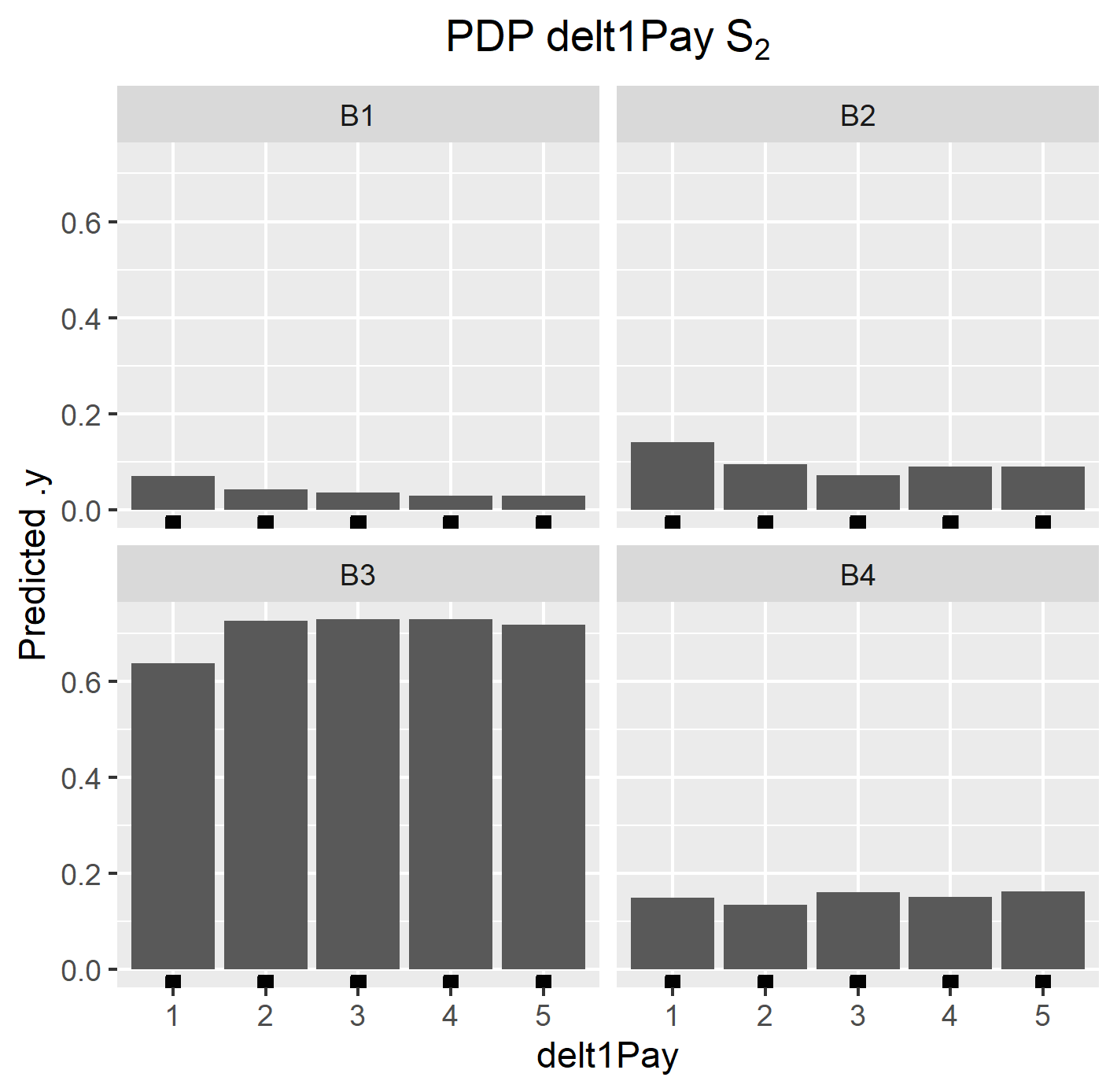}
    \caption{}
    \label{fig:pdpdeltpay_s2_pay}
\end{subfigure}
\newline
\begin{subfigure}{.45\textwidth}
    \centering
    \includegraphics[width=.85\linewidth]{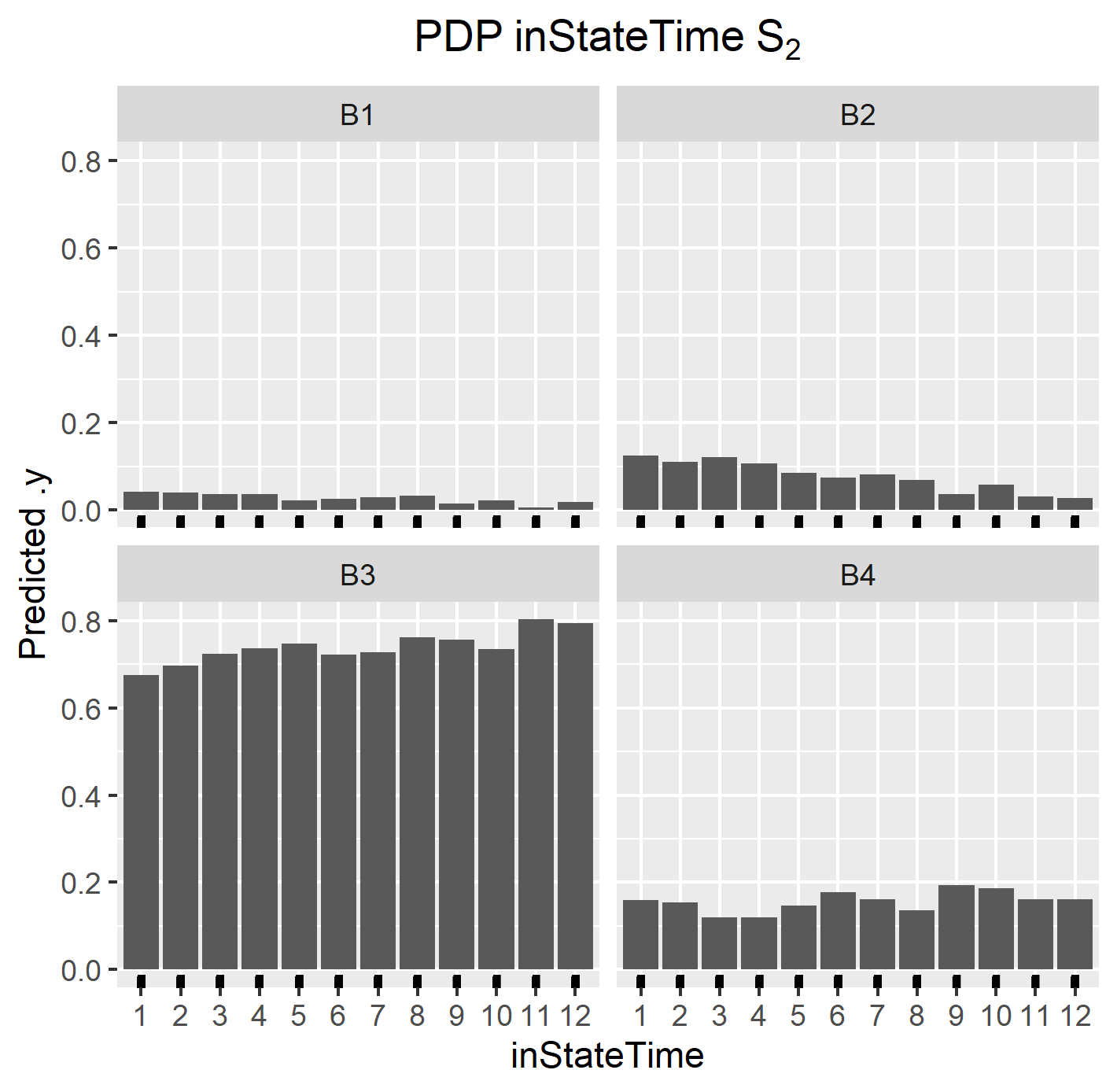}
    \caption{}
    \label{fig:pdp_statetime_s2_pay}
\end{subfigure}
\begin{subfigure}{.45\textwidth}
    \centering
    \includegraphics[width=.85\linewidth]{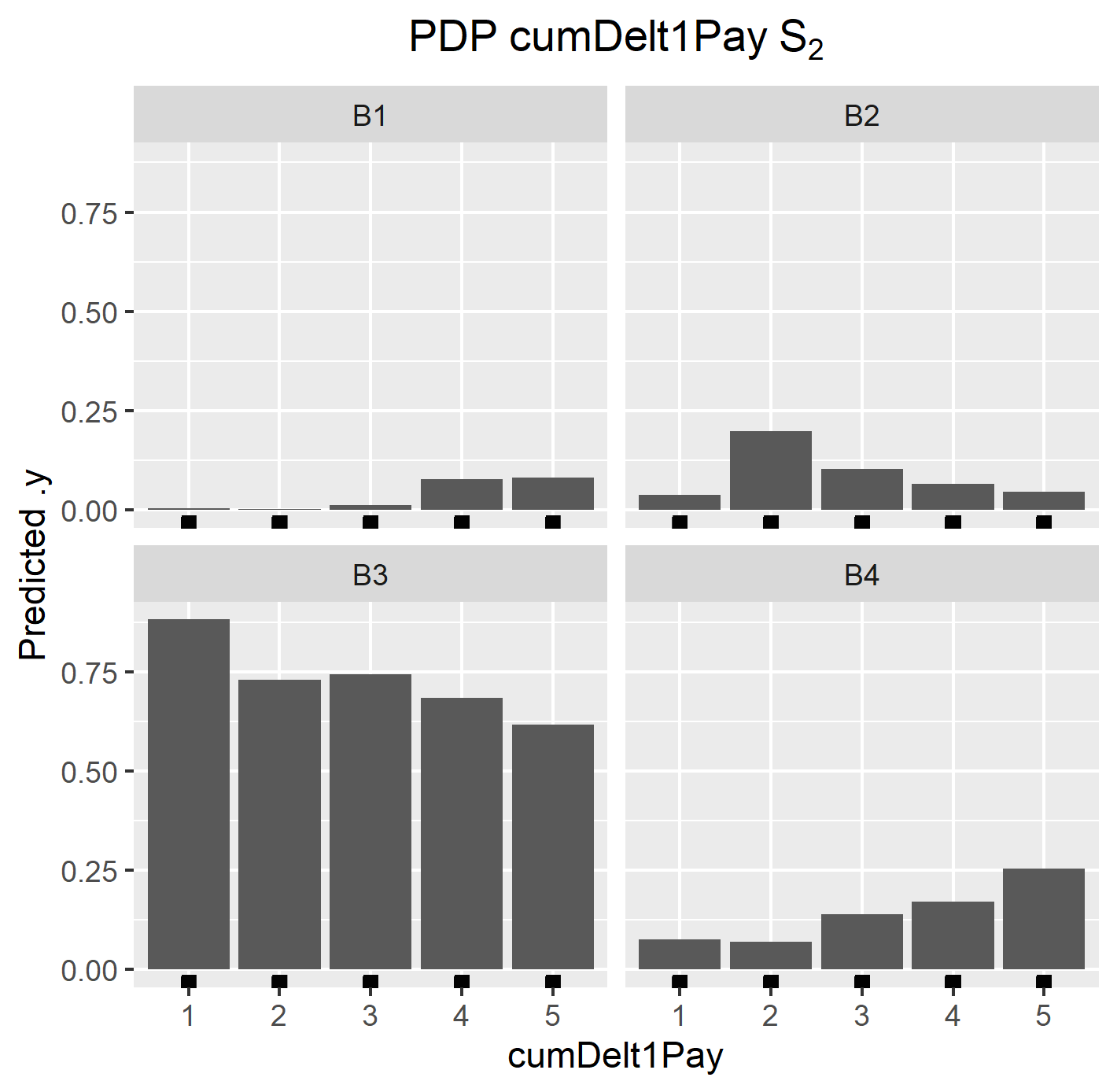}
    \caption{}
    \label{fig:pdp_cumdeltpay_s2_pay}
\end{subfigure}
\caption{Partial dependence plots representing the marginal effect on probabilities to belong to a payment bin of the time spent in the process (a), the previous payment size (b), the time spent in the state (c) and the cumulative previous payment size (d) for claims in $S_{2}$.}
\label{fig:pdp_s2_pay}
\end{figure}

From Figure \ref{fig:pdp_s2_pay}, we observe that claims that stay longer in the process tend to have a  higher probability of a positive payment. This relates to the fact that these are longer tailed claims that tend to cost more. We also observe that claims with a high cumulative previous payment tend to produce higher payments, hence claims that are costly early in development tend to stay costly as development progresses. The marginal effect of covariates for states $S_{3}$, $S_{4}$ and $S_{5+}$ are similar to those of state $S_{2}$ and are shown in Appendix G.

\subsection{ IBNR count comparison with the chain-ladder}

We start by computing the yearly IBNR claim counts based on the methodology presented in Section \ref{sec:IBNR}. As explained in this section, the four steps to obtain the predicted number of IBNR claims for a given accident and development year are repeated 1000 times. The mean  and 95\% quantiles of these IBNR predictions are shown in Table \ref{table:yearlyIBNRcounts}, where the quantiles for the chain-ladder are obtained with the bootstrapped strategy proposed by \cite{englandverrall2002} with an over-dispersed Poisson process distribution. For each accident year, the mean of these IBNR counts are used to build the IBNR  data set as explained in Appendix B.

\begin{table}[!htbp]
\centering
\begin{tabular}{c c c c c c c c c}
\hline\hline
 &2006 & 2007& 2008 &2009&2010&2011&2012&Total \\
\hline
Database&0 &0 &0 &0 &2 &29 &283& 314 \\
 CL mean & 0 &0 &1 &1 &3 &15 &173 &193\\
 CL 95\% & 0 &0 &4 &4 &8& 26& 213& 233\\
 mCube mean &0 &0 &1 &1 &2 &15 &175 & 194\\
 mCube 95\% & 0 & 0& 3 &3 &5 & 20 &197&228\\ 
 \hline\hline
\end{tabular}
\caption{Predicted yearly IBNR claim counts. \label{table:yearlyIBNRcounts} }
\end{table}

Clearly, most of the IBNR claims come from the last observed accident year, as these are bodily injury claims which are typically reported rather fast. As mentioned in Section \ref{sec:IBNR}, the average number of IBNR claims predicted by mCube corresponds to the number of IBNR claims predicted by Mack's chain-ladder. However, both the chain-ladder and mCube underestimate the number of IBNR claims in the later accident years.

\subsection{ Comparison with other micro-reserving models}

{In this section, we compare the performance of the proposed methodology on an individual level to the multi-state model of  \cite{Bettonville2020} and the hierarchical GLM of  \cite{CREVECOEUR2022}. The goal is to see how well the predictive distributions of the reserves capture the true reserves for the RBNS claims. Let $\hat{q}_{k}^{\alpha}$ denote the $\alpha$-quantile of the predictive distribution of ${R}_{k,\tau}$, consisting of $N_{sim}$ possible reserve values $\hat{R}_{k,\tau}^{1}, \ldots, \hat{R}_{k,\tau}^{N_{sim}}$, then we can define the following measures:
\begin{itemize}
    \item The Interval Score, IS := $mean(\hat{q}_{k}^{\alpha} - \hat{q}_{k}^{1-\alpha} )$, measures the width of the prediction intervals.
    \item The Prediction Interval Coverage Probability, PICP = $mean( \mathbbm{1}_{{R}_{k,\tau} \in [\hat{q}_{k}^{1-\alpha}, \hat{q}_{k}^{\alpha} ] })$, represents which fraction of the true reserves falls in the prediction intervals of the different reserving methods.
    \item The Continuous Ranked Probability Score (CRPS) of \cite{Gneiting2007} is a strictly proper scoring rule, which assesses the quality of probabilistic forecasts and rewards the forecaster for honest estimation of the predictive distribution. We use the formulation from  \cite{Grimit2006}, given by
\begin{equation*}
CRPS ({R}_{k,\tau}) = \frac{1}{N_{sim}} \sum_{i = 1}^{N_{sim}}  \mid \hat{R}_{k,\tau}^{i} - {R}_{k,\tau}  \mid  - \frac{1}{2N_{sim}^{2}} \sum_{i=1}^{N_{sim}} \sum_{r=1}^{N_{sim}}  \mid  \hat{R}_{k,\tau}^{i} - \hat{R}_{k,\tau}^{r} \mid ,   
\end{equation*}
{where a lower score for CRPS represents a more accurate probabilistic forecast.}
\end{itemize}
}


\begin{table}[ht]
\centering
\begin{tabular}{rrrr}
  \hline
 & mCube &  Bettonville & Crevecoeur  \\ 
  \hline
mean\_CRPS & $\mathbf{11,148.82}$  &  14,166.51 & 15,405.99 \\ 
  median\_CRPS & 4,077.85  & $\mathbf{2,377.34}$  &  4,251.77 \\ 
  PICP\_99 & $\mathbf{0.60}$ &  0.37 & 0.12 \\ 
  IS\_99 & {101,642.41} & $\mathbf{385,729.23}$ & 71,808.31 \\ 
  PICP\_95 & $\mathbf{0.57}$  & 0.29  & 0.11 \\ 
  IS\_95 & {66,168.42} & $\mathbf{66,284.85}$  & 61,955.80 \\ 
   \hline
\end{tabular}
\caption{CRPS, IS, PICP  for the RBNS claims of the 3 competing micro-reserving methods on the subset of Allianz bodily injury claims.} 
\label{tab:comparemicroBE}
\end{table}

{From Table \ref{tab:comparemicroBE}, we observe that the predictive distribution produced by mCube provides a better representation of the true observed reserves than the two other micro-reserving models. mCube obtains the lowest mean CRPS, representing that the predictive distribution for the RBNS claims are the most accurate. However, the method from  \cite{Bettonville2020} obtains the lowest median CRPS, representing the fact that it is more suited for the claims with a lower reserve amount. Moreover, the prediction interval coverage probabilities are the highest for mCube, although none of the methods have the required coverage of 95\% or 99\%. }

{We also compare the observed reserves with the mean of the predictive distributions of the methods. To this end, we use the following pointwise accuracy measures, where $R_{k,\tau}$ represents the true observed reserve and  $\hat{R}_{k,\tau}$ the mean of the predictive distribution obtained from the methods: 
\begin{itemize}
    \item bias := $\sum_{k}({R}_{k,\tau} - \hat{R}_{k,\tau})$
    \item Mean Absolute Error (MAE) := $mean( \mid  \hat{R}_{k,\tau} - R_{k,\tau} \mid )$
    \item Root Mean Square Error (RMSE) := $\sqrt{mean( \mid  \hat{R}_{k,\tau} - R_{k,\tau} \mid ^2)}$
    \item Symmetric Mean Absolute Percentage Error (sMAPE):= $mean(200 \times  \mid {R}_{k,\tau} - \hat{R}_{k,\tau}  \mid  / ({R}_{k,\tau} + \hat{R}_{k,\tau}))$
\end{itemize}}


\begin{table}[ht]
\centering
\begin{tabular}{rrrr}
  \hline
 & mCube &  Bettonville & Crevecoeur \\ 
  \hline
bias & -582.34  & 2,145.68 & $\mathbf{-130.53}$ \\ 
  MAE & $\mathbf{16,089.18}$  & 24,746.39 & 21,442.60 \\ 
  RMSE &  $\mathbf{41,119.29}$ &  100,108.94 &  48,911.23 \\ 
  sMAPE & $\mathbf{1.30}$  & 1.63 & 1.59 \\ 
   \hline
\end{tabular}
\caption{Pointwise accuracy measures for the individual RBNS reserves on the subset of Allianz bodily injury claims.} \label{tab:comparemicro}
\end{table}

{From Table \ref{tab:comparemicro}, we observe that mCube produces the best pointwise forecast for most of the metrics under consideration. We remark however that the method of  \cite{CREVECOEUR2022} has the lowest absolute bias. These findings highlights the superiority of mCube with respect to the other methods under consideration for the data set of Allianz bodily injury claims.}

\subsection{ Best estimate comparison with the chain-ladder and other micro-reserving models}

\begin{figure}[!htbp]
\centering
\includegraphics[width=0.8\linewidth]{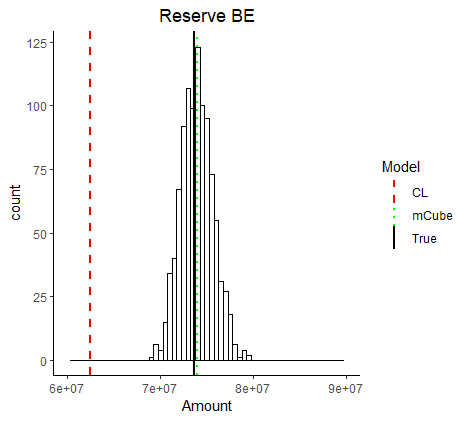}
\caption{Best estimate distribution for all claims (RBNS and IBNR)\label{fig:ibnr+rbnsreserves}}
\end{figure}

In this section, we compare the performance of mCube to the chain-ladder and other micro-reserving models for predicting best estimate reserves on the same Allianz data set.
To obtain the reserve predicted by mCube, we follow Sections \ref{sec:openclaims}  and \ref{sec:IBNRclaims}. Given that most claims are reported within one month of their occurrence as shown in Figure \ref{fig:repDel}, we use $\hat{\lambda}_{oc,0} = 1$ in Equation \eqref{eq:repDel}. 
From Figure \ref{fig:ibnr+rbnsreserves}, we observe the good performance of mCube on the prediction of the reserves as the true reserve is near the center of the best estimate distribution. In particular,  from Table \ref{table:meanBE}, we observe that we perform very well on the RBNS reserves, but less well on the IBNR reserves given that we do not model the number of IBNR claims sufficiently well as explained in the previous section. The chain-ladder however under performs on this data set, showing the added value of our proposed methodology. The predicted reserve by mCube is 73,916,247 giving a percentage error (PE) of 0.37\%, whereas the chain-ladder predicts 62,433,801 giving a percentage error of -15\%. {Given that the method from  \cite{CREVECOEUR2022} does not produce IBNR claims, we choose to simulate reserves of IBNR claims for  \cite{Bettonville2020} and  \cite{CREVECOEUR2022} using the methodology explained in Section \ref{sec:IBNR}. The methods produce reserves of 58,104,262 and 78,436,112, given a percentage error of respectively 21\% and 7\%.}

\begin{table}[!htbp]
\centering
\begin{tabular}{c c c c c c}
\hline\hline
 &Database & mCube & Chain-ladder & Bettonville & Crevecoeur  \\
\hline
RBNS reserve &69,837,157  & 72,689,539 &62,433,801 & 57,628,135& 68,725,695\\
 IBNR reserve &3,808,605 &1,226,708 & -& 476,127 & 9,710,417 \\
 Total reserve& 73,645,764& 73,916,247 &62,433,801& 58,104,262  & 78,436,112 \\
 PE & 0 & 0.37\% & -15\% & -21\% & 7\%\\ 
 \hline\hline
\end{tabular}
\caption{Observed reserves and mean of the predicted reserves for the subset of bodily injury claims. \label{table:meanBE} }
\end{table}

\section{Conclusion}
\label{sec:conclusion}
In this article we have presented a multinomial multi-state micro-level (mCube) model to estimate the reserves of IBNR and RBNS claims. We present a semi-parametric modelling of the payment distribution, taking into account claim specific information.  On a portfolio level, the proposed model is unbiased 
and produces a best estimate distribution that is centered around the true reserve. Moreover, the estimates on an individual level are very accurate in our real data analysis.
Future studies could replace the multinomial models used for the time and payment processes with more flexible machine learning models to obtain a higher predictive power.

\section*{Acknowledgements}
The authors gratefully acknowledge the financial support from the Allianz Research Chair \textit{Prescriptive business analytics in insurance} at KU Leuven.

\bigskip
\begin{center}
{\large\bf Appendix}
\end{center}

\subsection*{A. Variables in the data set}

The original data set obtained from the European insurer was further pre-processed by the following steps:
\begin{itemize}

    \item As discussed in Section \ref{sec:RBNS}, a transition occurs in the multi-state model when a payment was recorded higher than \textit{"minPayVal"} (or lower than -\textit{"minPayVal"}  in case of a reimbursement). Therefore, payments are lumped together when necessary.
    \item Several variables were created based on the time on which the claim occured, the time of reporting of the claim,   and the time on which payments were made. These variables include \textit{"fastRep"}, which is is an indicator of if the claim was reported less than 30 days from when it occured and \textit{"finYear"} which is the financial year in which a payment happens. Other variables that are also created are  \textit{"deltRep"} representing the reporting delay, \textit{"inStateTime"} which is the time a claim spends in a specific state and \textit{"inProcTime"} which is the time a claim spends in the entire multi-state process. The variable \textit{"delt1PayTimeTrans"} is the time since the previous payment. All these time variables are expressed in the equivalent number periods of 30 days. Moreover,  a maximum of 15 periods have been set such that if a  claim has more  than 15 periods, it is either forced to move to the next state or out of the multi-state process if it has reached the maximum number of transitions. 
    
    \item Variables are created from the payment amount: The variable "delt0Pay" is the  amount of the payment in the current state,  The variable "delt1Pay" is the payment amount of the previous state, and the variable cumDelt1Pay is the cumulative payment amount from all the previous states of the claim.  
\end{itemize}

\subsection*{B. Building the IBNR data set}

 Once the yearly number of IBNR claims have been estimated and the monthly reporting delay evaluated, an IBNR multi-state data set must be constructed and passed through the multi-state process starting from  the reporting state. The variables from Table \ref{table:variablesinDT} are constructed in the following way: The reporting delay (deltRep) depends on the accident year and the  estimated reporting month, fastRep is 0 as the claim is IBNR, procTime and stateTime are both 1, deltPay, deltPayTime and cumDeltPay are NA as there has been no  previous payment.

\subsection*{C. Algorithm for simulating open claim reserves trajectories}

This section presents the algorithm for simulating claims reserves as illustrated and explained in Section \ref{sec:openclaims}. For this algorithm, We need the following elements:
 
 \begin{itemize}
     \item \textbf{timeMods}: The fitted time models. This should be a list of length \textit{"maxMod"}. 
     \item \textbf{payMods}: The fitted payment models. This should be a list of length \textit{"maxMod"}.
     \item \textbf{testData}: The test data on which to simulate reserves. 
     \item \textbf{splits}: Splitting points for the numeric variables that were binned. This should also contain the levels for the time variables that are categorized.
     \item \textbf{fixedTimeMax}: Maximum amount of time a claim is allowed to stay in a state. This should be of length \textit{"maxMod"}.
     \item \textbf{nSims}: Number of trajectories to be simulated for each claim. 
     \item \textbf{npmax}: Maximum number of transitions we allow a claim to make. This is used to capture claims with longer developments. 
 \end{itemize}
 
 We note that when a claim has stayed for too long in a state as defined by \textit{"fixedStateTimeMax"}, we modify the estimated discrete-time hazard functions to 
 
  \begin{align}
 \label{timeprocmultinommodif}
        \tilde{\lambda}_{j,j+1}(t \mid \mathbf{x_{k,t}}) &= \hat{\lambda}_{j,j+1}(t \mid \mathbf{x_{k,t}}) + \frac{ 1 - \hat{\lambda}_{j,j+1}(t \mid \mathbf{x_{k,t}}) - \hat{\lambda}_{j,tp}(t \mid \mathbf{x_{k,t}}) - \hat{\lambda}_{j,tn}(t \mid \mathbf{x_{k,t}})}{3}   \nonumber,\\
         \tilde{\lambda}_{j,{j}}(t \mid \mathbf{x_{k,t}}) &= \hat{\lambda}_{j,tp}(t \mid \mathbf{x_{k,t}}) + \frac{ 1 - \hat{\lambda}_{j,j+1}(t \mid \mathbf{x_{k,t}}) - \hat{\lambda}_{j,tp}(t \mid \mathbf{x_{k,t}}) - \hat{\lambda}_{j,tn}(t \mid \mathbf{x_{k,t}})}{3}   \nonumber,\\
          \tilde{\lambda}_{j,{tn}}(t \mid \mathbf{x_{k,t}}) &= \hat{\lambda}_{j,tn}(t \mid \mathbf{x_{k,t}}) + \frac{ 1 - \hat{\lambda}_{j,j+1}(t \mid \mathbf{x_{k,t}}) - \hat{\lambda}_{j,tp}(t \mid \mathbf{x_{k,t}}) - \hat{\lambda}_{j,tn}(t \mid \mathbf{x_{k,t}})}{3}  \nonumber,\\
         \tilde{\lambda}_{j},{j}(t \mid \mathbf{x_{k,t}}) &= 0  \nonumber \\
    \end{align}

    Similarly, when a claim has reached state \textit{"npmax"}-1, we need to modify the transition probabilities in the following way: 
    
\begin{align}
 \label{timeprocmultinommodif2}
        \tilde{\tilde{\lambda}}_{npmax-1,npmax}(t \mid \mathbf{x_{k,t}}) & = 0   \nonumber,\\
          \tilde{\tilde{\lambda}}_{npmax-1,{tp}}(t \mid \mathbf{x_{k,t}}) &= \hat{\lambda}_{npmax-1,tp}(t \mid \mathbf{x_{k,t}}) + \hat{\lambda}_{npmax-1,npmax}(t \mid \mathbf{x_{k,t}}) \nonumber,\\
          \tilde{\tilde{\lambda}}_{npmax-1,tn}(t \mid \mathbf{x_{k,t}}) & = {\hat{\lambda}}_{npmax-1,tn}(t \mid \mathbf{x_{k,t}})  \nonumber,\\ 
          \tilde{\tilde{\lambda}}_{npmax-1,npmax-1}(t \mid \mathbf{x_{k,t}}) & = {\hat{\lambda}}_{npmax-1,npmax-1}(t \mid \mathbf{x_{k,t}})  \nonumber,\\ 
\end{align}

If a claim stays too long in state \textit{"npmax"}-1, we can also modify \eqref{timeprocmultinommodif2} using \eqref{timeprocmultinommodif}.

 \subsection*{D. Selecting hyper-parameters}
 During the pre-processing stage, we need to decide on the value for the hyper-parameters \textit{"nMinLev"} and \textit{"nGroups"} that are necessary for binning the continuous predictors. We suggest to set \textit{"nMinLev"} to at least 30 for statistical significance of estimated parameters. We suggest to set \textit{"nGroups"} between 5 and 15. We also have to define \textit{"minPayVal"}, which is the minimum amount paid for a non-terminal payment to be considered. Intermediate payments that are lower in absolute value than this amount, will be aggregated and considered as a single payment. We suggest to discuss with the business to determine what is a meaningful payment amount. We also define \textit{"perLen"}, which is the  number of days in one time period. We recommend to choose \textit{"perLen"} so that it represents either monthly, quarterly or yearly information.  
 For binning the continuous variable \textit{"inStateTime"}, representing the time a claim spends in a specific state, the minimum number of observations in each category should be \textit{"nMinTimeLev"}. Note that this hyper-parameter can differ from \textit{"nMinLev"}. We choose a value of 30 for statistical signifacnce of estimated parameters. Other hyper-parameters include \textit{"nMaxLevInstate"}, the maximum number of time periods a claim is allowed to stay in the same state and \textit{"nMaxLevInProc"}, the maximum number of periods a claim is allowed to stay in the whole multi-state process. These two hyper-parameters should be chosen so that only a small percentage, say 1 \% of the claims stay for more than these hyper-parameters in the state or in the process respectively. In order to have valid statistical models, we need a sufficient number of observations. Therefore, we define  \textit{"nMinModT"} as the minimum number of observations required to fit a multinomial model with predictors in the time process and we define \textit{"nMinNoModT"} similarly in case of no predictors. Note that in case the chosen value for \textit{"nMinModT"} is smaller than the number of predictors multiplied with \textit{"nTimesParamT"}, it is replaced by this product. However, it is quite likely that the number of required observations is not met for the time model in state $S_{npmax-1}$, since claims with a large number of payments are rare. Therefore, it is decided to construct \textit{"maxMod"} unique time models for states $S_1$, \ldots, $S_{maxMod}$. For the states $S_{maxMod+1}$,\ldots, $S_{npmax-1}$ the model of state $S_{maxMod}$ will be reused. This implies that the model of state $S_{maxMod}$ is based on payments that happened from the $maxMod^{th}$ payment on for each claim.


During the modelling of the payment process, we have to choose \textit{"nBins"} which equals $L+1$ and thus represents the  number of bins during the splicing procedure. Besides the number of bins, the splitting points $\mathbf{b}$ themselves need to be determined as well. Finally, we define  \textit{"nMinModP"} as the minimum number of observations to fit a multinomial model with predictors in the payment process and we define \textit{"nMinNoModP"} similarly in case of no predictors. Once again, when the chosen value for \textit{"nMinModP"} is smaller than the number of predictors multiplied with \textit{"nTimesParamP"}, it is replaced by this product. 

When simulating trajectories for open claims, we need to choose a value for \textit{"nSims"}, \textit{"fixedTimeMax"}, and \textit{"npmax"}. The value for \textit{"fixedTimeMax"} should make business sense, and should be such that only a small percentage of the claims stay in a state for longer than this, hence we set it to 24. The value for \textit{"npmax"} should be large enough to capture claims with longer developments, hence we set it to 50.

Taking into account the time necessary to fit our multi-state process, a cross validation strategy for hyper-parameter tuning proved to be extremely time consuming. We therefore propose to set values for these hyper-parameters based on actuarial experience and observed results.

\begin{table}[!htbp]
\centering
\begin{tabular}{cc}
\hline\hline
 & Hyper-parameters  \\
 \hline
    \multirow{ 1}{*}{Pre-processing} &  \textit{nMinLev} = 30; \textit{nGroups} = 5;\\
    &
    \textit{minPayVal} = 200;  \textit{perLen} = 30 \\
   \hline
   \multirow{3}{*}{Time process} &   \textit{nMinTimeLev} = 30; \textit{nMaxLevInState} = 12;\\
   & \textit{nMaxLevInProc} = 24 ; \textit{maxMod} = 6\\
   &  \textit{nMinModT} = 500; \textit{nMinNoModT} = 50    \\
   &    \textit{nTimesParamsT} = 5; $npmax$ = 30   \\
   \hline 
    \multirow{2}{*}{Payment process} & \textit{nBins} = 4;  \textit{nMinModP} = 500; \textit{nMinNoModP} = 50;\\
    &\textit{nTimesParamsP} = 5 \\
    \hline
   Open claims simulations &  $N_{sim}=100$; fixedTimeMax = 24; npmax = 50\\
  \hline\hline
\end{tabular}
\caption{Hyper-parameters for the multinomial multi-state model. \label{table:Hyperparameters} }
\end{table}

\subsection*{E. Splitting points for past payment information}

The split obtained after binning the continuous predictors deltPay and cumDeltPay are shown respectively in table \ref{table:splitpointprevpay}  and \ref{table:spliitpointcumdeltpay}.

\begin{table}[!htbp]
\centering
\begin{tabular}{ c c c c c c }
\hline\hline
& \multicolumn{5}{c}{\hspace{3ex} \textbf{State}}\\
\cline{2-6}
 & $S_{1}$  & $S_{2}$  & $S_{3}$  & $S_{4}$ & $S_{5+}$\\
\hline
 &586.20 & -4045.03&   -657.02 & 0.00 & 0.00 \\
 & 1,247.06 & -1,169.81& 1,179.08 & 965.04   &  611.74  \\
& 2,584.57&  1,669.23& 3,615.08 & 3,576.90  & 2,596.18  \\
 &  7,955.07&  3,805.08& 4,775.55  & 5,070.10  & 3,501.23 \\
  \hline\hline
\end{tabular}
\caption{Splitting points for the previous payment (deltPay) in each data set.\label{table:splitpointprevpay}}
\end{table}

\begin{table}[!htbp]
\centering
\begin{tabular}{c c c c c c c }
\hline\hline
& \multicolumn{5}{c}{\hspace{3ex} \textbf{State}}\\
\cline{2-6}
 & $S_{1}$  & $S_{2}$  & $S_{3}$  & $S_{4}$ & $S_{5+}$\\
\hline
 &80.00 & 24.62 & 146.26 & 249.78 & 761.56 \\
 &380.70 & 258.52 & 1,284.04 & 4,651.18   & 4,630.25  \\
& 1,416.29 & 634.86 & 5,048.92  & 8,626.82  & 10,292.00  \\
 &  2,631.85& 7285.79 & 6,954.32  & 12,356.65  & 18,322.42  \\
  \hline\hline
\end{tabular}
\caption{Splitting points for the cumulative previous payment (cumDeltPay) in each data set.\label{table:spliitpointcumdeltpay}}
\end{table}

\subsection*{F. Partial dependence plots of the time models}
\begin{figure}[!htbp]
\begin{subfigure}{.55\textwidth}
    \centering
    \includegraphics[width=.85\linewidth]{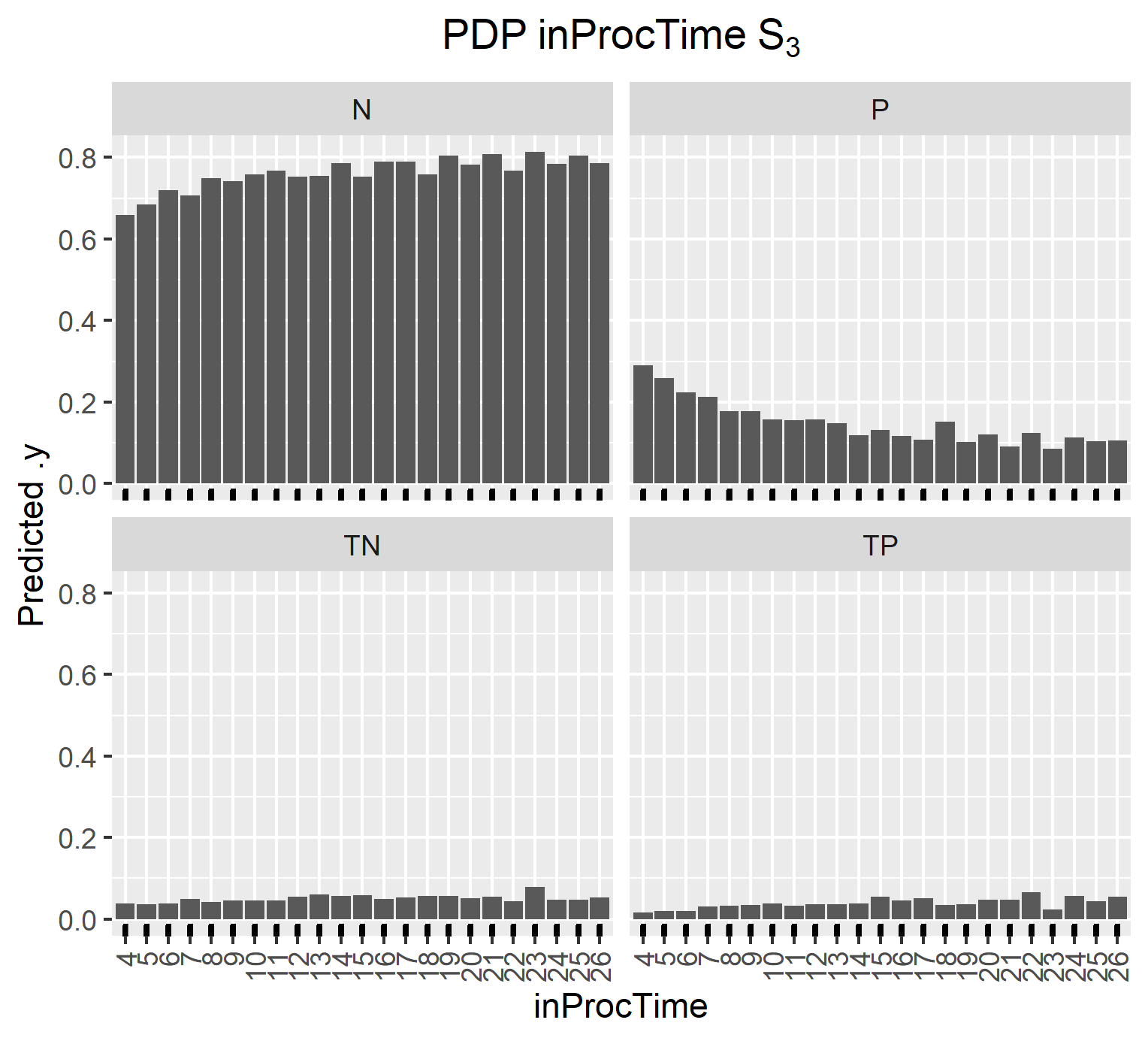}
    \caption{}
    \label{fig:pdp_proctime_s3}
\end{subfigure}
\begin{subfigure}{.55\textwidth}
    \centering
    \includegraphics[width=.85\linewidth]{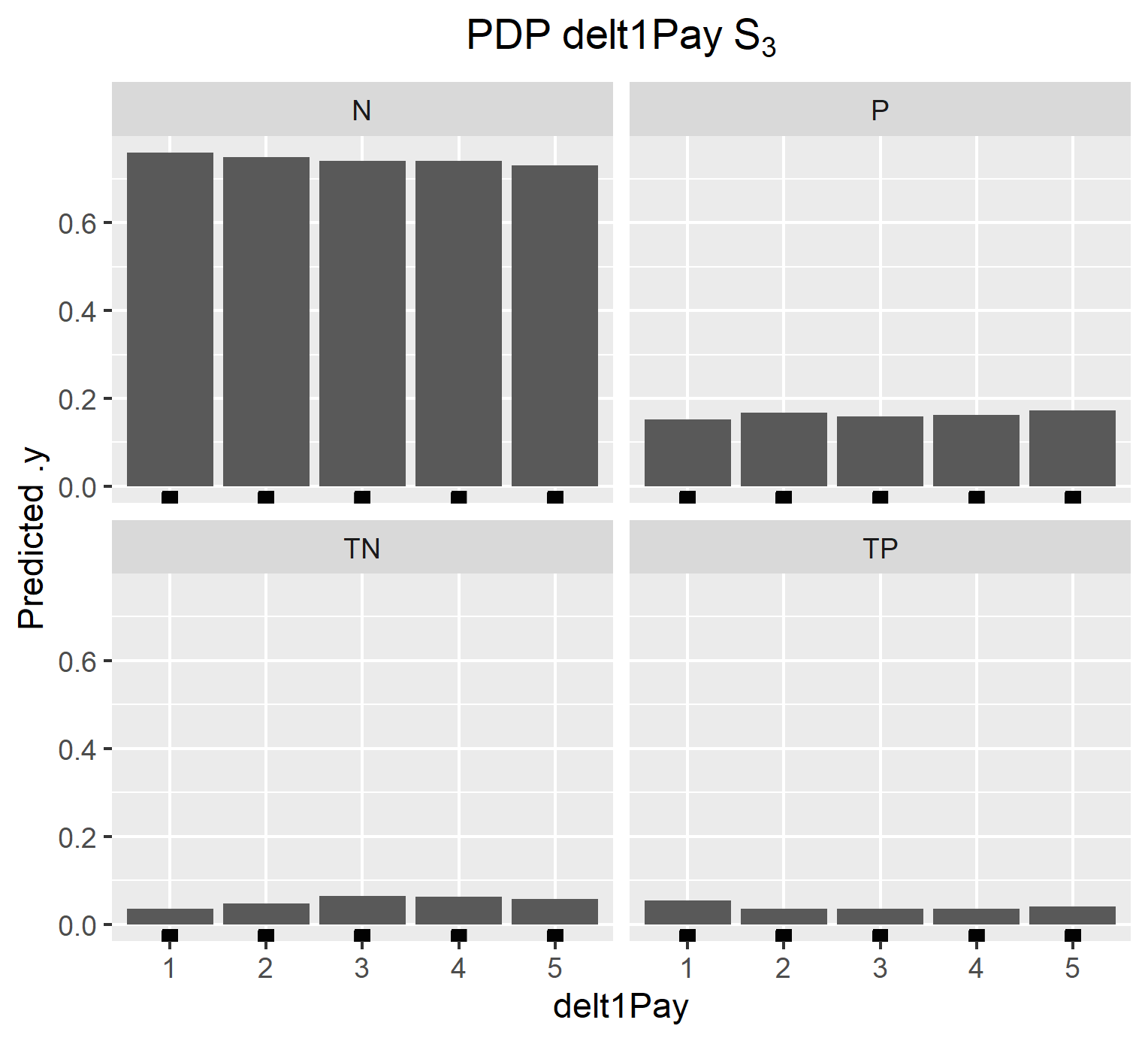}
    \caption{}
    \label{fig:pdpdeltpay_s3}
\end{subfigure}
\newline
\begin{subfigure}{.55\textwidth}
    \centering
    \includegraphics[width=.85\linewidth]{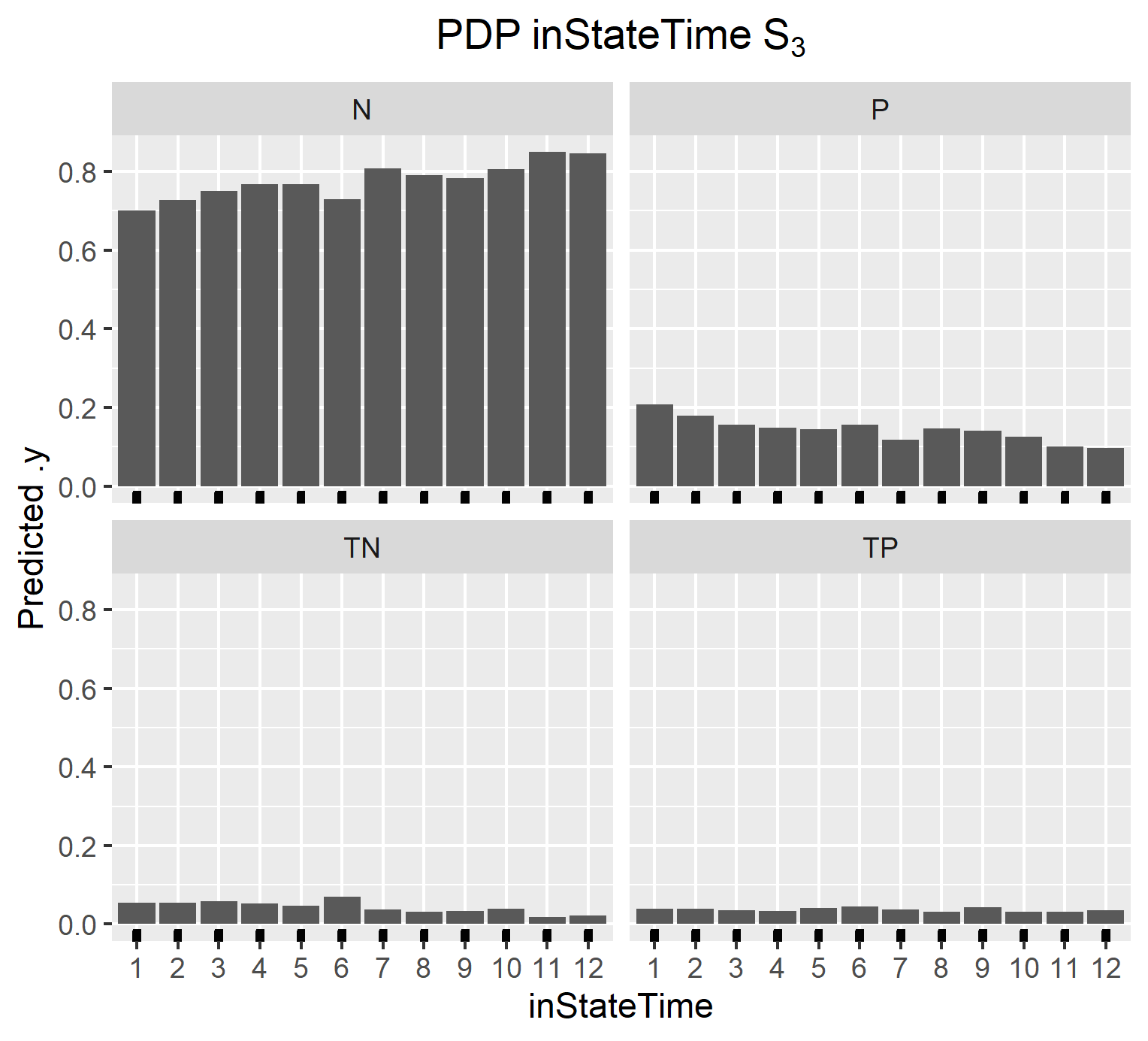}
    \caption{}
    \label{fig:pdp_statetime_s3}
\end{subfigure}
\begin{subfigure}{.55\textwidth}
    \centering
    \includegraphics[width=.85\linewidth]{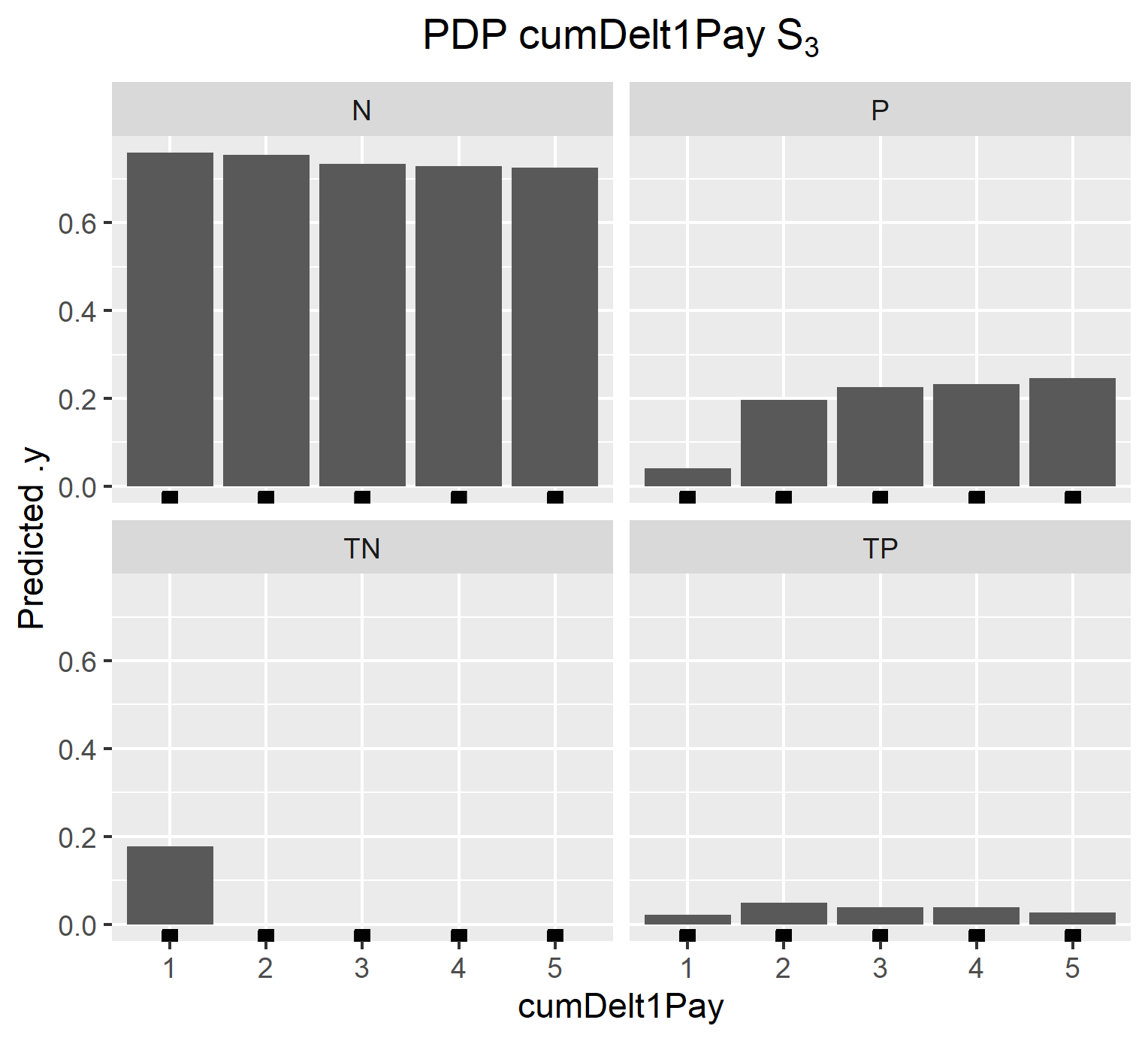}
    \caption{}
    \label{fig:pdp_cumdeltpay_s3}
\end{subfigure}
\caption{Partial dependence plots representing the marginal effect on transition probabilities of the time spent in the process (a), the previous payment size (b), the time spent in the state (c) and the cumulative previous payment size (d) for claims in $S_{3}$.}
\label{fig:pdp_s3}
\end{figure}

\begin{figure}[!htbp]
\begin{subfigure}{.55\textwidth}
    \centering
    \includegraphics[width=.85\linewidth]{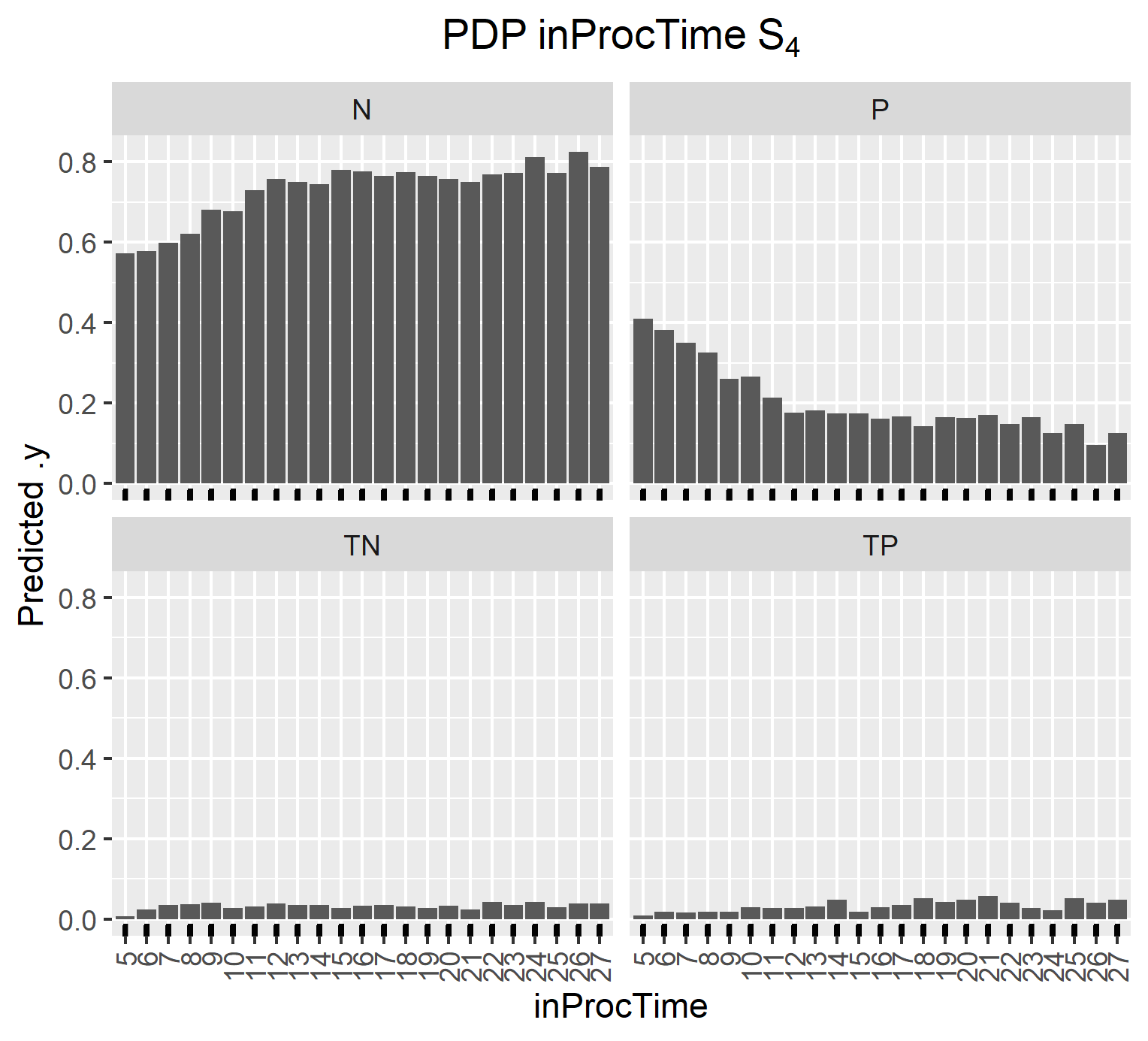}
    \caption{}
    \label{fig:pdp_proctime_s4}
\end{subfigure}
\begin{subfigure}{.55\textwidth}
    \centering
    \includegraphics[width=.85\linewidth]{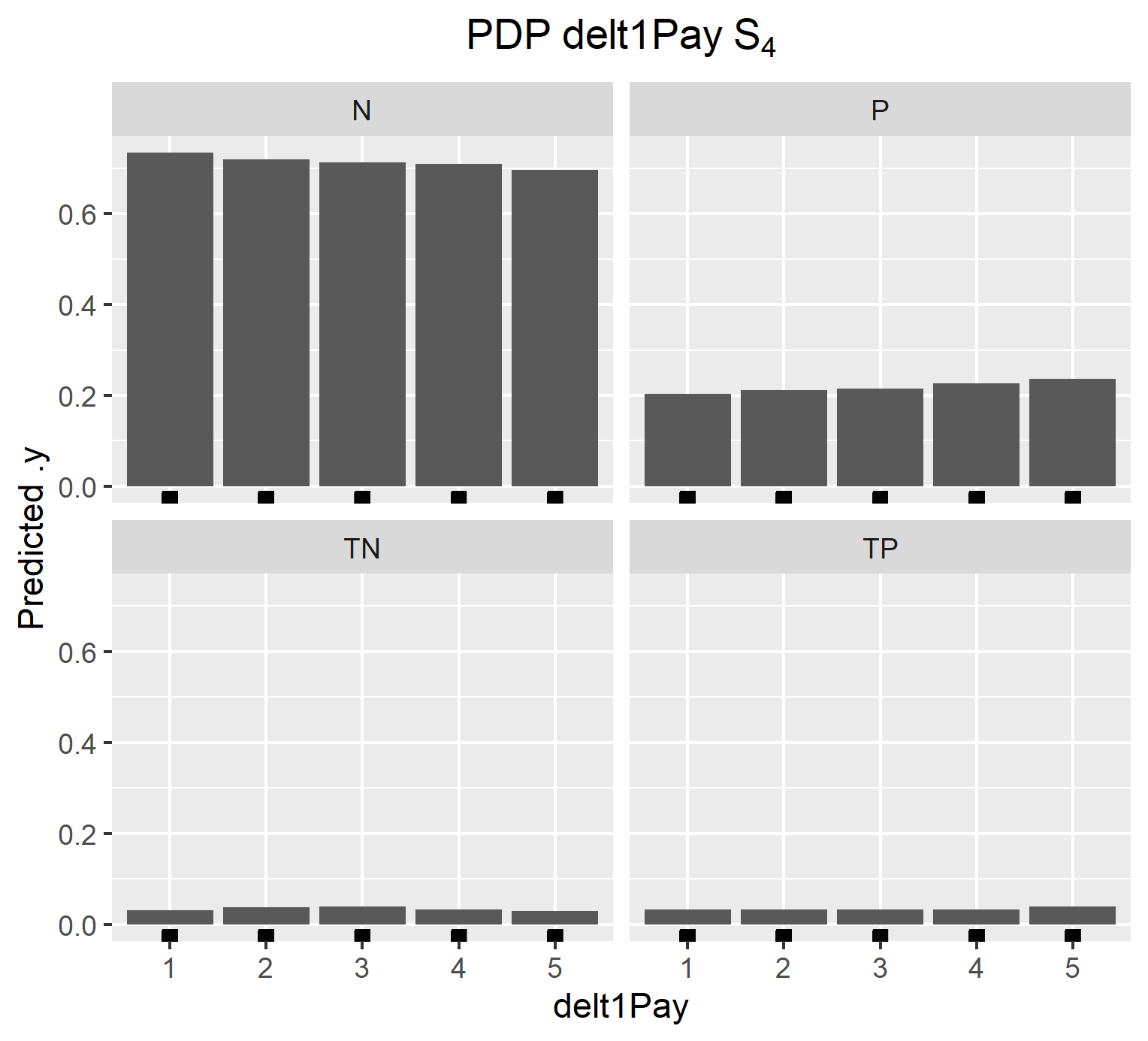}
    \caption{}
    \label{fig:pdpdeltpay_s4}
\end{subfigure}
\newline
\begin{subfigure}{.55\textwidth}
    \centering
    \includegraphics[width=.85\linewidth]{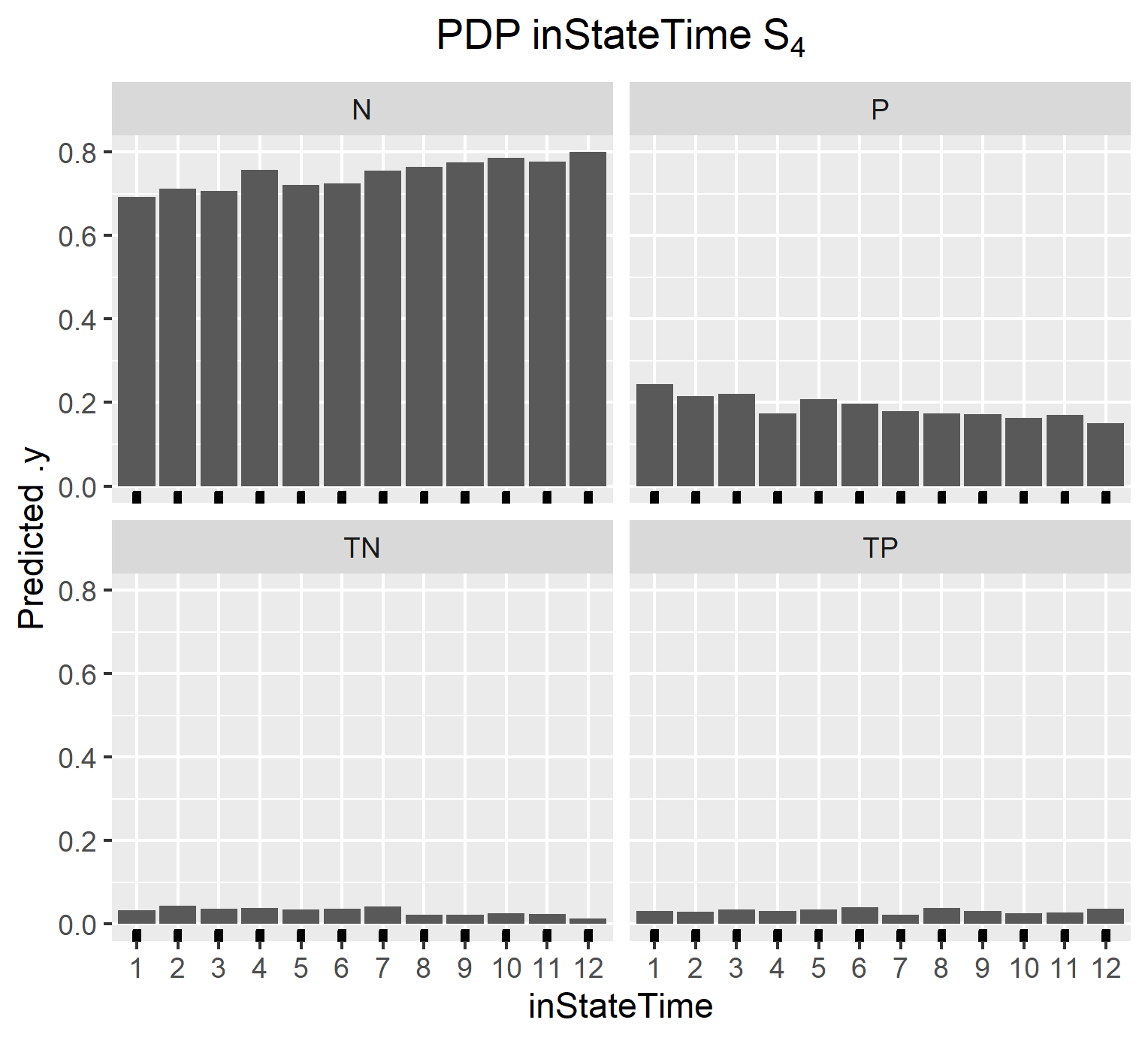}
    \caption{}
    \label{fig:pdp_statetime_s4}
\end{subfigure}
\begin{subfigure}{.55\textwidth}
    \centering
    \includegraphics[width=.85\linewidth]{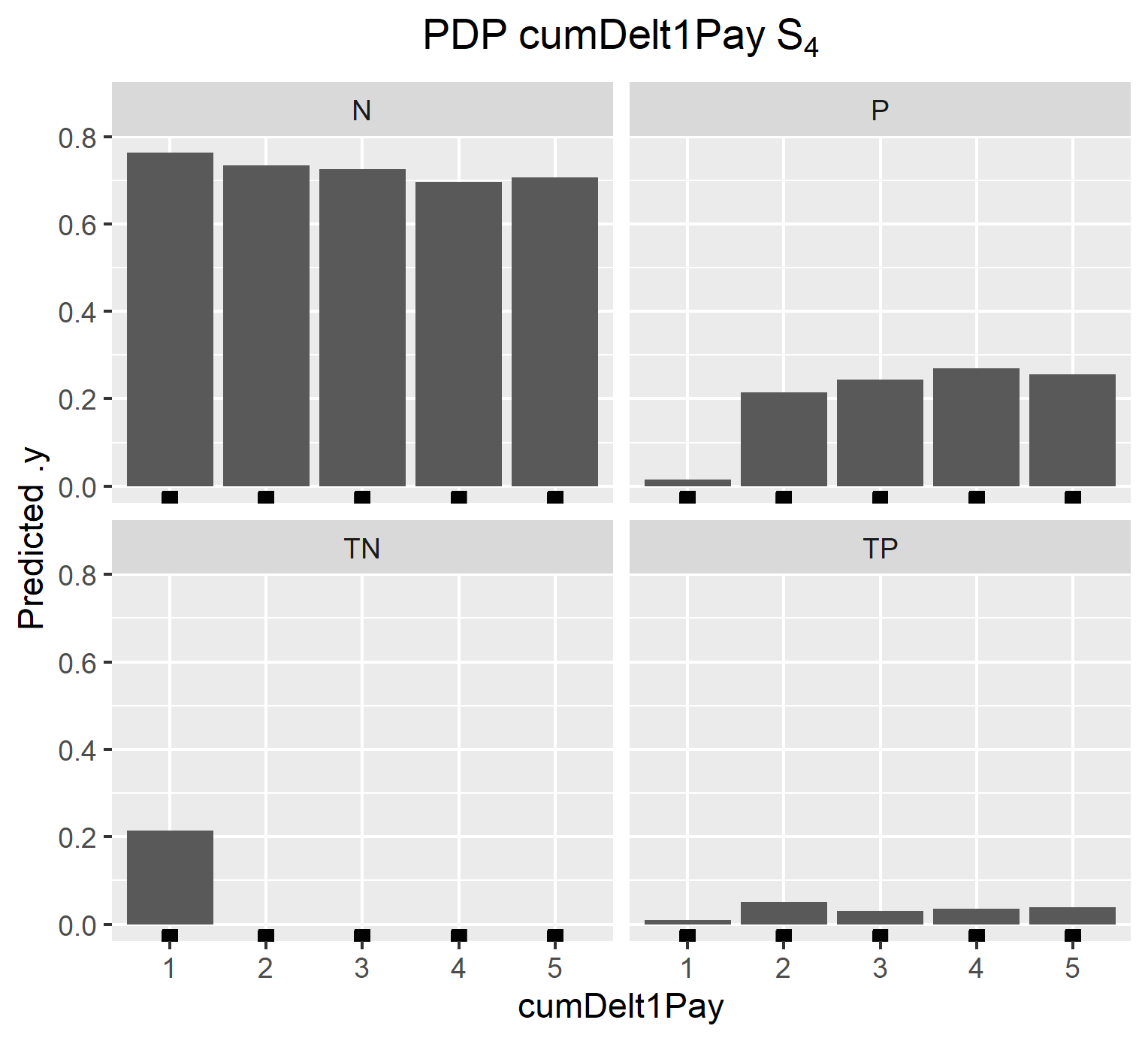}
    \caption{}
    \label{fig:pdp_cumdeltpay_s4}
\end{subfigure}
\caption{Partial dependence plots representing the marginal effect on transition probabilities of the time spent in the process (a), the previous payment size (b), the time spent in the state (c) and the cumulative previous payment size (d) for claims in $S_{4}$.}
\label{fig:pdp_s4}
\end{figure}

\begin{figure}[!htbp]
\begin{subfigure}{.55\textwidth}
    \centering
    \includegraphics[width=.85\linewidth]{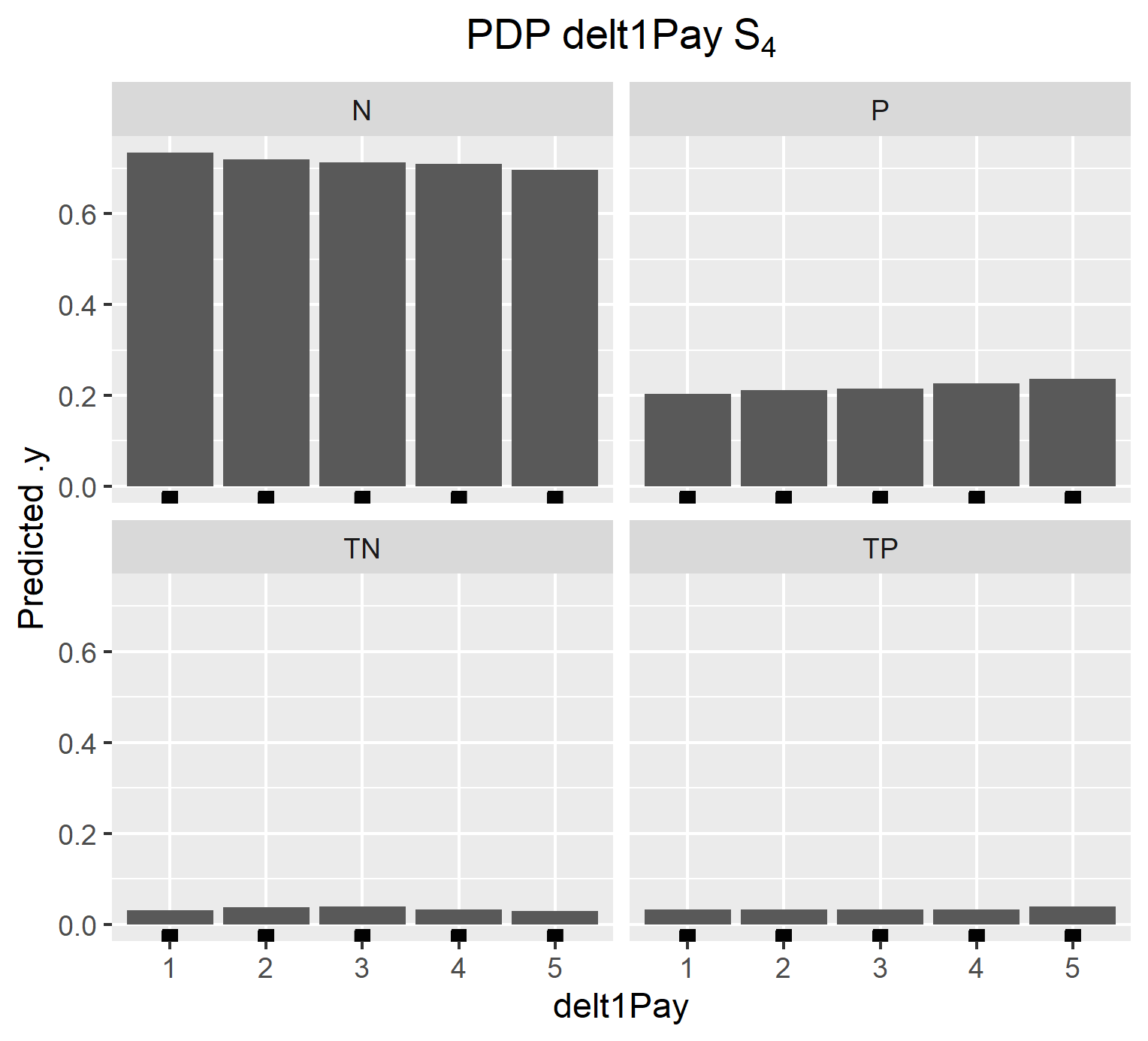}
    \caption{}
    \label{fig:pdp_proctime_s5}
\end{subfigure}
\begin{subfigure}{.55\textwidth}
    \centering
    \includegraphics[width=.85\linewidth]{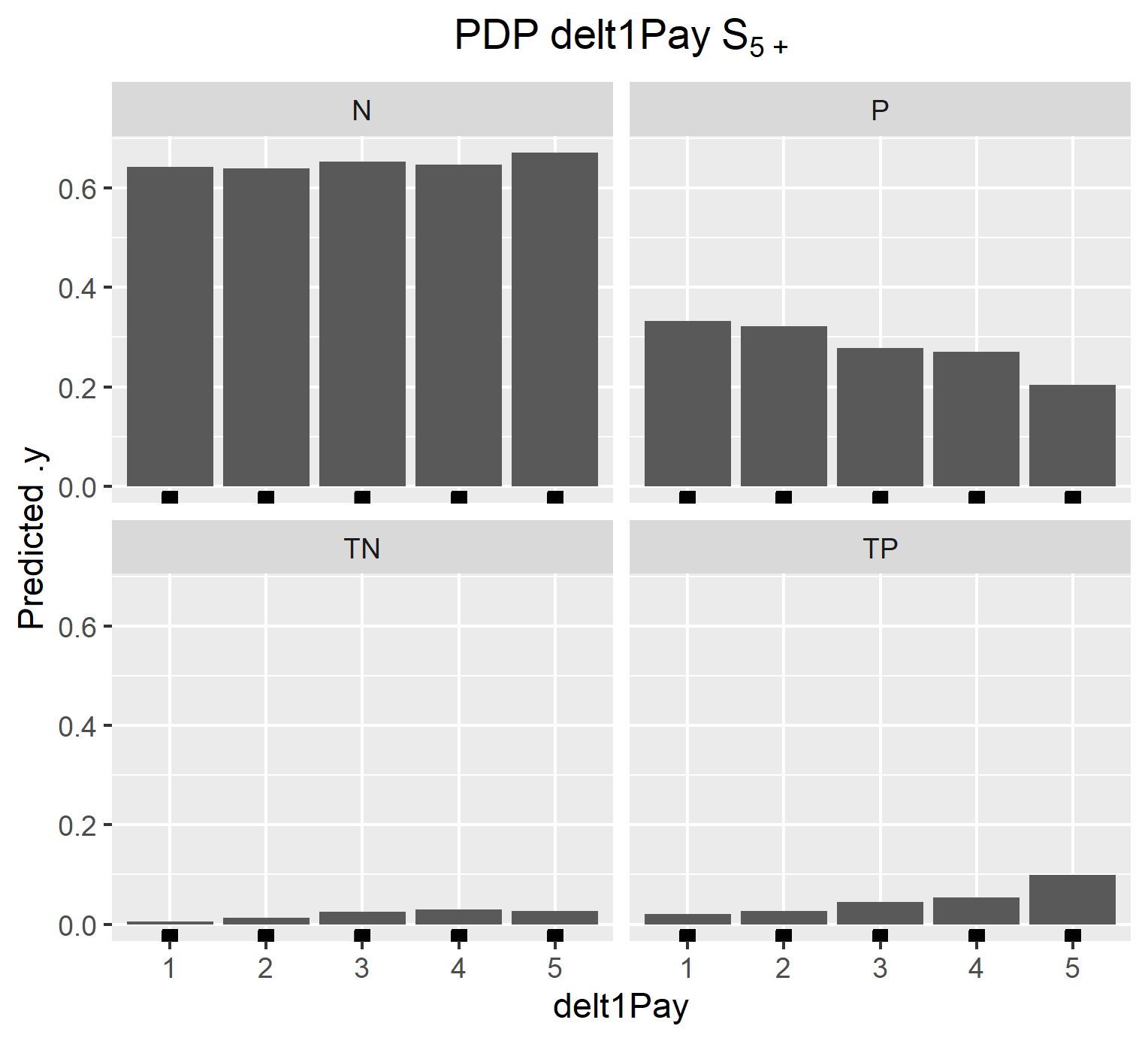}
    \caption{}
    \label{fig:pdpdeltpay_s5}
\end{subfigure}
\newline
\begin{subfigure}{.55\textwidth}
    \centering
    \includegraphics[width=.85\linewidth]{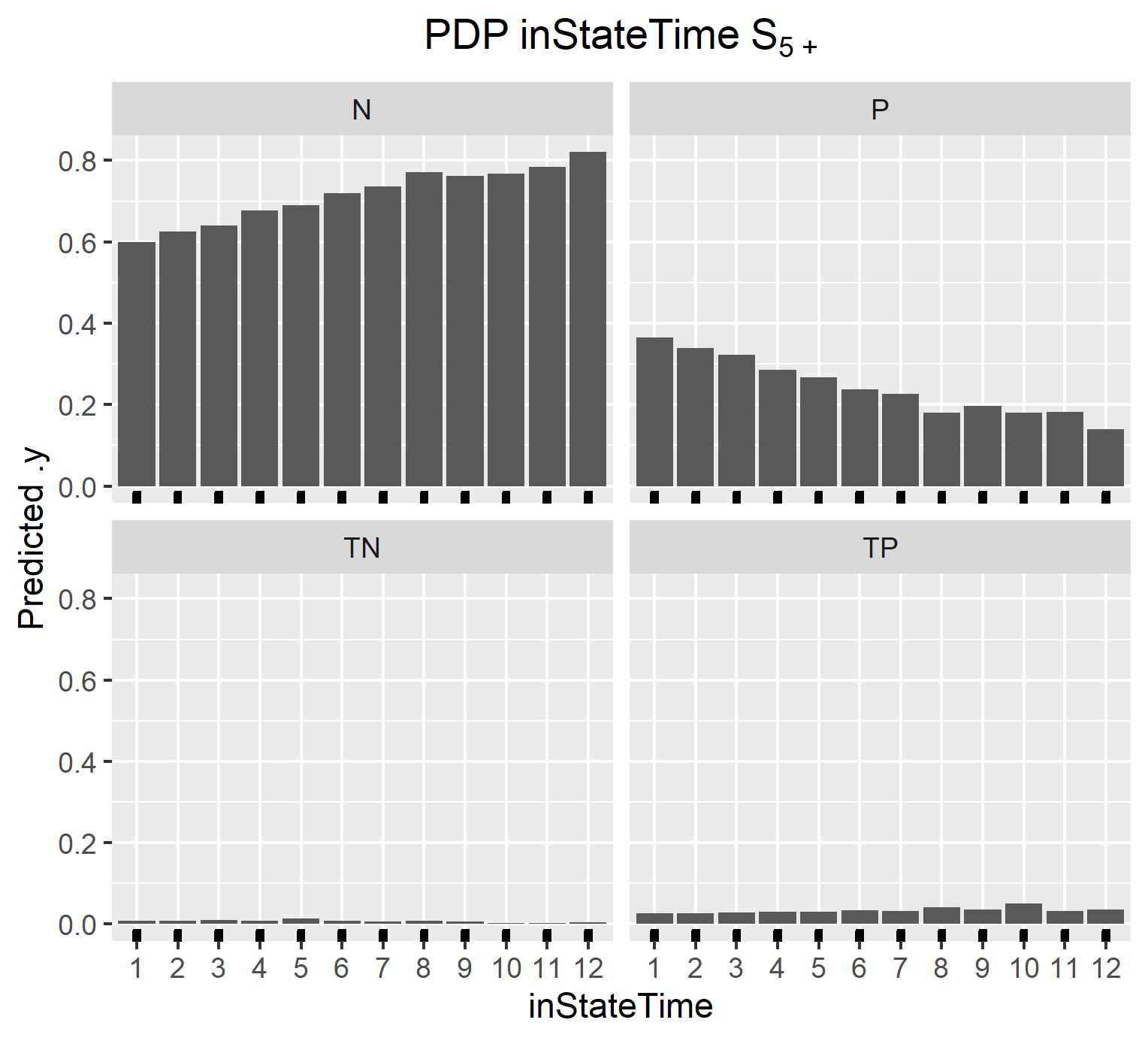}
    \caption{}
    \label{fig:pdp_statetime_s5}
\end{subfigure}
\begin{subfigure}{.55\textwidth}
    \centering
    \includegraphics[width=.85\linewidth]{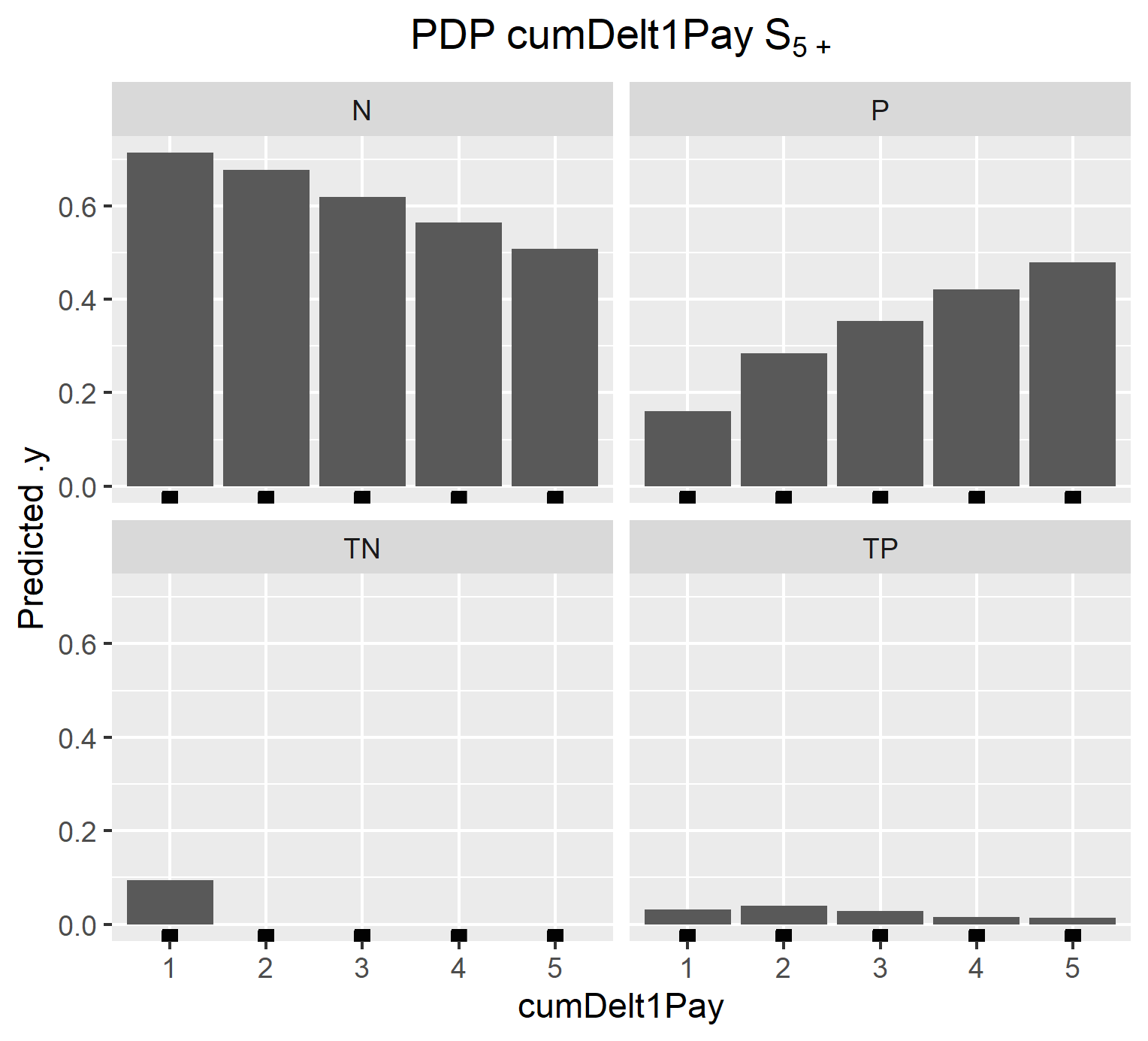}
    \caption{}
    \label{fig:pdp_cumdeltpay_s5}
\end{subfigure}
\caption{Partial dependence plots representing the marginal effect on transition probabilities of the time spent in the process (a), the previous payment size (b), the time spent in the state (c) and the cumulative previous payment size (d) for claims in $S_{5+}$.}
\label{fig:pdp_s5}
\end{figure}

In this section, we present the partial dependence plots of the time models for states $S_{3}$, $S_{4}$ and $S_{5+}$.  From figures \ref{fig:pdp_s3}, \ref{fig:pdp_s4}, \ref{fig:pdp_s5}, we observe similar marginal effects of covariates as in state $S_{2}$.

\subsection*{G. Partial dependence plots of the payment models}
\begin{figure}[!htbp]
\begin{subfigure}{.55\textwidth}
    \centering
    \includegraphics[width=.85\linewidth]{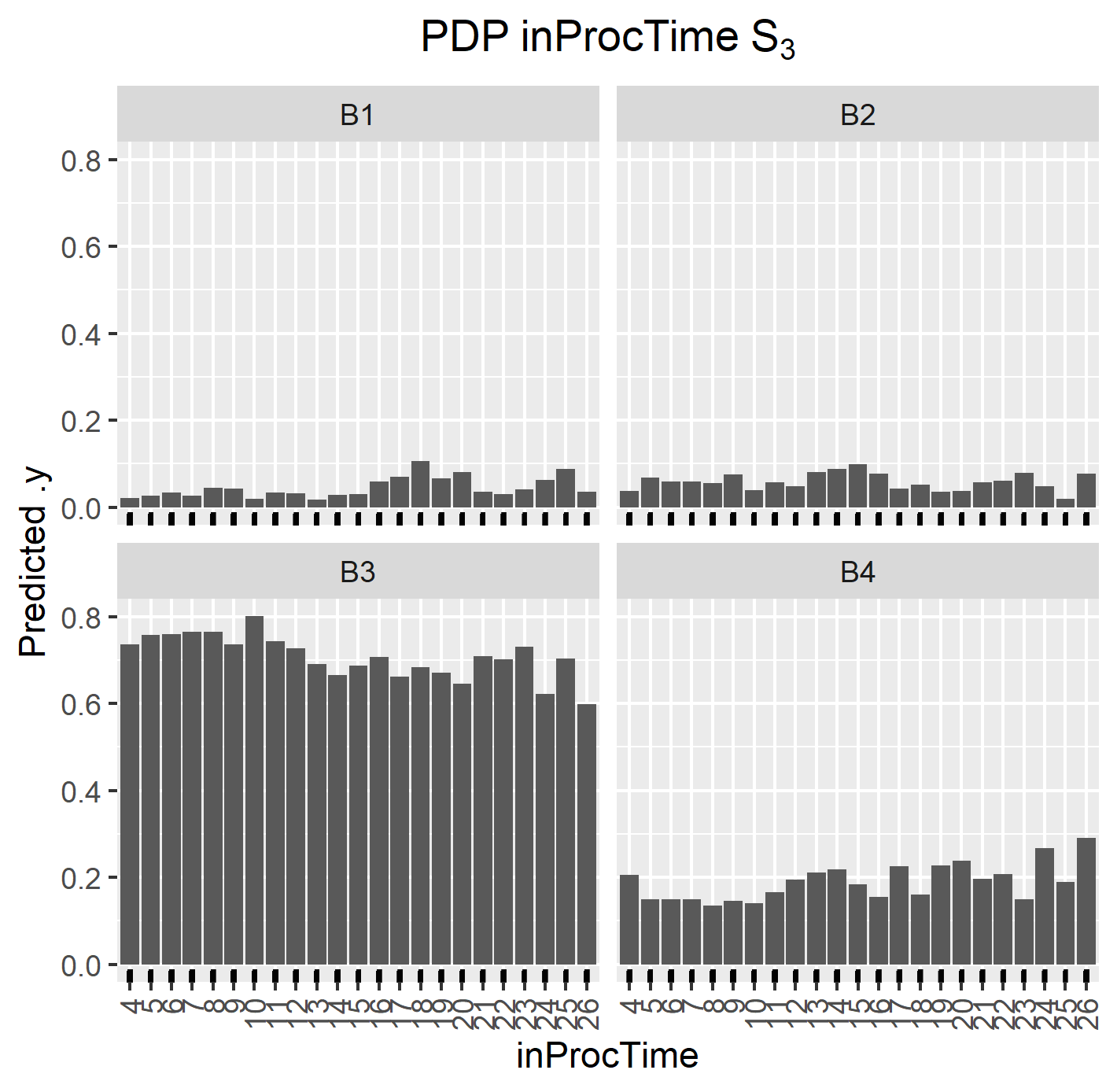}
    \caption{}
    \label{fig:pdp_proctime_s3_pay}
\end{subfigure}
\begin{subfigure}{.55\textwidth}
    \centering
    \includegraphics[width=.85\linewidth]{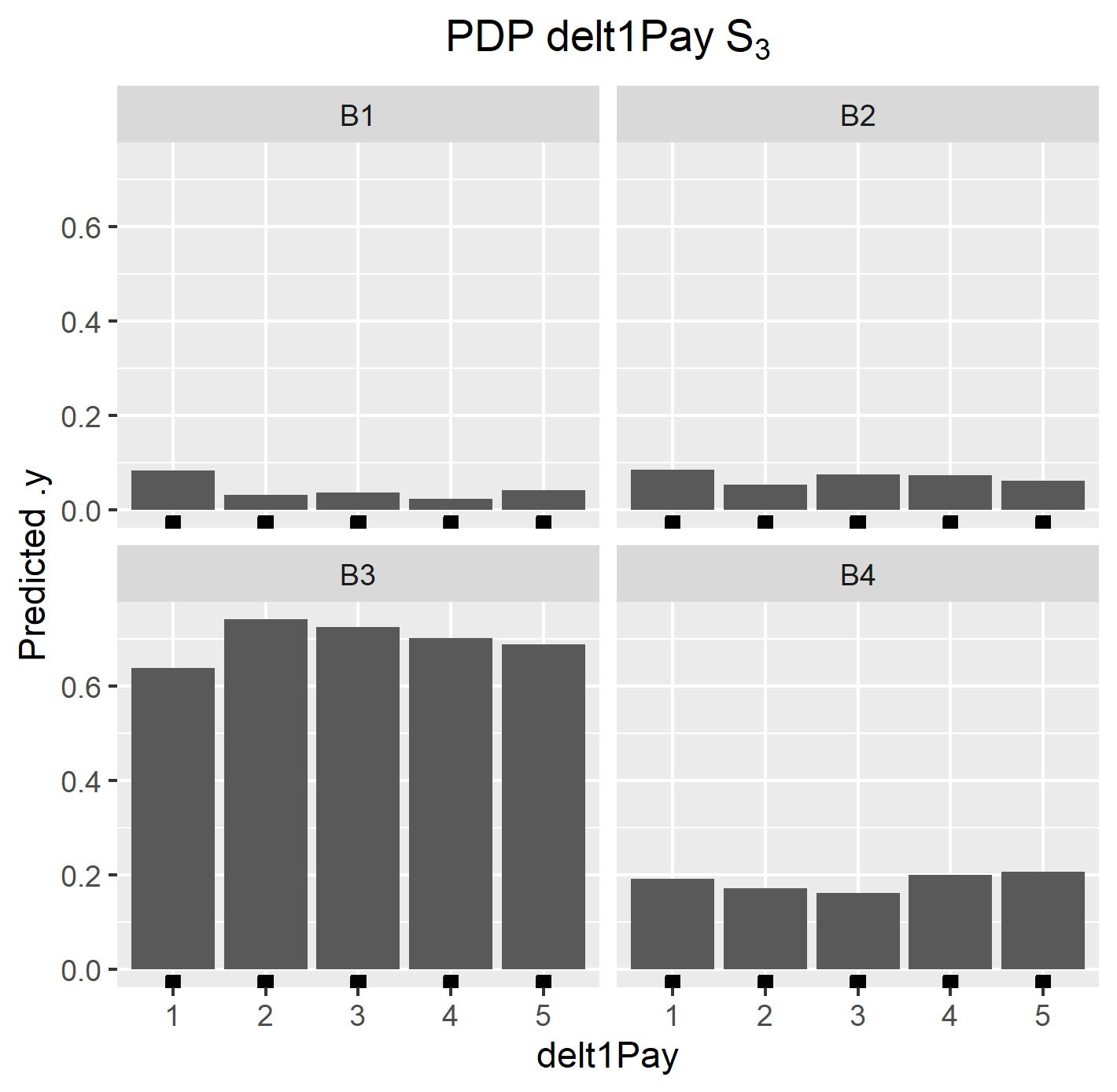}
    \caption{}
    \label{fig:pdpdeltpay_s3_pay}
\end{subfigure}
\newline
\begin{subfigure}{.55\textwidth}
    \centering
    \includegraphics[width=.85\linewidth]{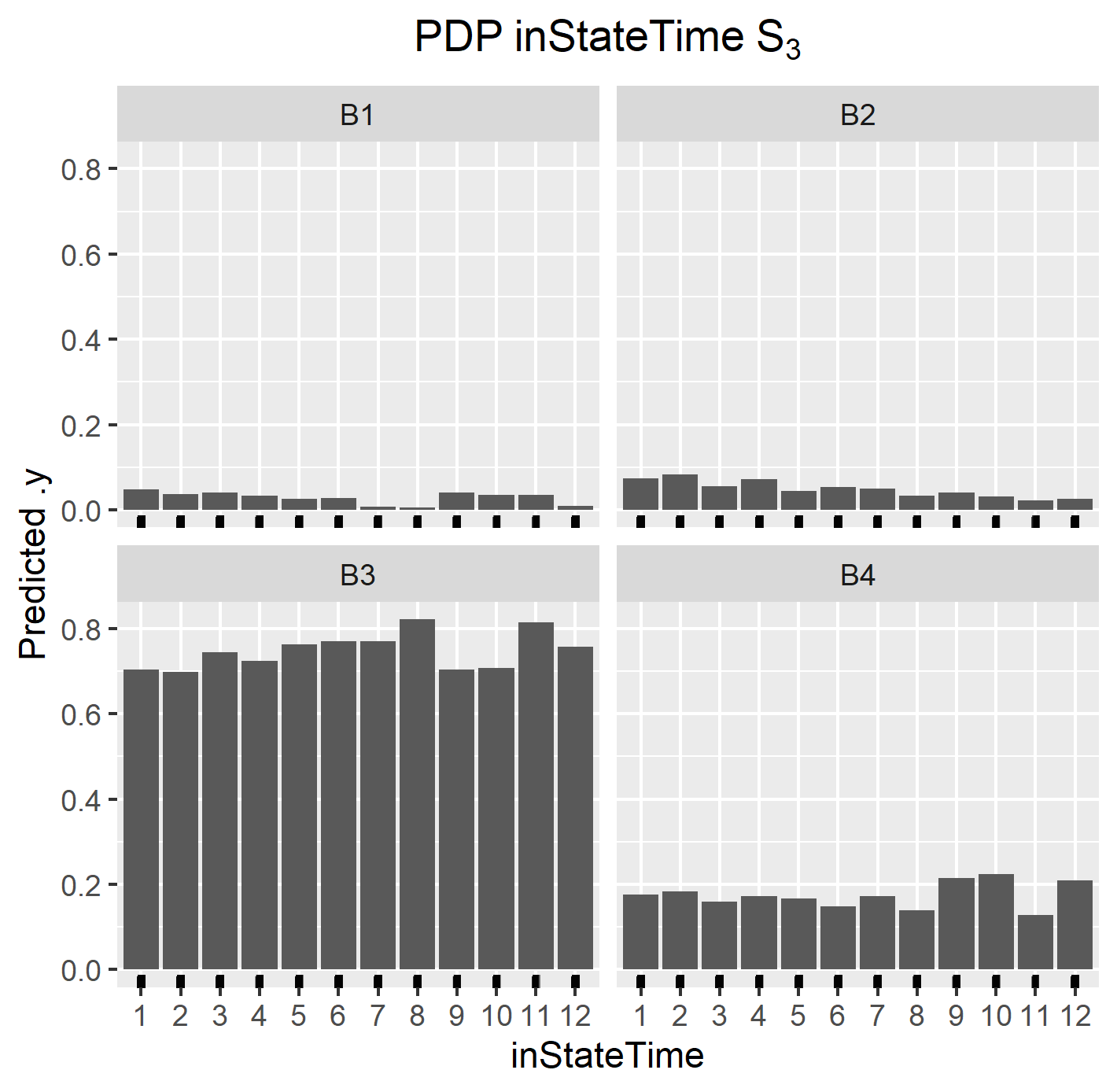}
    \caption{}
    \label{fig:pdp_statetime_s3_pay}
\end{subfigure}
\begin{subfigure}{.55\textwidth}
    \centering
    \includegraphics[width=.85\linewidth]{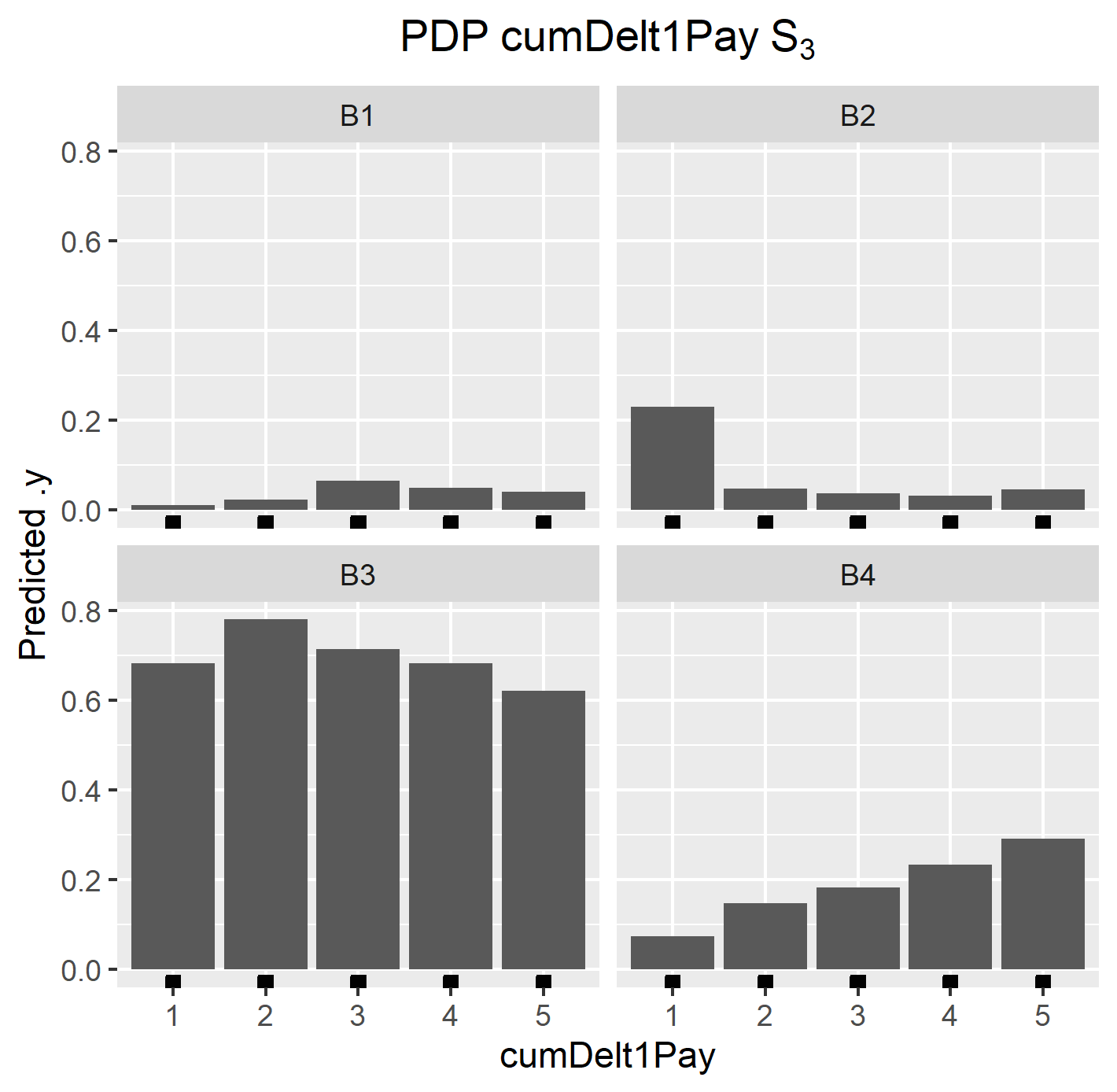}
    \caption{}
    \label{fig:pdp_cumdeltpay_s3_pay}
\end{subfigure}
\caption{Partial dependence plots representing the marginal effect on probabilities to belong to a payment bin of the time spent in the process (a), the previous payment size (b), the time spent in the state (c) and the cumulative previous payment size (d) for claims in $S_{3}$.}
\label{fig:pdp_s3_pay}
\end{figure}

\begin{figure}[!htbp]
\begin{subfigure}{.55\textwidth}
    \centering
    \includegraphics[width=.85\linewidth]{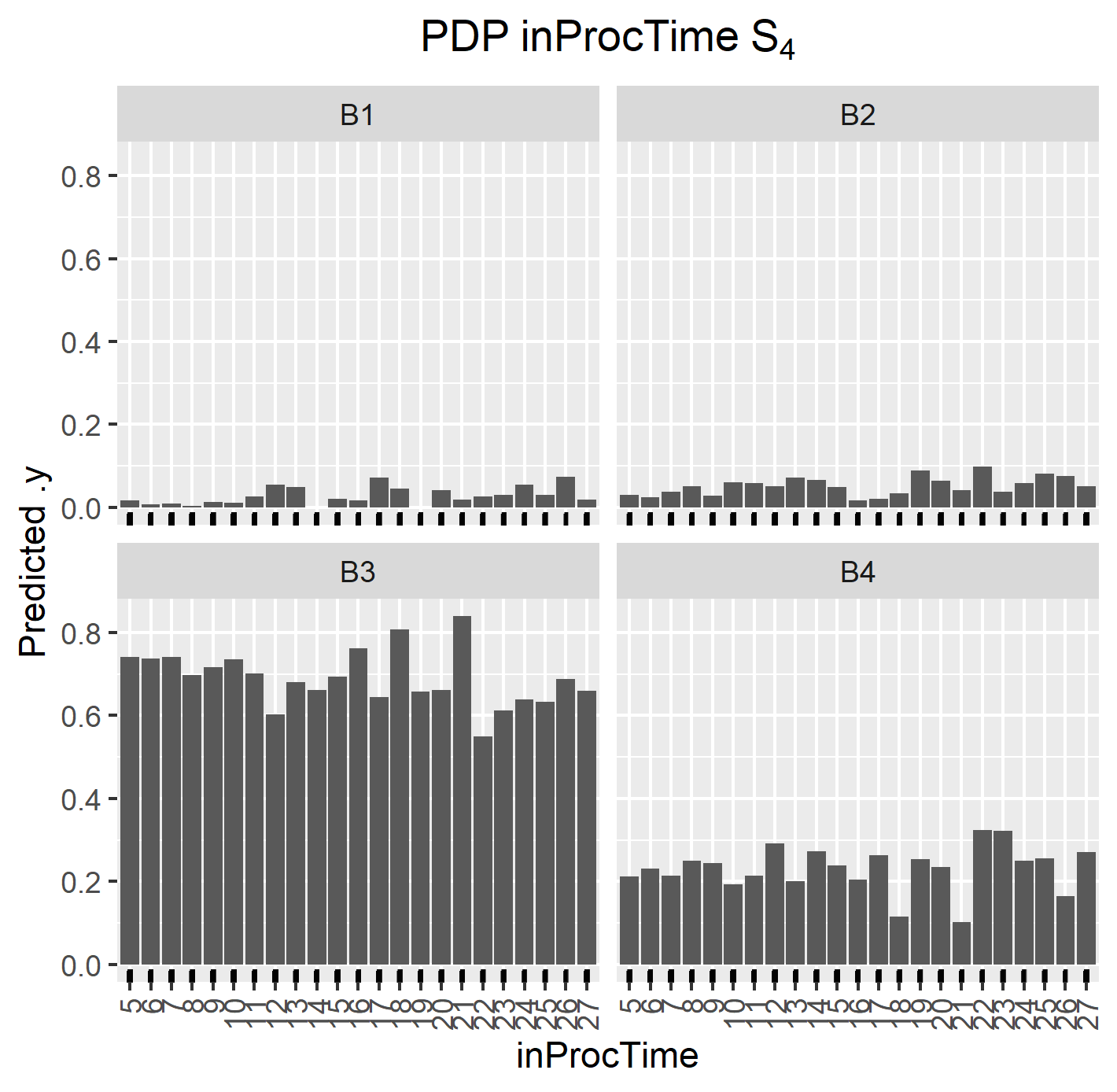}
    \caption{}
    \label{fig:pdp_proctime_s4_pay}
\end{subfigure}
\begin{subfigure}{.55\textwidth}
    \centering
    \includegraphics[width=.85\linewidth]{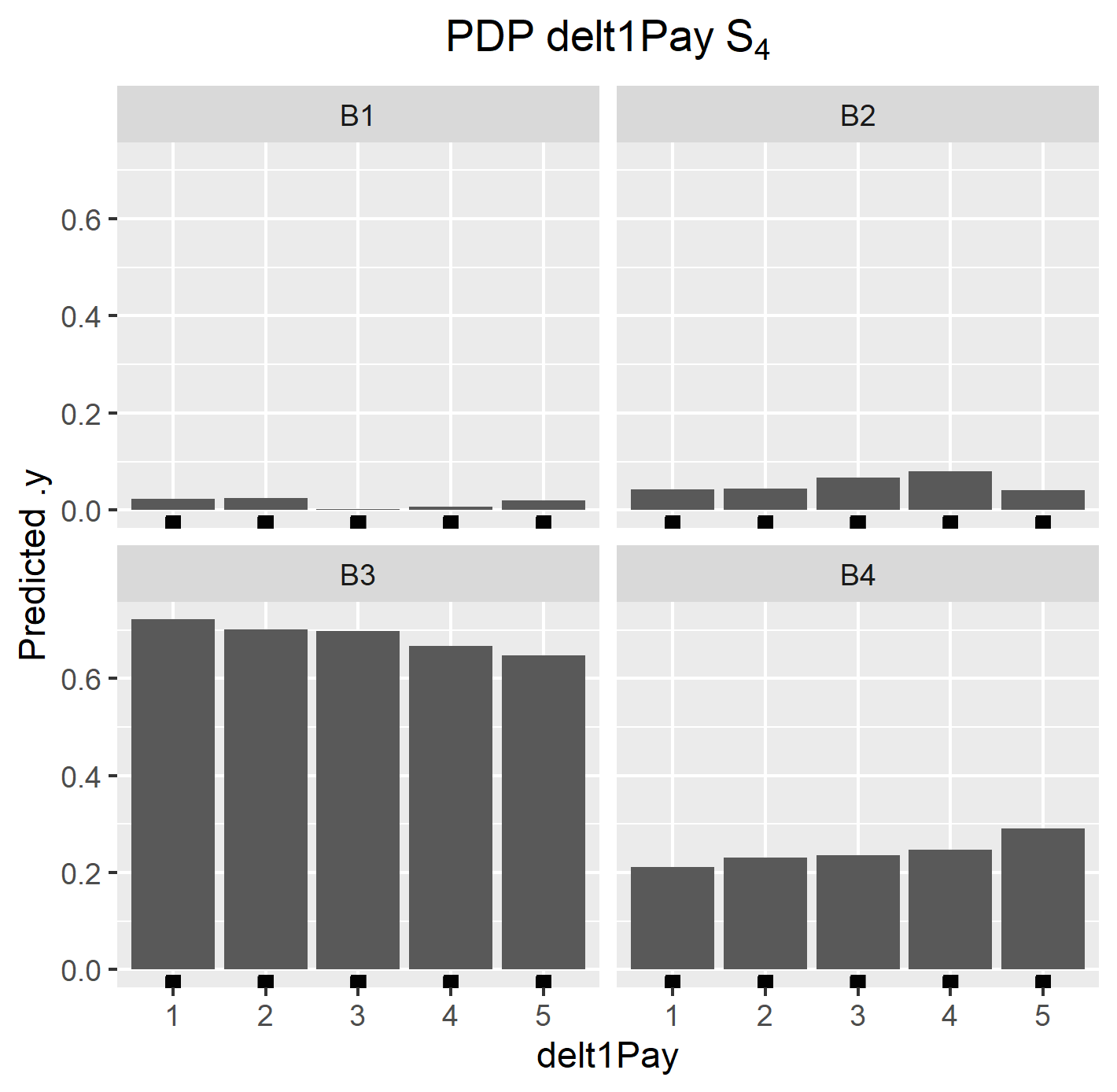}
    \caption{}
    \label{fig:pdpdeltpay_s4_pay}
\end{subfigure}
\newline
\begin{subfigure}{.55\textwidth}
    \centering
    \includegraphics[width=.85\linewidth]{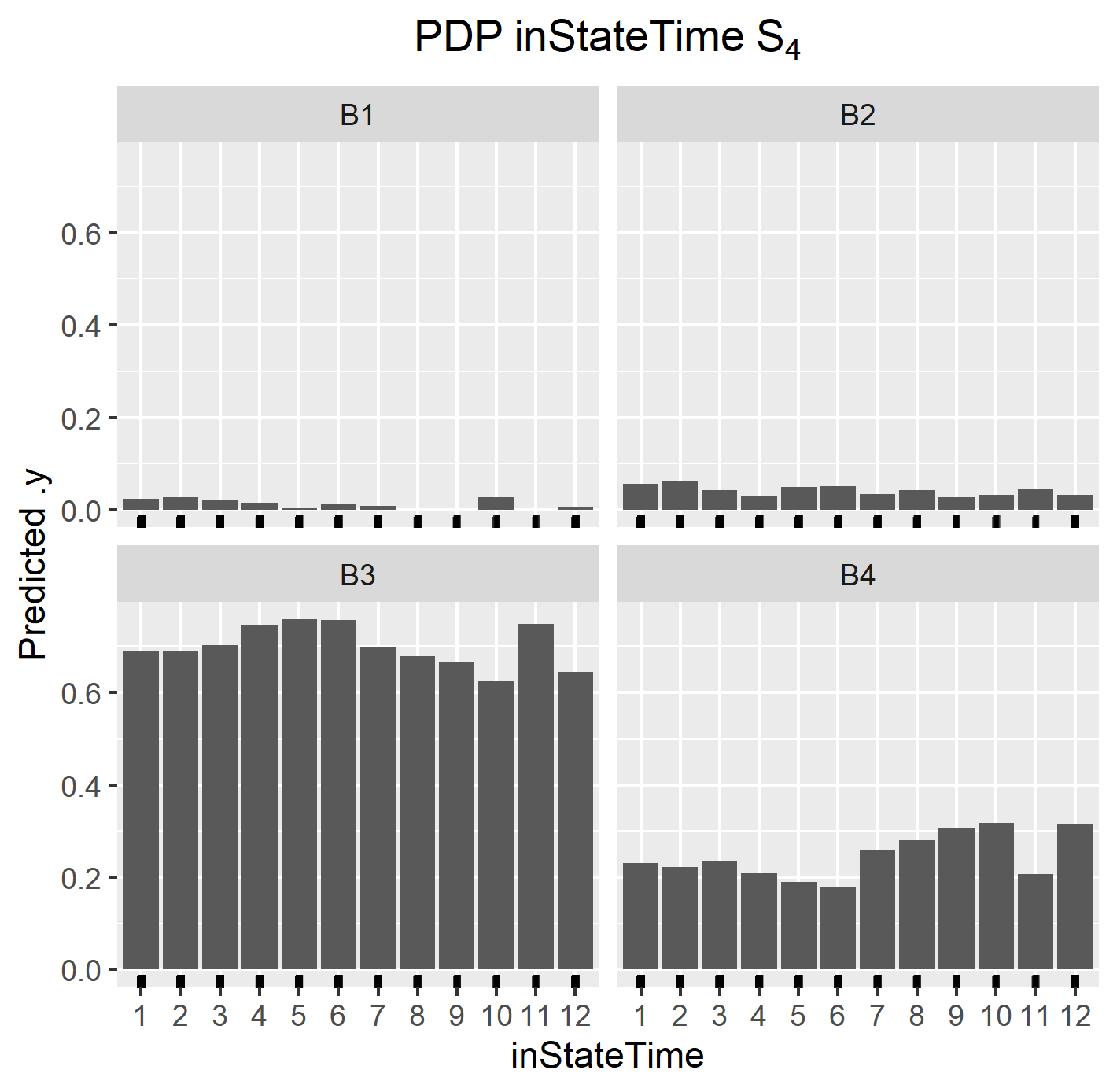}
    \caption{}
    \label{fig:pdp_statetime_s4_pay}
\end{subfigure}
\begin{subfigure}{.55\textwidth}
    \centering
    \includegraphics[width=.85\linewidth]{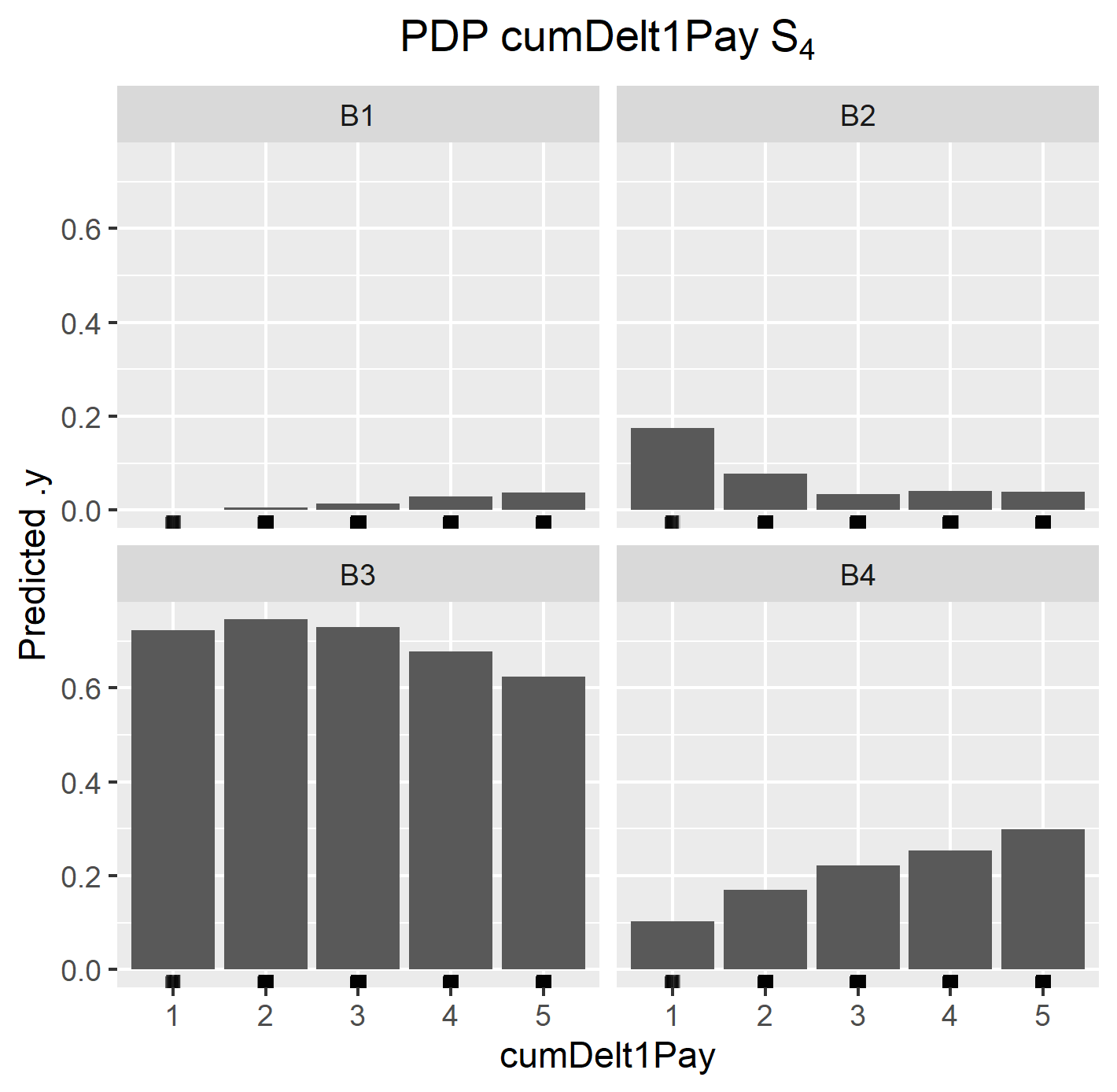}
    \caption{}
    \label{fig:pdp_cumdeltpay_s4_pay}
\end{subfigure}
\caption{Partial dependence plots representing the marginal effect on probabilities to belong to a payment bin of the time spent in the process (a), the previous payment size (b), the time spent in the state (c) and the cumulative previous payment size (d) for claims in $S_{4}$.}
\label{fig:pdp_s4_pay}
\end{figure}

\begin{figure}[!htbp]
\begin{subfigure}{.55\textwidth}
    \centering
    \includegraphics[width=.85\linewidth]{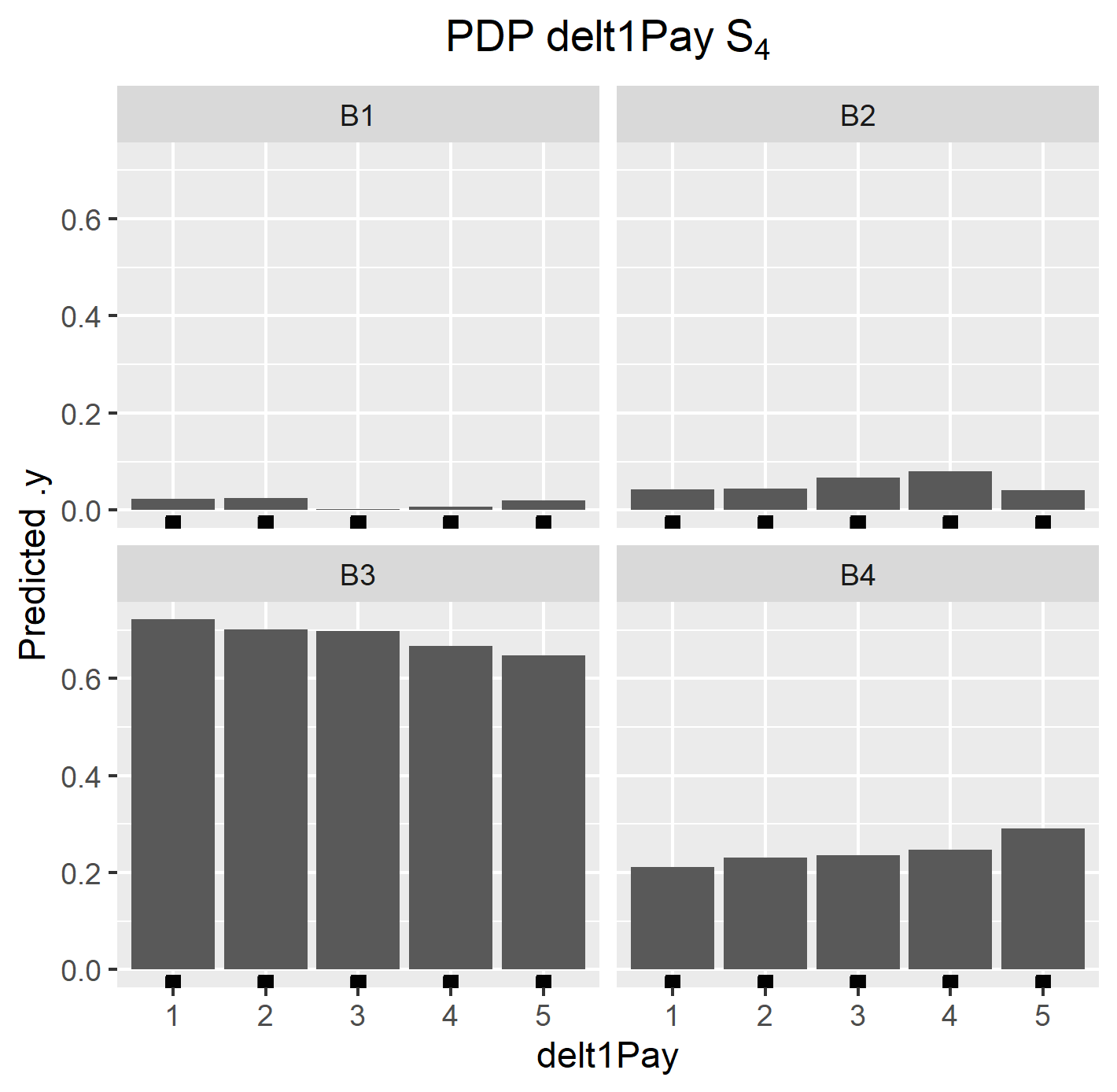}
    \caption{}
    \label{fig:pdp_proctime_s5_pay}
\end{subfigure}
\begin{subfigure}{.55\textwidth}
    \centering
    \includegraphics[width=.85\linewidth]{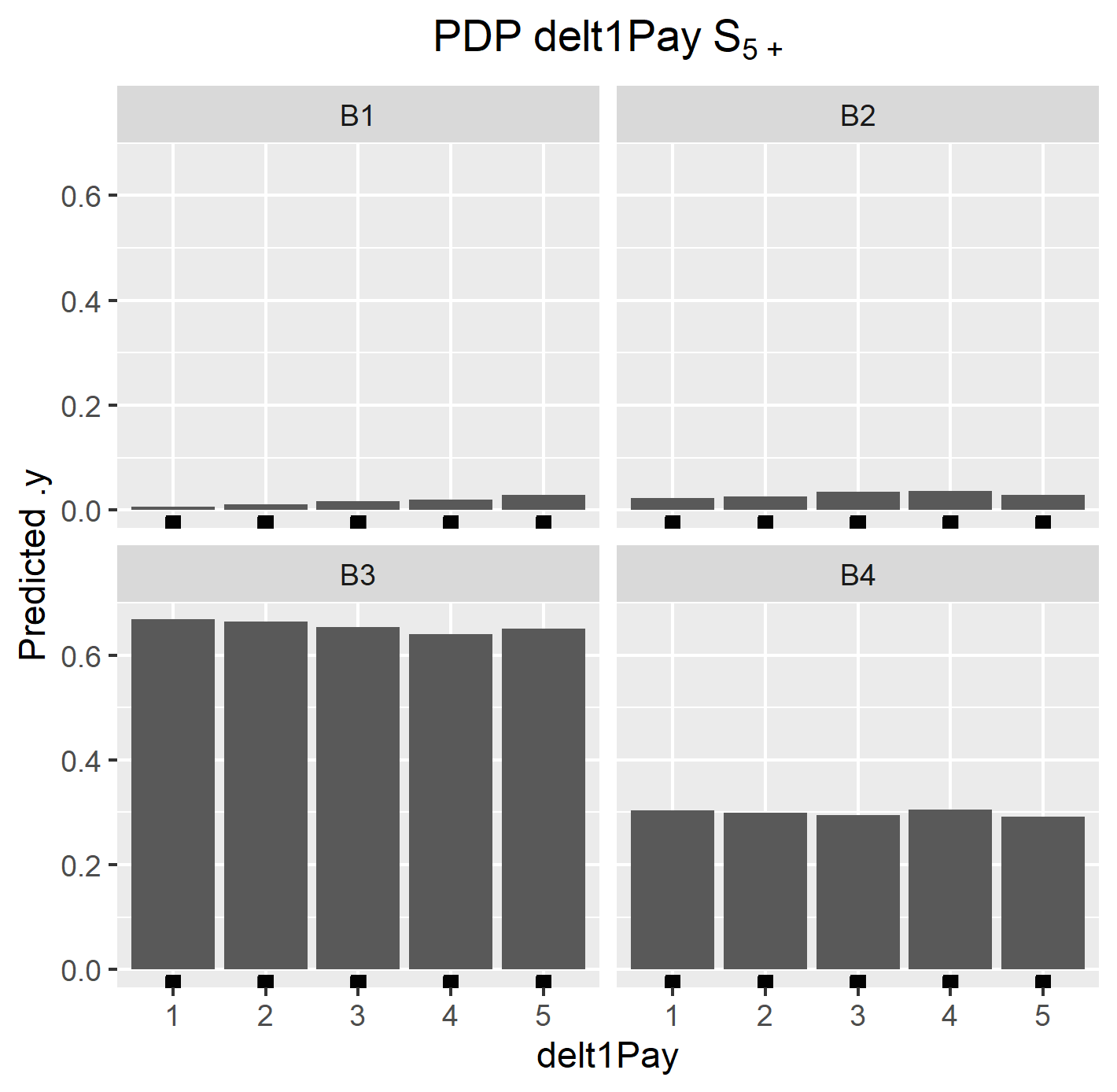}
    \caption{}
    \label{fig:pdpdeltpay_s5_pay}
\end{subfigure}
\newline
\begin{subfigure}{.55\textwidth}
    \centering
    \includegraphics[width=.85\linewidth]{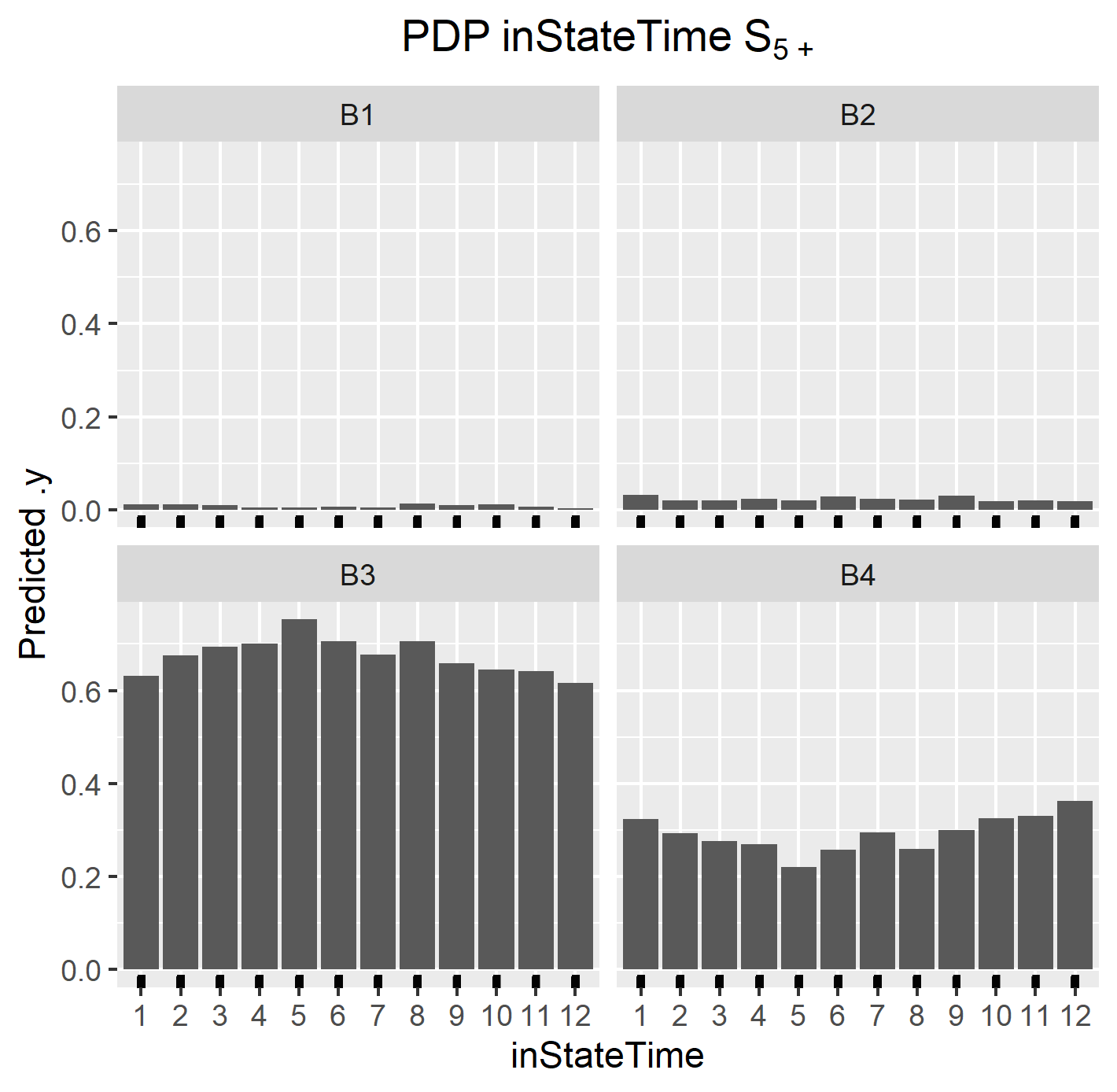}
    \caption{}
    \label{fig:pdp_statetime_s5_pay}
\end{subfigure}
\begin{subfigure}{.55\textwidth}
    \centering
    \includegraphics[width=.85\linewidth]{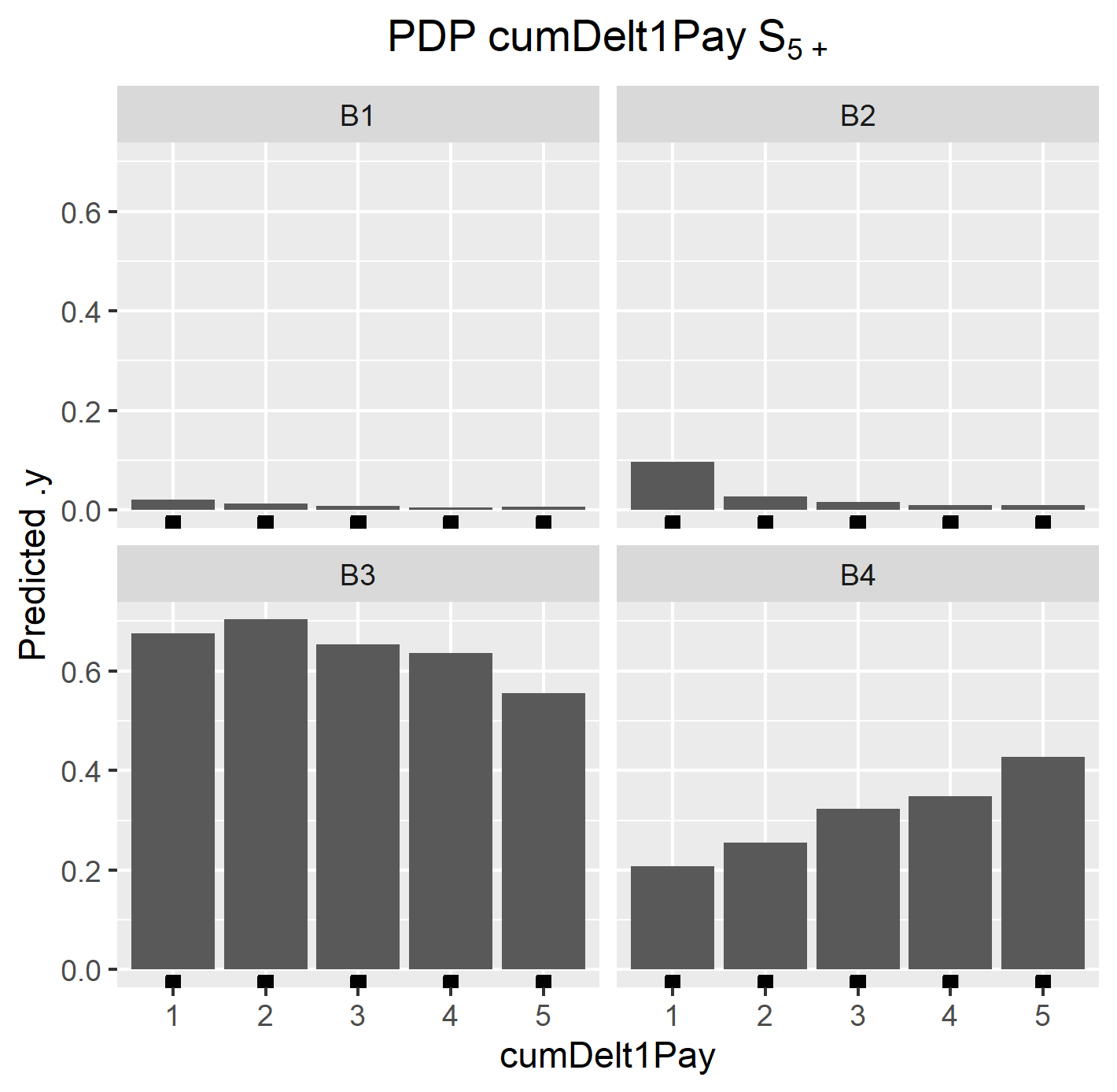}
    \caption{}
    \label{fig:pdp_cumdeltpay_s5_pay}
\end{subfigure}
\caption{Partial dependence plots representing the marginal effect on probabilities to belong to a payment bin of the time spent in the process (a), the previous payment size (b), the time spent in the state (c) and the cumulative previous payment size (d) for claims in $S_{5+}$.}
\label{fig:pdp_s5_pay}
\end{figure}

This section,shows the marginal effect of predictors in the payment models for states $S_{3}$, $S_{4}$ and $S_{5+}$ using partial dependence plots.  From figures \ref{fig:pdp_s3_pay}, \ref{fig:pdp_s4_pay}, \ref{fig:pdp_s5_pay}, we observe similar marginal effects of covariates as in state $S_{2}$.

\clearpage
\bibliography{bibsEAJ}


\end{document}